\documentclass[preprint2]{aastex6}

\newcommand\aastex{AAS\TeX}

\shorttitle{\aastex\ Obscured IRAGNs in COSMOS}
\shortauthors{Chang et al.}

\begin{document}


\title{Infrared Selection of Obscured Active Galactic Nuclei in the COSMOS Field}



\author{Yu-Yen~Chang\altaffilmark{1,2}}
\email{yuyenchang.astro@gmail.com}
\author{Emeric~Le~Floc'h\altaffilmark{1}}
\author{St\'ephanie~Juneau\altaffilmark{1}}
\author{Elisabete~da~Cunha\altaffilmark{3}}
\author{Mara~Salvato\altaffilmark{4}}
\author{Francesca~Civano\altaffilmark{5}}
\author{Stefano~Marchesi\altaffilmark{6}}
\author{Olivier~Ilbert\altaffilmark{7}}
\author{Yoshiki~Toba\altaffilmark{2}}
\author{Chen-Fatt~Lim\altaffilmark{2,8}}
\author{Ji-Jia~Tang\altaffilmark{2,8}}
\author{Wei-Hao~Wang\altaffilmark{2}}
\author{Nicholas~Ferraro\altaffilmark{2,9}}
\author{Megan~C.~Urry\altaffilmark{10}}
\author{Richard~E.~Griffiths\altaffilmark{11}}
\author{Jeyhan~S.~Kartaltepe\altaffilmark{12}}
\altaffiltext{1}{CEA Saclay, DSM/Irfu/Service d'Astrophysique, Orme des Merisiers, F-91191 Gif-sur-Yvette Cedex, France}
\altaffiltext{2}{Academia Sinica Institute of Astronomy and Astrophysics, PO Box 23-141, Taipei 10617, Taiwan}
\altaffiltext{3}{The Australian National University, Mt Stromlo Observatory, Cotter Rd, Weston Creek, ACT 2611, Australia}
\altaffiltext{4}{Max Planck Institut f\"ur Plasma Physik and Excellence Cluster, D-85748 Garching, Germany}
\altaffiltext{5}{Harvard-Smithsonian Center for Astrophysics, 60 Garden Street, Cambridge, MA 02138, USA}
\altaffiltext{6}{Department of Physics \& Astronomy, Clemson University, Clemson, SC 29634, USA}
\altaffiltext{7}{Aix-Marseille Universit\'e, CNRS, LAM (Laboratoire d'Astrophysique de Marseille) UMR 7326, F-13388, Marseille, France}
\altaffiltext{8}{Graduate Institute of Astrophysics, National Taiwan University, No.1 Section 4 Roosevelt Rd., Taipei 10617, Taiwan}
\altaffiltext{9}{Department of Astronomy, University of Virginia, 530 McCormick Rd., Charlottesville, VA 22904, USA}
\altaffiltext{10}{Yale Center for Astronomy and Astrophysics, 260 Whitney Avenue, New Haven, CT 06520, USA}
\altaffiltext{11}{Department of Physics \& Astronomy, University of Hawaii at Hilo, 200 W. Kawili Street, Hilo, HI 96720, USA}
\altaffiltext{12}{School of Physics and Astronomy, Rochester Institute of Technology, 84 Lomb Memorial Drive, Rochester, NY 14623, USA}

\begin{abstract}
We present a study of the connection between black hole accretion, star formation, and galaxy morphology at $z\leq2.5$. We focus on  active galactic nuclei (AGNs) selected by their mid-IR power-law emission. By fitting optical to far-IR photometry with state-of-the-art spectral energy distribution (SED) techniques, we derive stellar masses, star formation rates, dust properties, and AGN contributions in galaxies over the whole COSMOS field. We find that obscured AGNs lie within or slightly above the star-forming sequence. We confirm our previous finding about compact host galaxies of obscured AGNs at $z\sim 1$, and find that galaxies with 20-50\% AGN contributions tend to have smaller sizes, by $\sim$25-50\%, compared to galaxies without AGNs. Furthermore, we find that a high merger fraction of up to 0.5 is appropriate for the most luminous ($\log (L_{IR}/L_\odot)\sim 12.5$) AGN hosts and non-AGN galaxies, but not for the whole obscured AGN sample. Moreover, merger fraction depends on the total and star-forming infrared luminosity, rather than the decomposed AGN infrared luminosity. Our results suggest that major mergers are not the main driver of AGN activity, and therefore obscured AGNs might be triggered by internal mechanisms, such as secular processes, disk instabilities, and compaction in a particular evolutionary stage. We make the SED modeling results publicly available.
\end{abstract}

\keywords{galaxies: active --- galaxies: star formation --- infrared: galaxies}



\section{Introduction}
\label{sec1}

One of the most important issues in studies of galaxy evolution is the connection between the formation of stars in galaxies and the fueling of nuclear activity. Star formation arises  from the collapse of cold molecular clouds while active galactic nucleus (AGN) activity is caused by the accretion of matter onto super-massive black holes. Though star formation and AGN activity occur on very different scales, AGNs could  be involved in the process of quenching star formation.  Previous results revealed that X-ray and optically selected AGNs reside in galaxies harboring sustained activity of star formation \citep{2012ApJ...760L..15H,2012A&A...540A.109S,2012MNRAS.419...95M,2013ApJ...764..176J,2013ApJ...771...63R,2015MNRAS.453..591S,2017arXiv170700254M}. Furthermore, although  high luminosity AGNs seem to be connected to violent events and may reside in  galaxies more often involved in major mergers \citep{2012ApJ...757...23K,2012ApJ...758L..39T,2016ApJ...822L..32F}, the majority of X-ray selected AGNs appear to live mostly in disk-dominated isolated systems \citep[e.g.,][]{2011ApJ...726...57C,2012ApJ...744..148K}, implying that the bulk of super-massive black hole accretion is likely driven by internal processes and not by major mergers.

While the results reported above have been mostly drawn from populations of unobscured AGNs according to X-ray observations, it has recently been suggested that obscured AGNs, which are not well sampled by X-ray surveys, may reside in different environments. For instance, Compton-thick AGNs hidden by extreme column densities ($N_H>10^{24}$ cm$^{-2}$) tend to show a fraction of disturbed morphologies increasing with obscuration \citep{2015ApJ...814..104K,2015A&A...578A.120L}. This suggests that Compton-thick AGNs can be in a phase of obscured super-massive black hole growth following a merger event. Current samples of obscured AGNs are mostly limited to moderate obscuration by X-ray detections, which are selected from gas column density measurements \citep{2011A&A...534A.110L,2013ApJ...777...86L,2014MNRAS.443.2077B,2014MNRAS.437.3550M,2015MNRAS.446.2394B}. 
Compton-thick AGNs are found by deep X-ray observations \citep{2014MNRAS.443.1999B}, but the limited field sizes also limit detections.

Another approach to explore obscured AGNs is from infrared (IR) observations, which can provide complementary samples compared to X-ray data \citep{2010ApJ...713..503C,2016MNRAS.456.2105D}. 
IR-selected AGNs can be identified from their mid-infrared (MIR) power-law emission using Spitzer/IRAC photometry \citep{2004ApJS..154..166L,2005ApJ...631..163S,2012ApJ...748..142D}. This technique can be applied up to $z\sim2.5$, or even higher redshifts \citep{2012ApJ...754..120M}.
Since this approach only relies on the detection of the dust continuum emission in the infrared,  it can unveil AGNs suffering very strong obscuration in the X-rays, including AGNs even missed by the deepest X-ray observations obtained to date. 
A proxy for obscuration can be inferred using the infrared to X-ray luminosity ratio (or lower limit if there is no X-ray detection), where the AGN infrared luminosity can be estimated using state-of-the-art spectral energy distribution (SED) fitting techniques as discussed in \citet{2011MNRAS.414.1082M,2012ApJ...759..139K,2013ApJ...763..123K,2014ApJ...784...83D,2015ApJ...803..109H,2015ApJ...814....9K}. Obscured AGN can then be selected as sources that are under-luminous in the X-rays relative to their bolometric luminosity inferred from the infrared \citep{2014A&A...562A.144M}.

The morphologies of host galaxies have been characterized using two-dimentional surface brightness modeling and nonparametric morphology methods, and visual classification. 
\citet{2000AJ....120.1739V} found that there are no significant differences between their control and active Seyfert host galaxies in terms of light asymmetries. 
\citet{2005ApJ...627L..97G} showed that $\sim$ 150 X-ray AGNs have bulge-dominated morphologies from their concentration and asymmetry up to $z\sim 1.3$. \citet{2007ApJ...660L..19P} found that 94 X-ray selected AGNs with $0.2<z<1.2$ mostly reside in E/S0/Sa galaxies, while infrared selected AGNs showed no clear preference for host morphology from nonparametric measures. \citet{2008MNRAS.385.2049G} studied Compton-thick AGN with $0.4<z<0.9$ and suggested that a large fraction of post-starbursts and red cloud galaxies have evidence for at least moderate levels of AGN obscuration. In order to avoid AGN contamination, \citet{2009ApJ...691..705G} decomposed $\sim$400 X-ray selected AGNs into AGN point source and galaxy light at $0.3<z<1.0$, and found that X-ray-selected AGN host morphologies span a substantial range that peaks between those of bulge-dominated and disk-dominated systems. \citet{2013A&A...549A..46B} also decomposed $\sim$200 type-1 X-ray AGN images at $z\sim0.7$ into nucleus and host components, and found that active and inactive galaxies show similar distributions in nonparametric measures space after decomposition. \citet{2014MNRAS.439.3342V} added simulated AGNs to a stellar mass-matched control sample at $0.5<z<0.8$ to show that X-ray selected AGN hosts and control sample galaxies have comparable asymmetries, S\'ersic indices and ellipticities. 
At higher redshift, \citet{2011ApJ...743L..37S} considered point source components for 57 X-ray AGN hosts with $1.25<z<2.67$ and found that half of the sample are disk-dominated. 
\citet{2012ApJ...761...75S} also chose one or two components models according to their fitting residuals and suggested the majority of X-ray AGN host galaxies are disk-dominated. \citet{2014ApJ...784L...9F} analyzed 35 X-ray selected AGNs at $z\sim2$ based on point-source-subtracted images and suggested that all the distributions of morphological parameters of AGN hosts are consistent with their control sample. \citet{2015A&A...573A..85R} showed that X-ray AGN hosts are slightly diskier and more disturbed than massive inactive galaxies at $z\sim1$, and show a red central light enhancement at $z\sim2$. They also demonstrated that the central excess is likely due to the bulge and the two-component model may underestimate S\'ersic index, driving the main stellar components towards disk-dominated profiles. \citet{2016MNRAS.458.2391B} found that the structural parameters of X-ray selected AGN hosts are indistinguishable from the general galaxy population, but have significantly higher bulge fractions beyond $z\sim1.5$. \citet{2017MNRAS.466..812V} created a control sample of mock AGN to compare with 20 optically and X-ray selected luminous AGNs at $z\sim0.6$, and found no enhanced disturbance in the AGNs relative to the control sample. 
According to these previous studies, the relation between morphology and AGN activity is till unclear, largely because it depends on sample selection, redshift, and techniques used to measure morphology and AGN activity.

Recently, we explored the optical-light radial profile from a IR-selected sample and found that obscured AGN hosts at $z\sim 1$ are more compact than a control sample of star-forming galaxies at the same stellar mass and redshift bins \citep{2017MNRAS.466L.103C}. 
Internal secular processes over long time scales could play a dominant role, but the potential impact of violent disk instabilities (VDI)  in gas-rich disks has also been discussed recently in the literature. Indeed, these VDI are believed to result in  highly clumpy galaxies as frequently observed at z$>$1. The possible migration of their giant star-forming clumps  toward their central region could lead to the quick formation of a dense stellar component \citep{2009Natur.457..451D,2010MNRAS.404.2151C}, which may also be accompanied by a phase of gas compaction \citep{2016MNRAS.457.2790T}, producing blue nuggets \citep{2015MNRAS.450.2327Z}, and  potentially affecting the fueling of central AGNs \citep{2015MNRAS.452.1502D}.

In this paper, we extend our previous study to a general comparison with star formation, AGN fraction, and obscuration in a complete MIR selected sample.
We compare physical properties of AGN hosts with normal star-forming galaxies by using both spectral energy distribution and morphology fitting techniques.
Following the morphology studies in \citet{2017MNRAS.466L.103C}, we quantify the structural parameters of obscured IR-AGN hosts, and present their visual classification result. 
The structure of this paper is as follows.
In  \S~\ref{sec2} we describe the data and our selection.  
We present our SED fitting results as well as the comparison between normal star-forming and AGN host galaxies in \S~\ref{sec3}. 
We present a morphology analysis of AGN hosts, based on two-dimensional fitting, nonparametric method, and visual classification in \S~\ref{sec4}.
The discussion is presented in \S~\ref{sec5}, followed by summary in \S~\ref{sec6}.
We use AB magnitudes and the cosmological parameters
($\Omega_M$,$\Omega_\Lambda$,$h$)=(0.30,0.70,0.70) and adopt the
\citet{2003PASP..115..763C} stellar initial mass function.

\section{Data and Sample Selection}
\label{sec2}

\begin{figure}
\centering
\includegraphics[width=1.00\columnwidth]{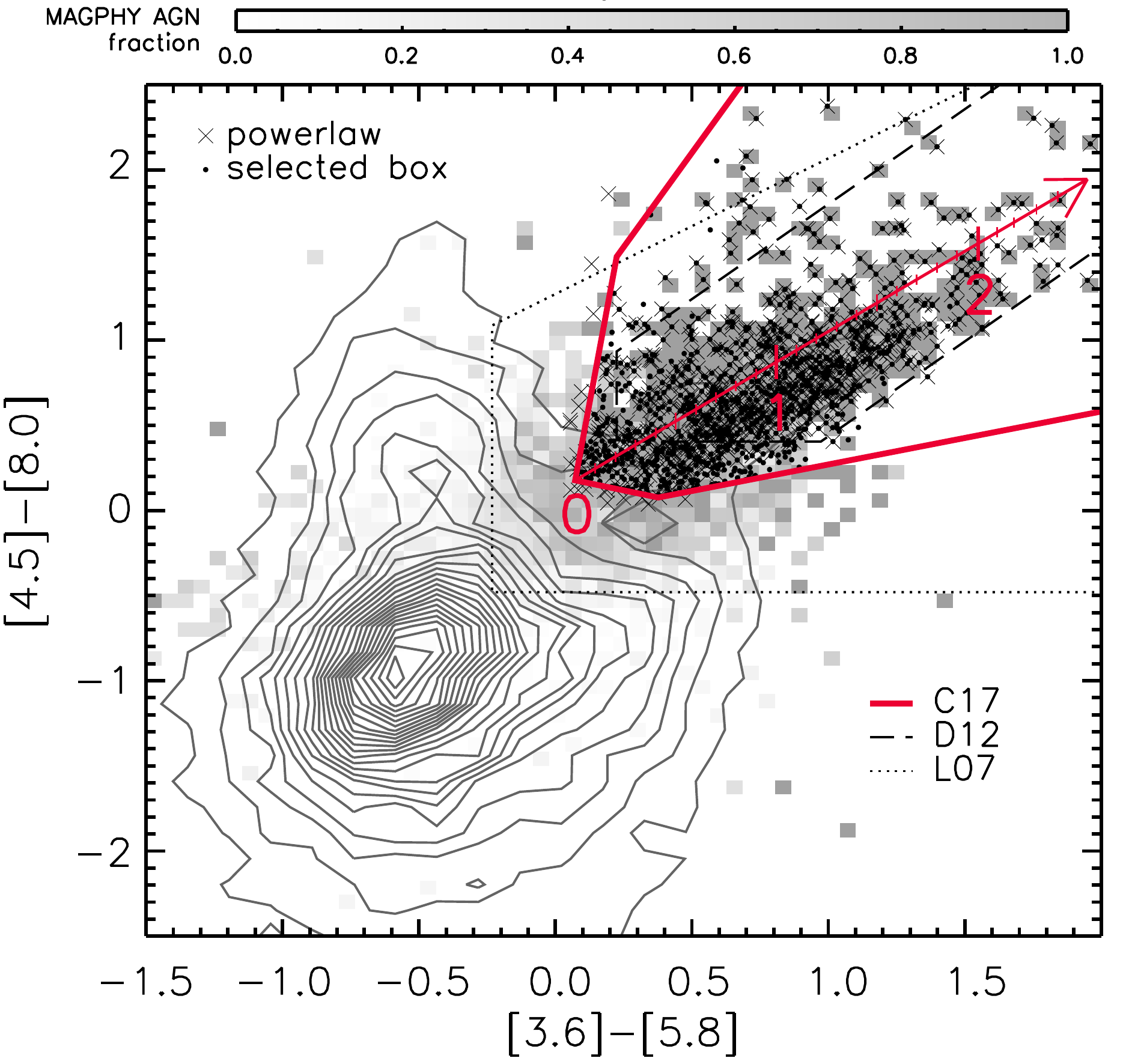} 
\includegraphics[width=1.00\columnwidth]{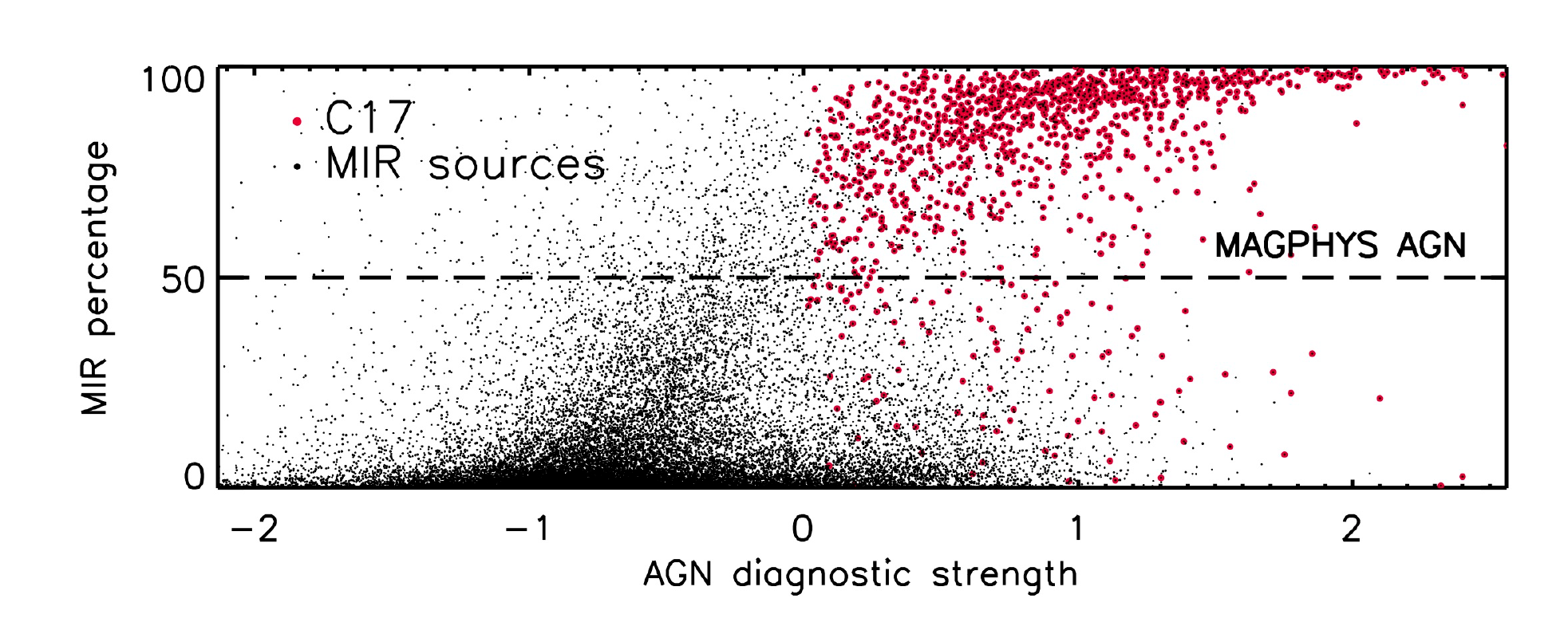} 
\caption[]{Upper: IR-AGN selection in the color-color plot at $z\leq2.5$. The cross symbols represent galaxies with monotonically rising MIR SEDs. The black dots are our IR-selected AGN according to the red box. The contour shows the major population. The gray coding shows the AGN fractions of dominant MIR AGNs according to SED fitting.  Lower: The correlation between AGN diagnostic strength (red arrow in the upper plot) and the MIR contribution of AGNs. We adopt the red box in this paper. This technique is similar to previous color-color selections (L07, D12), and consistent with power law or MAGPHYS AGNs.}
\label{cosp_irac}
\end{figure}

\subsection{MIR galaxies and IR AGNs}
\label{sec2_1}

Our sample is based on a MIR 24 $\mu$m selection \citep[$S_{24 \mu m}$ $\gtrsim$80 $\mu$Jy;][]{2009ApJ...703..222L} and contains 36,670 MIR galaxies in the Cosmic Evolution Survey \citep[COSMOS,][]{2007ApJS..172....1S} field. 
We match all the MIR galaxies with the COSMOS2015 catalog \citep{2013A&A...556A..55I,2016ApJS..224...24L} to get photometry from optical to far-infrared (FIR). 
There are 30,212 redshifts, which are firstly taken from the Chandra Legacy Survey \citep[6.66\%;][]{2011ApJ...742...61S,2016ApJ...817...34M,2016ApJ...819...62C}, then available spectroscopic redshifts (30.14\%), otherwise the COSMOS2015 photometric redshifts \citep[63.19\%;][]{2016ApJS..224...24L}. 
We found that the fraction of outliers ($\eta$=2.9\%) and accuracy ($\sigma_{NAMD}$=0.002) are reliable and good \citep[see][for more details]{2009ApJ...690.1236I,2011ApJ...742...61S} by comparing spectroscopic and photometric redshifts.

We selected IR power-law AGN candidates as sources according to a color-color selection, which was first introduced by \citet{2004ApJS..154..166L}. 
In Figure\,~\ref{cosp_irac}, the cross symbols show objects with IRAC F$_\nu$  flux densities monotonically rising from 3.6 to 8$\mu$m. We selected most ($>$99\%) of these power-law AGNs by avoiding contamination from normal galaxies and defined a box\footnote{We adopted the Classification and Regression Trees method \citep{cart84} by scikit-learn \citep{scikit-learn}, and simplified the decision surface according to the contour plot. As a result, the mean accuracy score (fraction of correct predictions computed by scikit-learn) is very high ($\sim$0.99).}:
\begin{eqnarray}
y<    2.22 \times x+1.01 \\
y<    8.67 \times x-0.28 \\
y> -3.33 \times x+0.17 \\
y> 0.31 \times x-0.06
\end{eqnarray}
where $x=m_{5.8\mu m}-m_{3.6\mu m}$, $y=m_{8\mu m}-m_{4.5\mu m}$, and $m_{x\mu m}$ is in AB magnitude. 
There are 1,085 infrared-selected AGNs (IR-AGNs) inside the box at $z\leq2.5$ as shown in Figure\,~\ref{cosp_irac}. 
670 of these 1,085 IR-AGNs are in the Chandra Legacy Survey catalog \citep{2016ApJ...817...34M,2016ApJ...819...62C}.
For the remaining 460 IR-AGNs, spectroscopic redshifts are available for 148 of them, and photometric redshifts are adopted for 267 of them.
Our selection is comparable to previous selections (\citet{2004ApJS..154..166L}, L07; \citet{2012ApJ...748..142D}, D12) and provides an up-to-date box according to the latest IRAC measurements. 
Here we define `AGN diagnostic strength' along the red arrow in the upper panel of Figure\,~\ref{cosp_irac}. The values of AGN diagnostic strength are calculated by the projected value along the peak population of the selected box. The scales are labeled in the upper panel and shown in the lower panel of Figure\,~\ref{cosp_irac}.
The gray scale shows that the AGN fractions of dominant MIR AGNs (MAGPHYS AGN; AGN contribution of MIR luminosity $>$50\% MIR from SED fitting as shown in the lower panel in Figure\,~\ref{cosp_irac}; see Section \ref{sec2_2} for more details about SED fitting) agree with our selection. 
The lower plot of Figure\,~\ref{cosp_irac} shows a good correlation with the MIR percentage of the AGN contribution estimated with SED fitting. 
We show the sample size with a less strict selection by L04, a stricter selection by D12, and the X-ray selection from Chandra Legacy Survey with $L_X$(2-10 keV)$>10^{42}$ ergs/s in Table~\ref{tab_1}. There are 631 AGNs selected by both X-ray ($\sim$41\%) and our infrared ($\sim$58\%) criterion.
In order to have a simple and consistent selection, we adopt the color-color selection in this paper. An updated technique by power law or MAGPHYS AGN will be considered in future works.

\begin{table}
\centering
\caption[]{Sample size of AGN host galaxies of X-ray selected AGNs ($L_X$(2-10 keV)$>10^{42}$ ergs/s), MAGPHYS SED AGNs (AGN contribution of MIR luminosity $>$50\% MIR from SED fitting), and MIR selected AGNs (this paper, D12, and L07). The columns represent the sample size of all AGN host and obscured ($L_{IR,AGN}/L_{X,AGN} > 20$) IR-AGN host galaxies at $z\leq2.5$.}
\label{tab_1}
\begin{tabular}{ccccccc}
\hline
\hline 
selection & X-ray & SED & C17 & D12 & L07 \\ 
\hline
All & 1538 & 2288 & 1085 & 656 & 3340 \\
obscured  & 370 & 620 & 469 & 342 & 612 \\
\hline 
\end{tabular} 
\end{table}

\subsection{SED fitting}
\label{sec2_2}

\begin{figure}[ht]
\centering
\includegraphics[width=1.00\columnwidth]{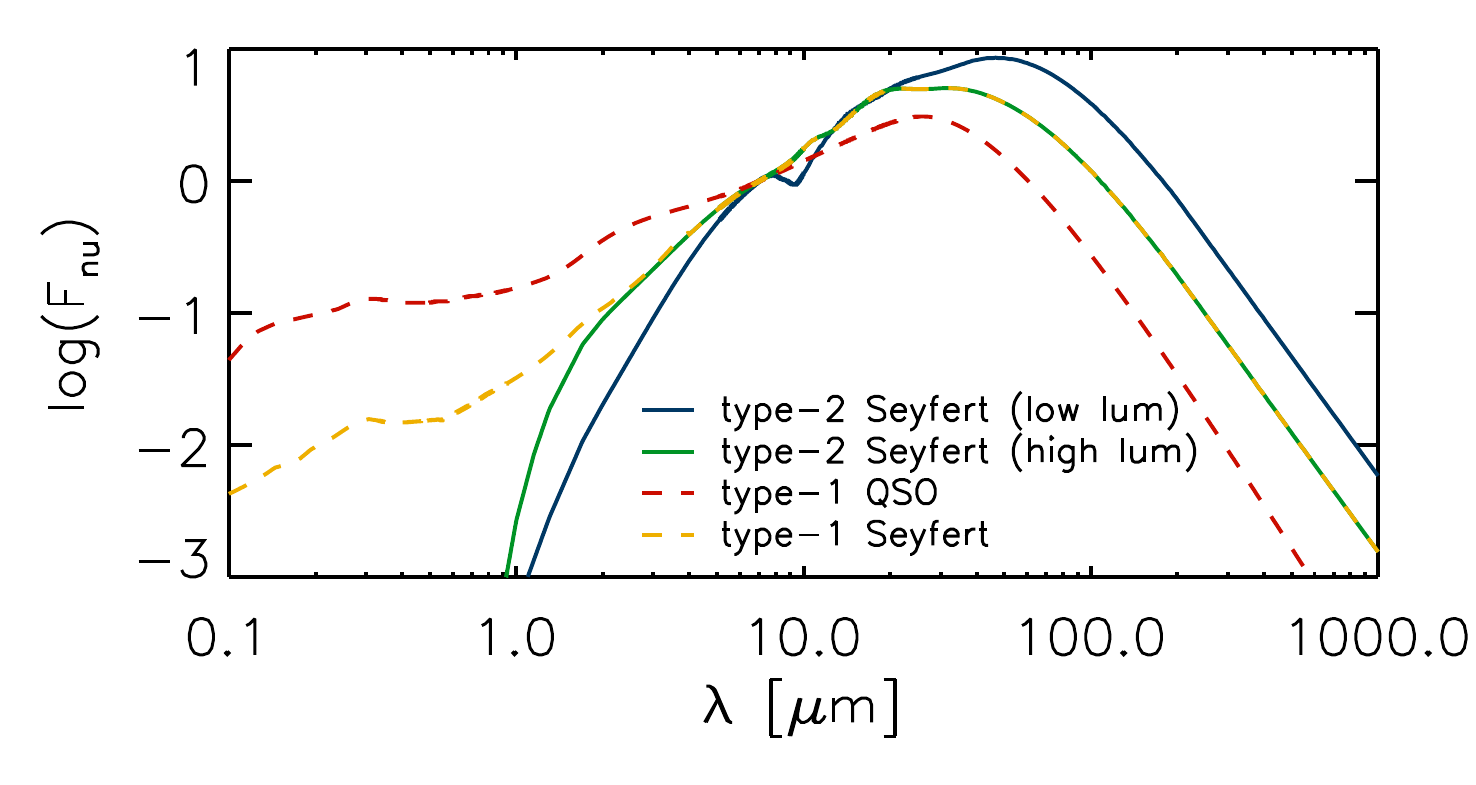} 
\caption[]{Four AGN empirical templates adopted in the MAGPHYS+AGN SED fitting: low luminosity type-2 Seyfert \citep[template=1,][]{2011MNRAS.414.1082M}, high luminosity type-2 Seyfert \citep[template=2,][]{2011MNRAS.414.1082M}, type-1 QSO \citep[template=4,][]{2006ApJS..166..470R,2010MNRAS.402..724P}, tyep-1 Seyfert \citep[template=5,][]{2007ApJ...663...81P}.}
\label{cosp_template}
\end{figure}
\begin{figure*}[ht]
\centering
\includegraphics[width=0.8\textwidth]{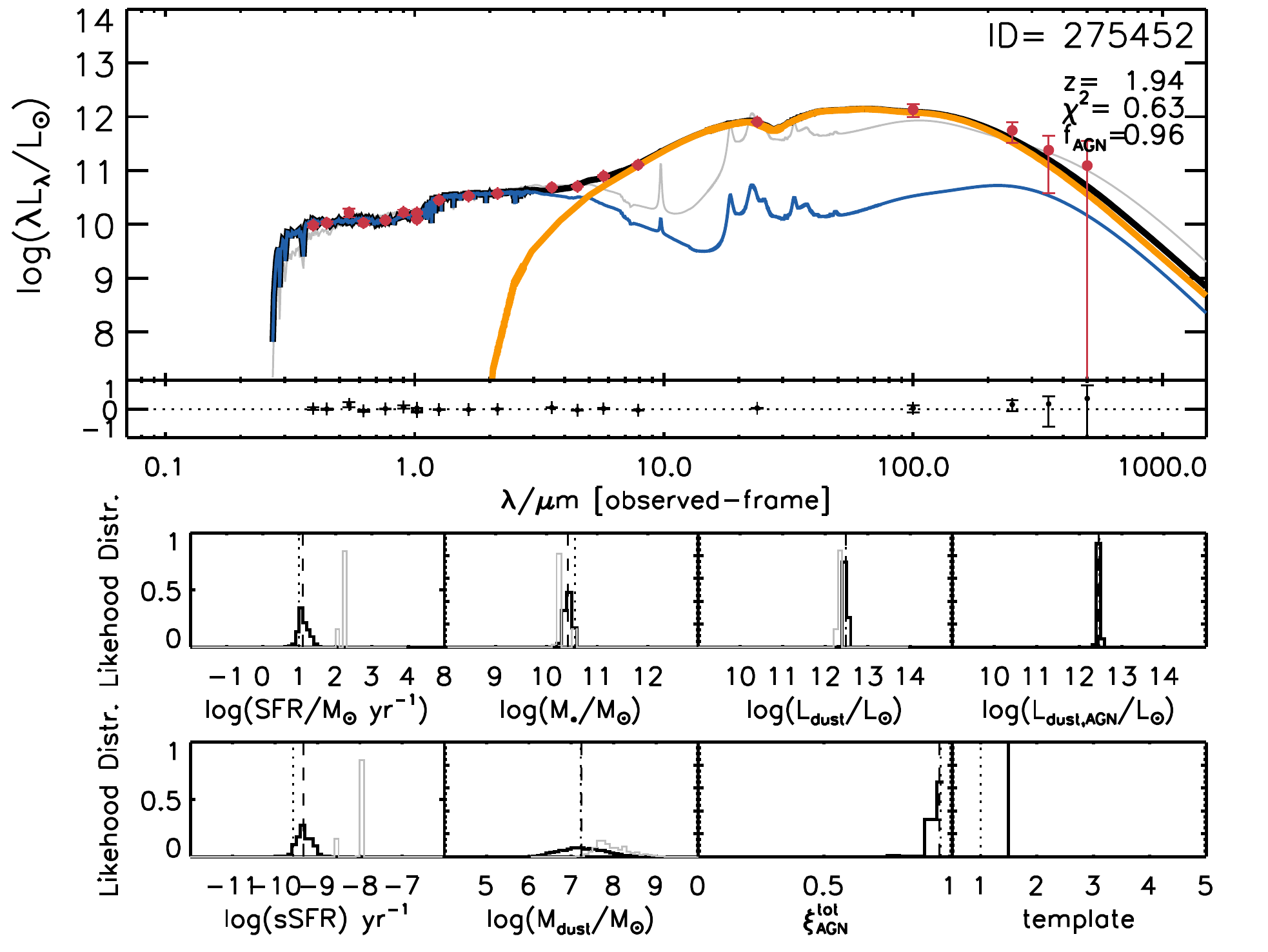} 
\includegraphics[width=0.8\textwidth]{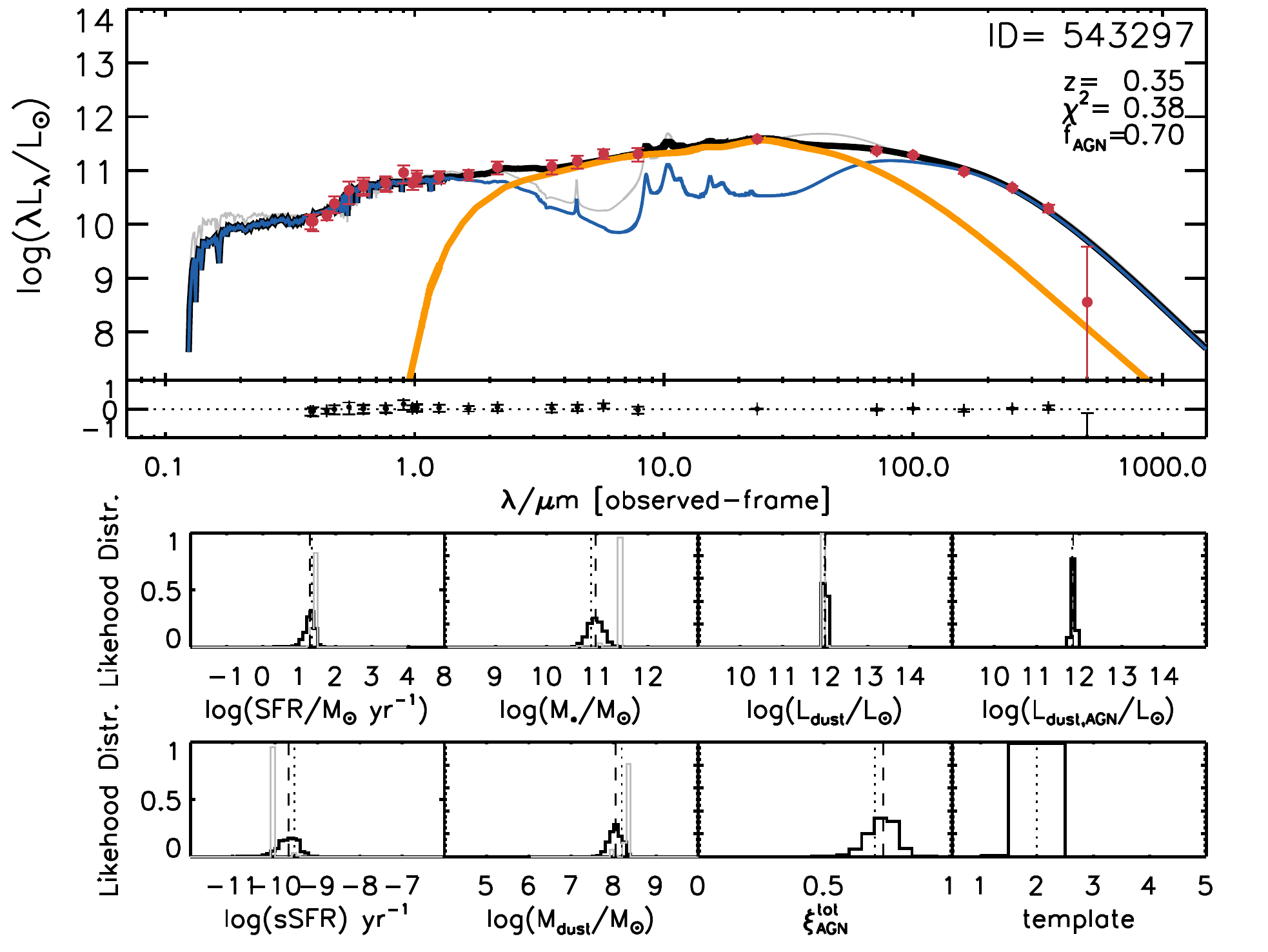} 
\caption[]{SED fitting with AGN component (orange line). The red points are the photometry and the black lines are the best-fitting model. The orange line is the AGN component and blue line is the star formation component. The residuals and histogram of the physical parameters are also shown for the three models. The dash lines are the median values and the dotted lines are the best-fitting values. They gray lines show SED fitting results without any AGN component. As described in Figure\,~\ref{cosp_template}, the numbers of the templates are: 1=low luminosity type-2 Seyfert; 2=high luminosity type-2 Seyfert; 4=type-1 QSO; 5=type-1 Seyfert.}
\label{cosp_sed_1}
\end{figure*}
\begin{figure*}[ht]
\centering
\includegraphics[width=0.8\textwidth]{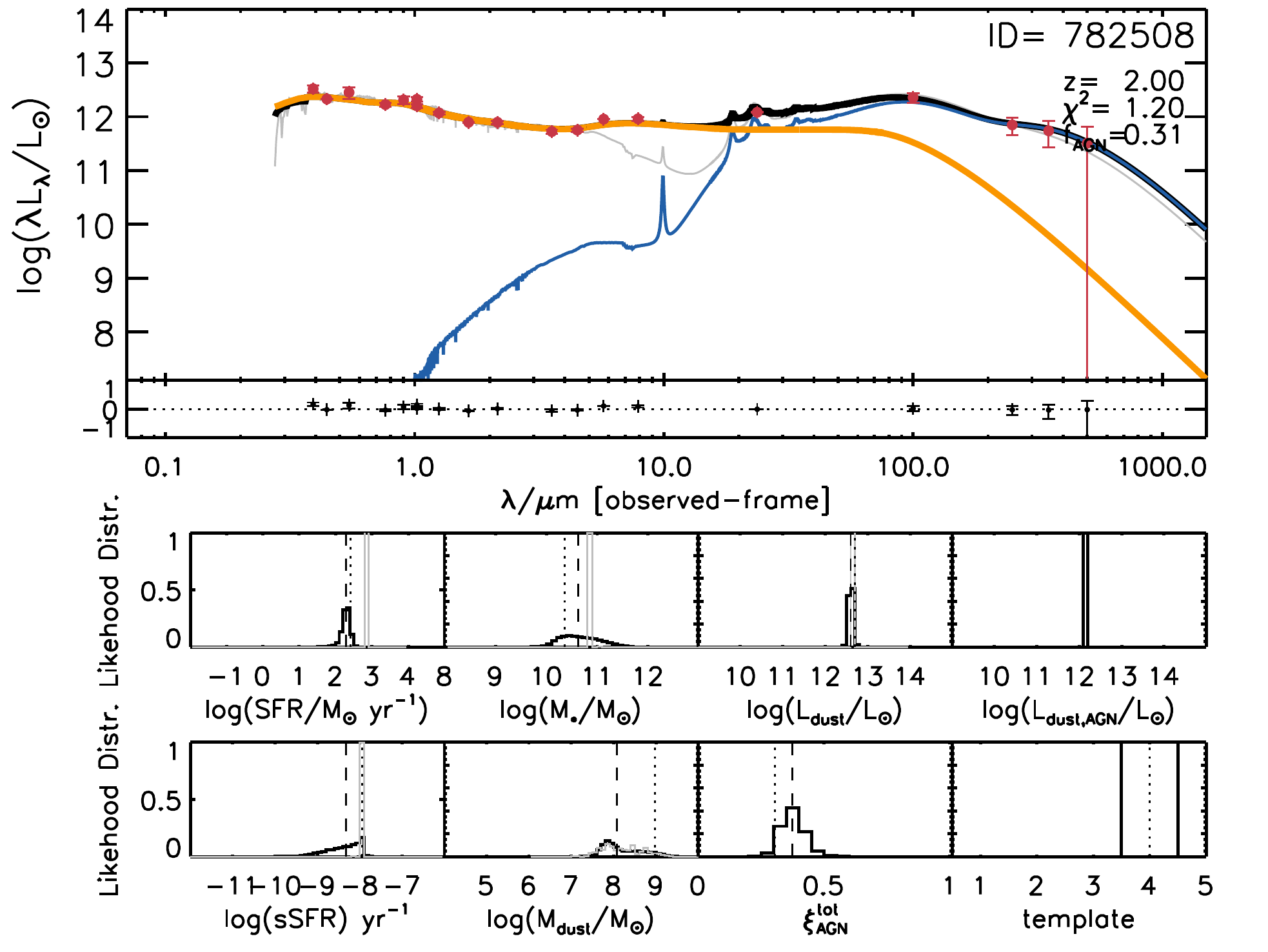} 
\includegraphics[width=0.8\textwidth]{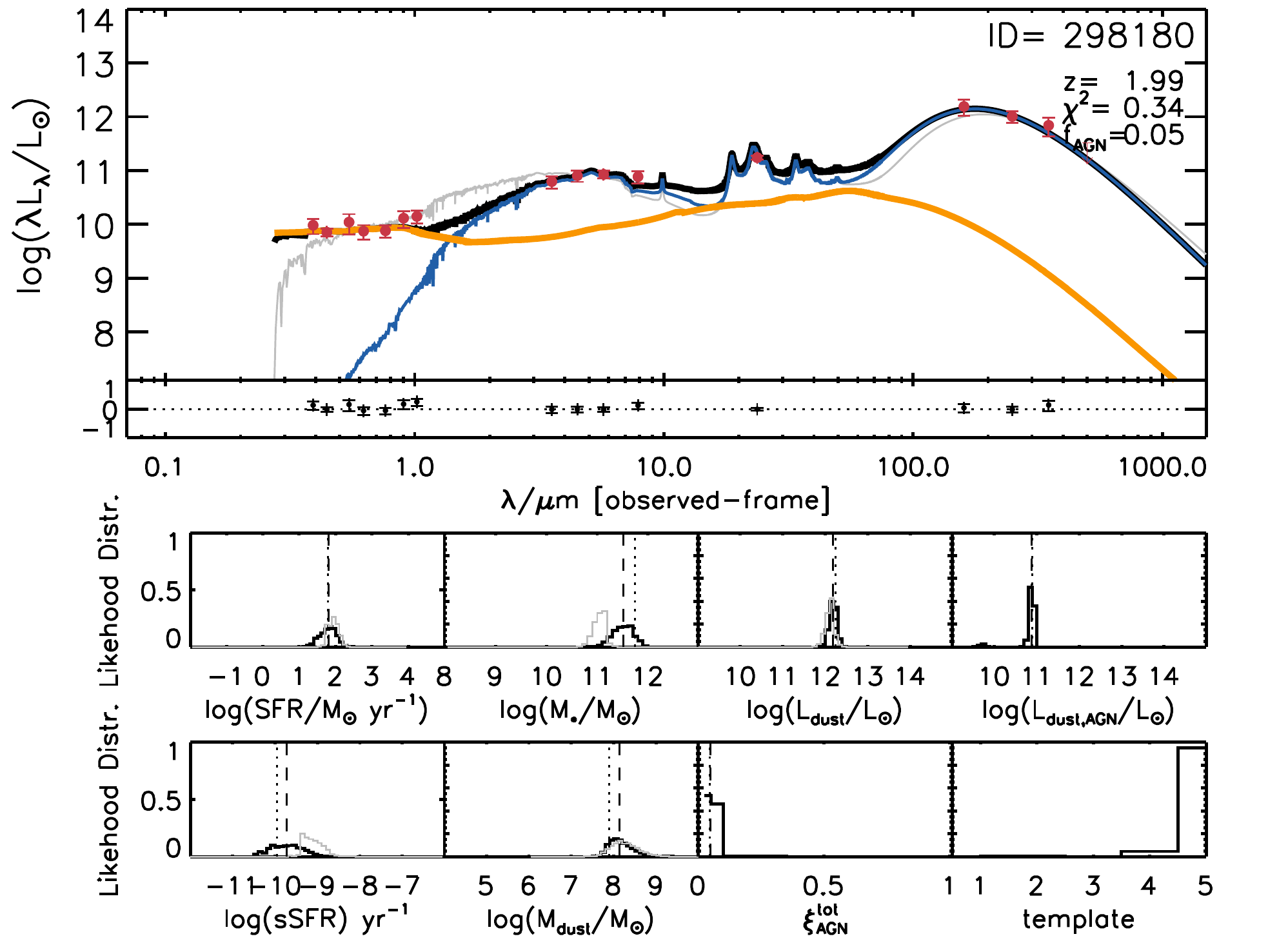} 
\caption[]{SED fitting with AGN component (orange line). The red points are the photometry and the black lines are the best-fitting model. The orange line is the AGN component and blue line is the star formation component. The residuals and histogram of the physical parameters are also shown for the three models. The dash lines are the median values and the dotted lines are the best-fitting values. They gray lines show SED fitting results without any AGN component. As described in Figure\,~\ref{cosp_template}, the numbers of the templates are: 1=low luminosity type-2 Seyfert; 2=high luminosity type-2 Seyfert; 4=type-1 QSO; 5=type-1 Seyfert.}
\label{cosp_sed_2}
\end{figure*}

We fit the spectral energy distributions (SEDs) of all the MIR-selected galaxies with a custom version of the MAGPHYS code \citep{2015ApJ...806..110D}. MAGPHYS computes the emission by stellar populations in galaxies from the UV to near-IR consistently with the emission from dust at mid-IR and far-IR wavelengths using an energy balance technique. Our version is a modification of the high-z extension \citep{2008MNRAS.388.1595D} that includes the contribution by AGN emission to the SEDs (MAGPHYS+AGN; da Cunha et al. in prep., Juneau et al. in prep.). The AGN emission is reproduced using a set of empirical templates from \citet{2011MNRAS.414.1082M} (type-2),  \citet{2006ApJS..166..470R}, \citet{2010MNRAS.402..724P} (QSO), and \citet{2007ApJ...663...81P} (Seyfert 1) as shown in Figure\,~\ref{cosp_template}.
These four templates span in a representative way the global range of AGN known SEDs; a small but representative set of templates is chosen to avoid degeneracies in the SED fitting.
In our AGN sample, most of them ($>$80\%) can find good fitting results (0 $\leq$ $\chi^2_{AGN}$ $\leq$ 3).
The contribution of the AGN template to the total infrared luminosity, $\xi_{\text{AGN}}=L_{\text{dust}}^{\text{AGN}}/(L_{\text{dust}}^{\text{AGN}}+L_{\text{dust}}^{\text{SF}})$, is allowed to vary between 0 and 1 for each of the templates, and we allow the fitting code to decide which template best fits the observations, while marginalizing all parameters over different contributions of the AGN component and the different AGN emission templates.
Figure\,~\ref{cosp_sed_1} and \ref{cosp_sed_2} show examples of an obscured AGN host and a QSO with high AGN fraction. The SEDs without AGN contribution (gray lines) show the difficulty to fit the MIR part, which can be well fitted by the SEDs with AGN component (black lines, which is often plotted beneath other lines). 
The estimates of physical parameters of AGN hosts and the control sample of star-forming (non-AGN) galaxies are also based on our SED fitting results. 

\subsection{Obscuration}
\label{sec2_3}

\begin{figure*}[]
\centering
\includegraphics[width=1.00\textwidth]{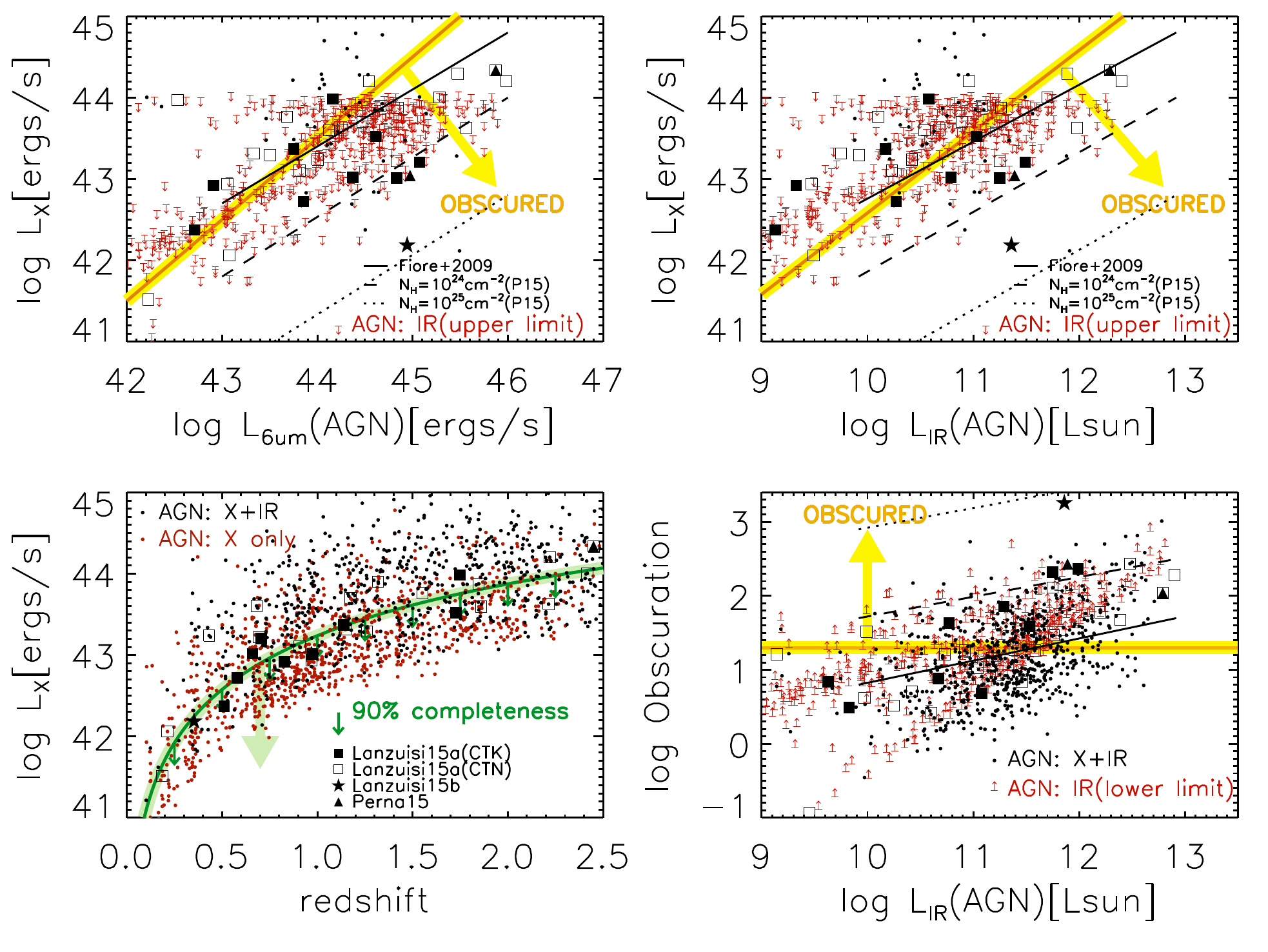} 
\caption[]{Obscured AGNs selection. 
Upper left: X-ray luminosity to 6 $\mu$m luminosity from the AGN component. This plot is comparable to \citet{2009ApJ...693..447F,2015A&A...573A.137L,2015A&A...578A.120L,2015A&A...583A..72P}. The red arrows are those sources where the upper limit has been calculated using the 90\% completeness threshold for X-ray detection (green line) illustrated in the lower left panel.
Upper right: X-ray luminosity to infrared luminosity from the AGN component. Lower left: X-ray luminosity to redshift plot. Here we define the upper limit (green line and arrows) of the X-ray luminosity (90\% completeness) for IR selected AGNs. 
The red points represent AGNs that are only selected by X-ray, and black points are selected both by X-ray and IR. Lower right: The ratio between IR luminosity and X-ray luminosity from AGN to IR luminosity plot. We choose $L_{IR,AGN}/L_{X,AGN} > 20$ as IR obscured AGNs (yellow and orange lines).}
\label{cosp_obscured}
\end{figure*}

In order to focus on obscured AGNs by infrared selection, we use the ratio of AGN luminosity between IR and X-ray to define obscuration: 
\begin{equation}
\frac{L_{IR,AGN}}{L_{X,AGN}},
\label{obscuration}
\end{equation}
where $L_{IR,AGN}$ is the AGN luminosity in the IR range (3 - 2000 $\mu$m) and $L_{X,AGN}$ is the AGN luminosity from X-ray observations. 
This definition represents the bolometric luminosity inferred from the infrared relative to the luminosity in the X-rays. 
In the upper plot in Figure\,~\ref{cosp_obscured}, we adopt the 90\% completeness of the X-ray luminosity as our upper limit in the X-ray luminosity to redshift plot \citep[see Figure 7 in][]{2016ApJ...817...34M}. 
Therefore, we can derive the lower limit of the obscuration and select obscured AGNs as in Equation~\ref{obscuration}. Here we define obscured AGNs by $L_{IR,AGN}/L_{X,AGN} > 20$ as an arbitrary choice according to Compton-thick AGNs in Figure\,~\ref{cosp_obscured}. There are 469 obscured AGNs at $z\leq2.5$. This is another approach to get the obscuration beside pure X-ray spectral measurement. We also compare our selection technique with previous obscured or Compton-thick AGNs in Figure\,~\ref{cosp_obscured}. This shows that our selection is reasonable, and we have a large obscured sample (red arrow) without X-ray detections.


\section{SED Results}
\label{sec3}

\subsection{Public Catalog: MIR-selected sources}
\label{sec3_1}

We provide a public catalog {\footnote{\url{www.asiaa.sinica.edu.tw/~yychang/ca.html}}} for all 36,670 MIR selected galaxies, including AGN information.  
In Table~\ref{tab_sed}, we include ID, Ra, Dec, redshift, modeling results from the preliminary MAGPHYS+AGN, and the public version of MAGPHYS which does not include an AGN component. Note that our analyses in this paper are based on the best redshifts, including private spectroscopic redshifts from the COSMOS team; the public version here is based on the photometric redshifts in the COSMOS2015 catalog \citep{2016ApJS..224...24L}. 
We recommend using the MAGPHYS modeling results for objects with FLAG=1 (20,311 out of 36,670).  These are all galaxies with good-quality SED fits (0 $\leq$ $\chi^2_{AGN}$ $\leq$ 3; 0 $\leq$ z $\leq$ 2.5).

In parallel, we provide SED fitting results, based on MAGPHYS \citep[see][]{2015ApJS..219....8C}, in galaxies with available photometric redshifts (537,173 out of 1,182,108 sources) over the whole COSMOS field. 

\begingroup 
\setlength{\tabcolsep}{12pt} 
\renewcommand{\arraystretch}{0.9} 
\begin{table*}
\centering
\caption[]{MAGPHYS+AGN output catalog in the COSMOS field.}
\label{tab_sed}
\resizebox{1.0\textwidth}{!}{
\begin{tabular}{p{5cm}p{2.0cm}p{2.0cm}p{14cm}}
\hline 
\hline
Column Name & Format & Unit  & Column Description \\ 
\hline
NUMBER & LONG & - & COSMOS2015 index\\ 
ALPHA\_J2000 & DOUBLE & deg &  J2000 R.A. [deg] from COSMOS2015\\
DELTA\_J2000 & DOUBLE & deg & J2000 Dec. [deg] from COSMOS2015\\
PHOTO\_Z & DOUBLE & - & photometric redshift from COSMOS2015 \\
TEMPLATE & INT & - & AGN template; 1: low lum type-2; 2:high lum type-2; 4: QSO type-1 5: Seyfert type-1\\
MASS\_2\_5\_AGN & FLOAT & $\log M_\odot$ & log stellar mass (2.5th percentile) [MAGPHYS+AGN]\\
MASS\_16\_AGN & FLOAT & $\log M_\odot$ & log stellar mass (16th percentile) [MAGPHYS+AGN]\\
MASS\_50\_AGN & FLOAT & $\log M_\odot$ & log stellar mass (50th percentile) [MAGPHYS+AGN]\\
MASS\_84\_AGN & FLOAT & $\log M_\odot$ & log stellar mass (84th percentile) [MAGPHYS+AGN]\\
MASS\_97\_5\_AGN & FLOAT & $\log M_\odot$ & log stellar mass (97.5th percentile) [MAGPHYS+AGN]\\
SFR\_2\_5\_AGN & FLOAT & $\log M_\odot/yr$ & log SFR (2.5th percentile) [MAGPHYS+AGN]\\
SFR\_16\_AGN & FLOAT & $\log M_\odot/yr$ & log SFR (16th percentile) [MAGPHYS+AGN]\\
SFR\_50\_AGN & FLOAT & $\log M_\odot/yr$ & log SFR (50th percentile) [MAGPHYS+AGN]\\
SFR\_84\_AGN & FLOAT & $\log M_\odot/yr$ & log SFR (84th percentile) [MAGPHYS+AGN]\\
SFR\_97\_5\_AGN & FLOAT & $\log M_\odot/yr$ & log SFR (97.5th percentile) [MAGPHYS+AGN]\\
AV\_2\_5\_AGN & FLOAT & - & dust attenuation parameter (2.5th percentile) [MAGPHYS+AGN]\\
AV\_16\_AGN & FLOAT & - & dust attenuation parameter (16th percentile) [MAGPHYS+AGN]\\
AV\_50\_AGN & FLOAT & - & dust attenuation parameter (50th percentile) [MAGPHYS+AGN]\\
AV\_84\_AGN & FLOAT & - & dust attenuation parameter (84th percentile) [MAGPHYS+AGN]\\
AV\_97\_5\_AGN & FLOAT & - & dust attenuation parameter (97.5th percentile) [MAGPHYS+AGN]\\
AGNF\_2\_5\_AGN & FLOAT & - & AGN IR fraction (2.5th percentile) [MAGPHYS+AGN]\\
AGNF\_16\_AGN & FLOAT & - & AGN IR fraction (16th percentile) [MAGPHYS+AGN]\\
AGNF\_50\_AGN & FLOAT & - & AGN IR fraction (50th percentile) [MAGPHYS+AGN]\\
AGNF\_84\_AGN & FLOAT & - & AGN IR fraction (84th percentile) [MAGPHYS+AGN]\\
AGNF\_97\_5\_AGN & FLOAT & - & AGN IR fraction (97.5th percentile) [MAGPHYS+AGN]\\
LDUST\_2\_5\_AGN & FLOAT & $\log L_\odot$ & log dust luminosity (2.5th percentile) [MAGPHYS+AGN]\\
LDUST\_16\_AGN & FLOAT & $\log L_\odot$ & log dust luminosity (16th percentile) [MAGPHYS+AGN]\\
LDUST\_50\_AGN & FLOAT & $\log L_\odot$ & log dust luminosity (50th percentile) [MAGPHYS+AGN]\\
LDUST\_84\_AGN & FLOAT & $\log L_\odot$ & log dust luminosity (84th percentile) [MAGPHYS+AGN]\\
LDUST\_97\_5\_AGN & FLOAT & $\log L_\odot$ & log dust luminosity (97.5th percentile [MAGPHYS+AGN]\\
LDUSTAGN\_2\_5\_AGN & FLOAT & $\log L_\odot$ & log dust AGN luminosity (2.5th percentile) [MAGPHYS+AGN]\\
LDUSTAGN\_16\_AGN & FLOAT & $\log L_\odot$ & log dust AGN luminosity (16th percentile) [MAGPHYS+AGN]\\
LDUSTAGN\_50\_AGN & FLOAT & $\log L_\odot$ & log dust AGN luminosity (50th percentile) [MAGPHYS+AGN]\\
LDUSTAGN\_84\_AGN & FLOAT & $\log L_\odot$ & log dust AGN luminosity (84th percentile) [MAGPHYS+AGN]\\
LDUSTAGN\_97\_5\_AGN & FLOAT & $\log L_\odot$ & log dust AGN luminosity (97.5th percentile) [MAGPHYS+AGN]\\
MASS\_2\_5\_0 & FLOAT & $\log M_\odot$ & log stellar mass (2.5th percentile) [MAGPHYS]\\
MASS\_16\_0 & FLOAT & $\log M_\odot$ & log stellar mass (16th percentile) [MAGPHYS]\\
MASS\_50\_0 & FLOAT & $\log M_\odot$ & log stellar mass (50th percentile) [MAGPHYS]\\
MASS\_84\_0 & FLOAT & $\log M_\odot$ & log stellar mass (84th percentile) [MAGPHYS]\\
MASS\_97\_5\_0 & FLOAT & $\log M_\odot$ & log stellar mass (97.5th percentile) [MAGPHYS]\\
SFR\_2\_5\_0 & FLOAT & $\log M_\odot/yr$ & log SFR (2.5th percentile) [MAGPHYS]\\
SFR\_16\_0 & FLOAT & $\log M_\odot/yr$ & log SFR (16th percentile) [MAGPHYS]\\
SFR\_50\_0 & FLOAT & $\log M_\odot/yr$ & log SFR (50th percentile) [MAGPHYS]\\
SFR\_84\_0 & FLOAT & $\log M_\odot/yr$ & log SFR (84th percentile) [MAGPHYS]\\
SFR\_97\_5\_0 & FLOAT & $\log M_\odot/yr$ & log SFR (97.5th percentile) [MAGPHYS]\\
AV\_2\_5\_0 & FLOAT & - & dust attenuation parameter (2.5th percentile) [MAGPHYS]\\
AV\_16\_0 & FLOAT & - & dust attenuation parameter (16th percentile) [MAGPHYS]\\
AV\_50\_0 & FLOAT & - & dust attenuation parameter (50th percentile) [MAGPHYS]\\
AV\_84\_0 & FLOAT & - & dust attenuation parameter (84th percentile) [MAGPHYS]\\
AV\_97\_5\_0 & FLOAT & - & dust attenuation parameter (97.5th percentile) [MAGPHYS]\\
LDUST\_2\_5\_0 & FLOAT & $\log L_\odot$ & log dust luminosity (2.5th percentile) [MAGPHYS]\\
LDUST\_16\_0 & FLOAT & $\log L_\odot$ & log dust luminosity (16th percentile) [MAGPHYS]\\
LDUST\_50\_0 & FLOAT & $\log L_\odot$ & log dust luminosity (50th percentile) [MAGPHYS]\\
LDUST\_84\_0 & FLOAT & $\log L_\odot$ & log dust luminosity (84th percentile) [MAGPHYS]\\
LDUST\_97\_5\_0 & FLOAT & $\log L_\odot$ & log dust luminosity (97.5th percentile) [MAGPHYS]\\
FLAG & INT & - & flag (1=good fits; 0=others) \\
\hline 
\end{tabular} 
}
\raggedright  {\scriptsize Note: we recommend using the MAGPHYS modeling results for objects with FLAG=1. \\
These are all 0 $\leq$ z $\leq$ 2.5 galaxies with good-quality SED fits (0 $\leq$ $\chi^2_{AGN}$ $\leq$ 3).}
\end{table*}
\endgroup

\subsection{AGN Fraction and Obscuration}
\label{sec3_2}

\begin{figure}
\centering
\includegraphics[width=1.0\columnwidth]{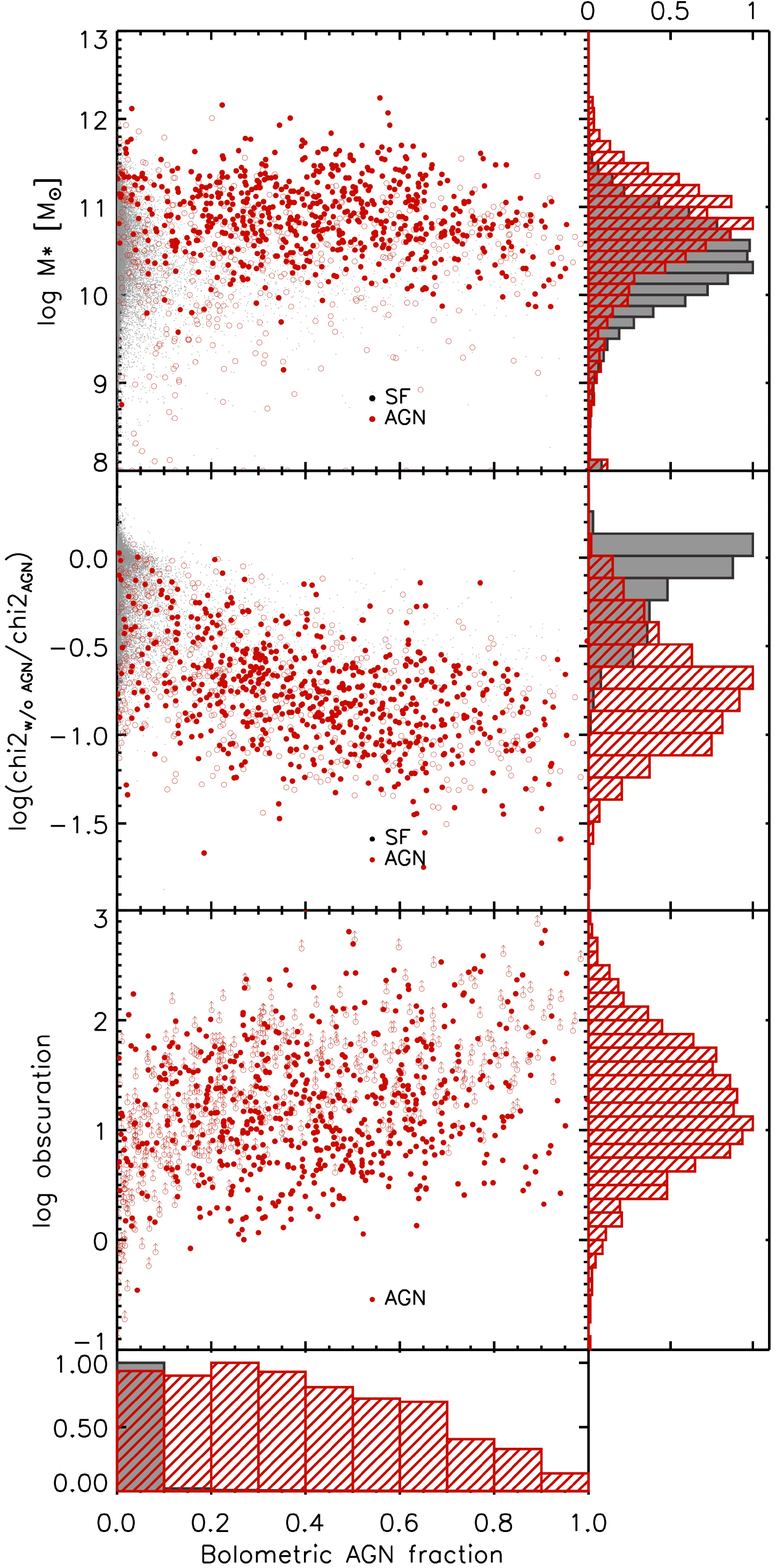} 
\caption[]{Upper: AGN bolometric fraction to stellar mass plot for normal star-forming galaxies (gray), and infrared selected AGNs (red). The open circles represent AGNs without X-ray detection ($L_X$(2-10 keV)$<10^{42}$ ergs/s). Middle: AGN bolometric fraction to the ratio of chi-square value plot. The ratio represents the chi-square values between the SED fitting results without and with AGN templates. The AGN templates improve the SED fitting, especially for high AGN fraction objects. Lower: AGN bolometric fraction to the obscuration plot. For non X-ray detected sources, we show the lower limit of the obscuration (red arrow). There is a slight correlation between AGN fraction and obscuration}
\label{cosp_agnf}
\end{figure}

\begin{figure}
\centering
\includegraphics[width=0.9\columnwidth]{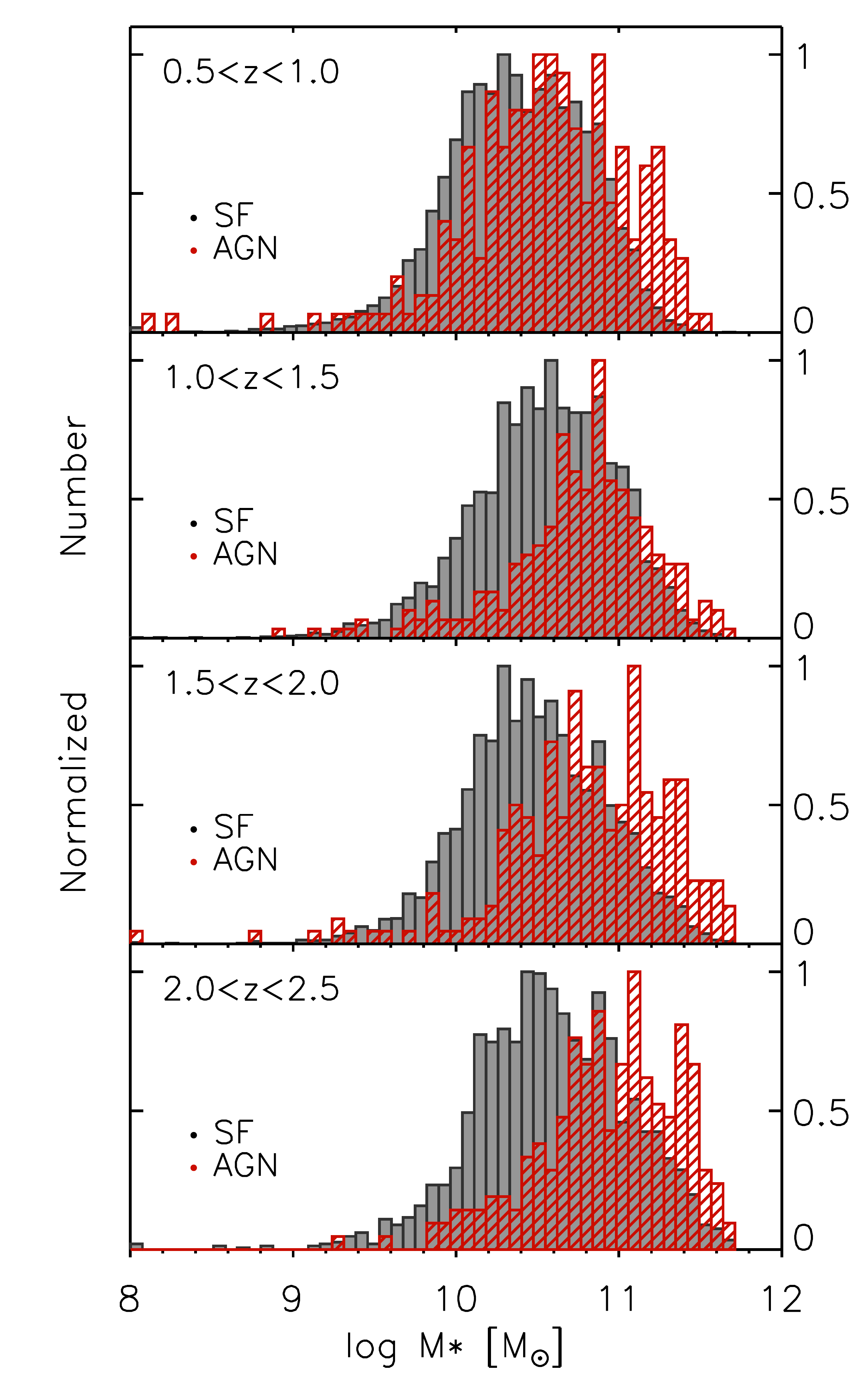} 
\caption[]{The mass distribution in different redshift bins. AGNs are in massive galaxies at all redshift bins.}
\label{cosp_m}
\end{figure}

Figure\,~\ref{cosp_agnf} shows stellar mass, difference of reduced chi-square values between SEDs of AGN and non-AGN consideration, obscuration as a function of AGN bolometric fraction (AGN to total luminosity integrated over the whole SED from optical to far-infrared). The AGNs are IR selected by the color-color diagram in Section \ref{sec3_1}. For the control sample, we excluded IR-AGN, MAGPHYS AGN, X-ray selected AGN host galaxies in the star-forming galaxies hereafter.
Some star-forming galaxies are dominated by an AGN bolometrically in Figure\,~\ref{cosp_agnf}. It is possible that AGNs were particularly luminous in the optical, rather than in the IR or X-ray as shown in Figure\,~\ref{cosp_sed_2}. In this case, we only observed the accretion disk, but not the corona or torus. Nevertheless, most of the star-forming galaxies ($\sim$96\%) contain little AGN contributions ($<$10\% bolometric AGN fraction).
There are more massive AGN hosts compared to control sample. 
Figure\,~\ref{cosp_m} also shows that AGNs are in higher mass galaxies at all redshifts as also found in the literature \citep{2012ApJ...746...90A,2016A&A...588A..78B}.
The middle panel shows the reduced chi-square value comparison between the SED fitting without and with AGN components. There is a clear separation between AGN hosts and normal star-forming galaxies. In general, our MAGPHYS+AGN SED fitting results work better than the standard fitting by considering the MIR contribution.
The small reduced chi-square values of high AGN fraction objects show the importance of including AGN components. 
The lower panel shows a slight correlation between AGN bolometric fraction and obscuration, but it is difficult to quantify the relation due to the lower limit of obscuration of non X-ray detected sources. 
This could be due to the correlation with AGN luminosity. This implies that the emission of heavily obscured AGNs is dominated by the AGN component.
The open circle and lower limit symbols show that many obscured objects with high AGN fraction are not detected in the X-ray. 
In our obscured IR-AGN sample, $\sim$54\% of them are also defined as X-ray AGNs. This is consistent with \citet{2013ApJ...770...40M}, which shows that the fraction of X-ray AGN that are IR-AGN selected depends on the X-ray depth.
This suggests that our infrared selection provides an obscured sample ($\sim$46\%) which is hidden in the X-ray observations at the depth of the Chandra data in COSMOS. 

\subsection{Comparison with the Star-forming Main Sequence}
\label{sec3_3}

\begin{figure*}[h]
\centering
\includegraphics[width=1.00\columnwidth]{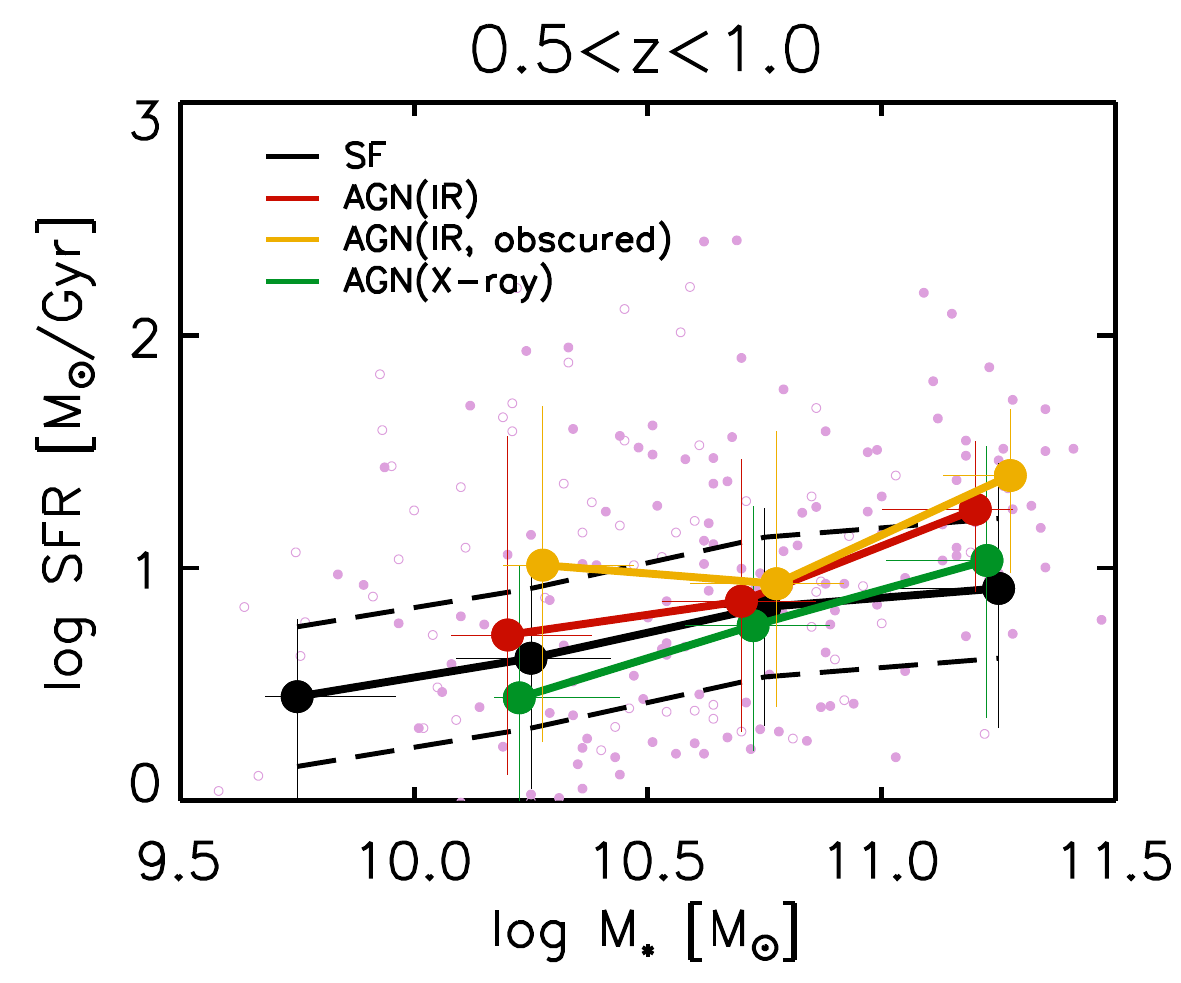} 
\includegraphics[width=1.00\columnwidth]{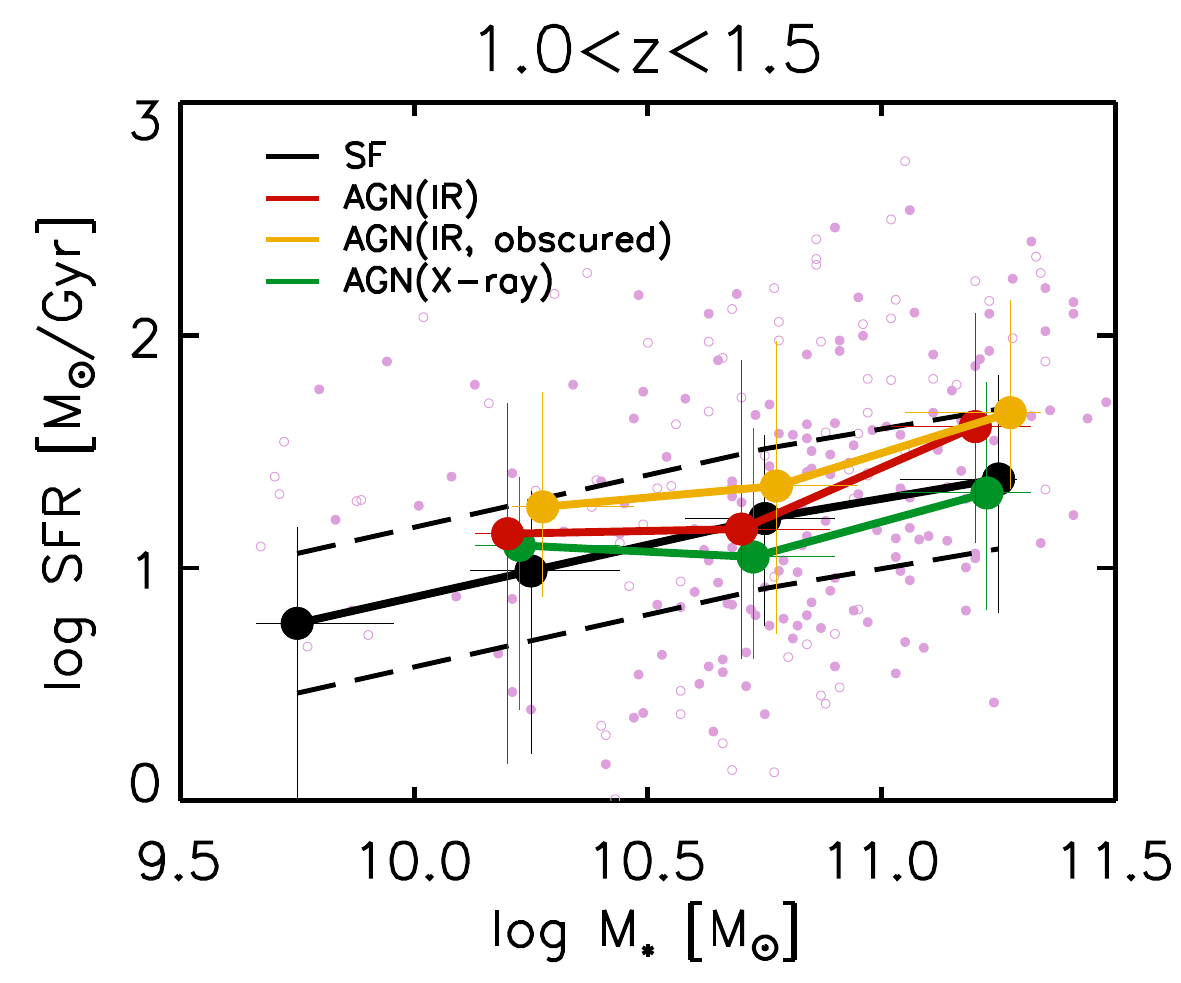} 
\includegraphics[width=1.00\columnwidth]{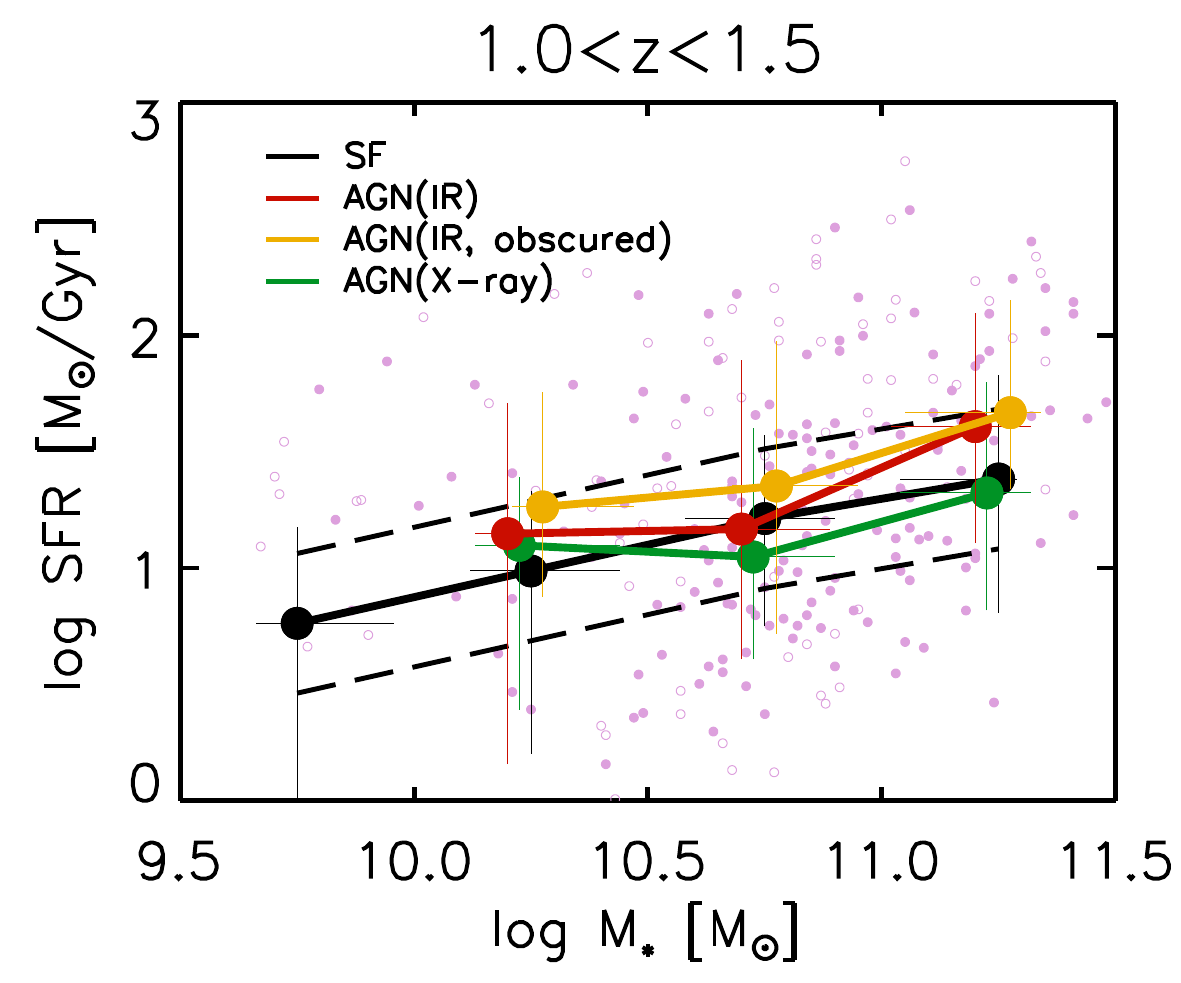} 
\includegraphics[width=1.00\columnwidth]{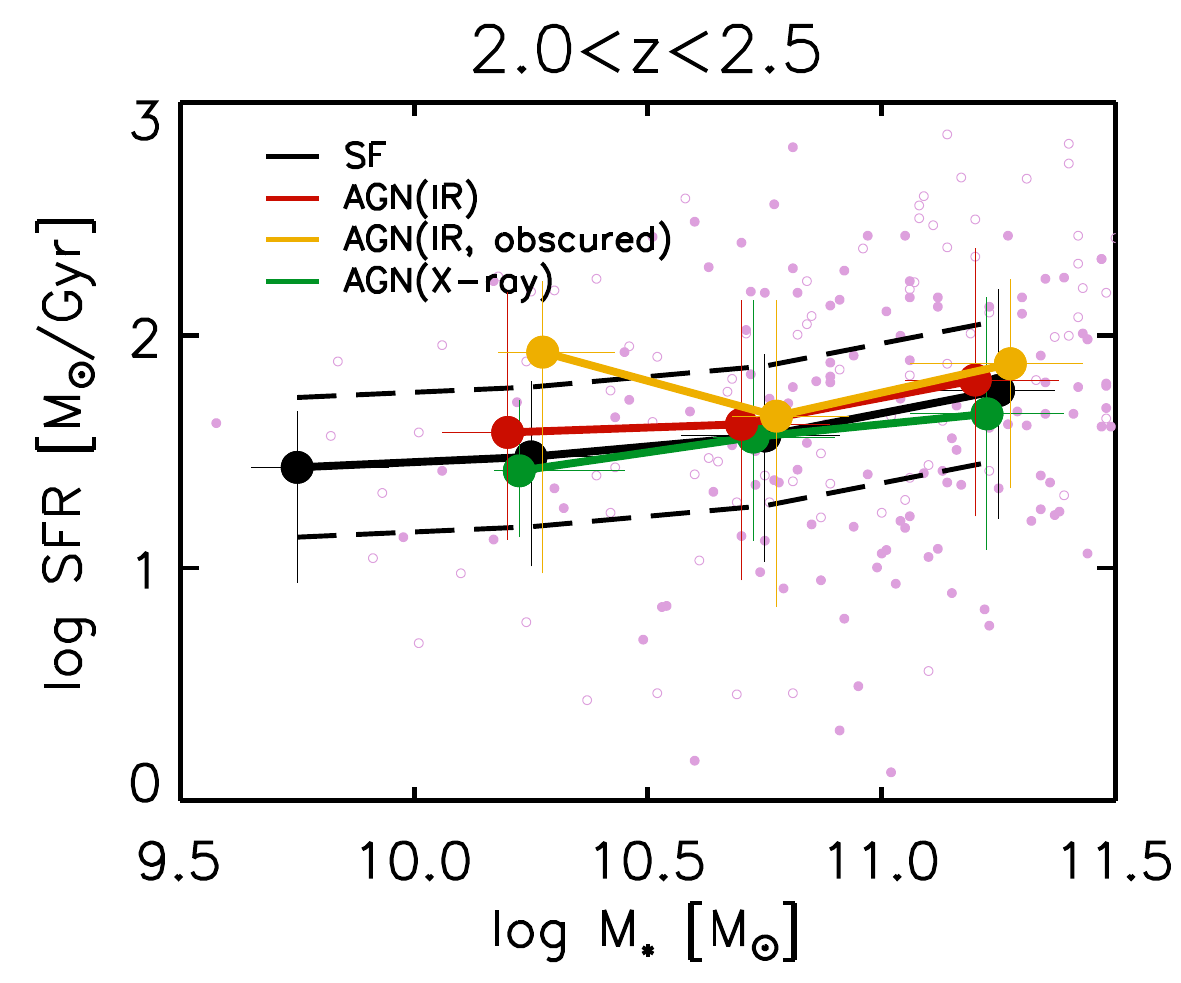} 
\caption[]{Star forming sequence (main sequence) plot. The black solid and dash lines show the star-forming sequence. Here we show the median and the 84\%-16\% values of IR-AGN hosts (red), obscured IR-AGN hosts (orange), X-ray AGN hosts (green), and star-forming galaxies (black). The plum dots show individual IR-AGN hosts. The open circles represent AGNs without X-ray detection. Obscured AGNs lie within or slightly above the star-forming sequence.}
\label{cosp_ms_z}
\end{figure*}

\begin{figure*}
\centering
\includegraphics[width=1.00\columnwidth]{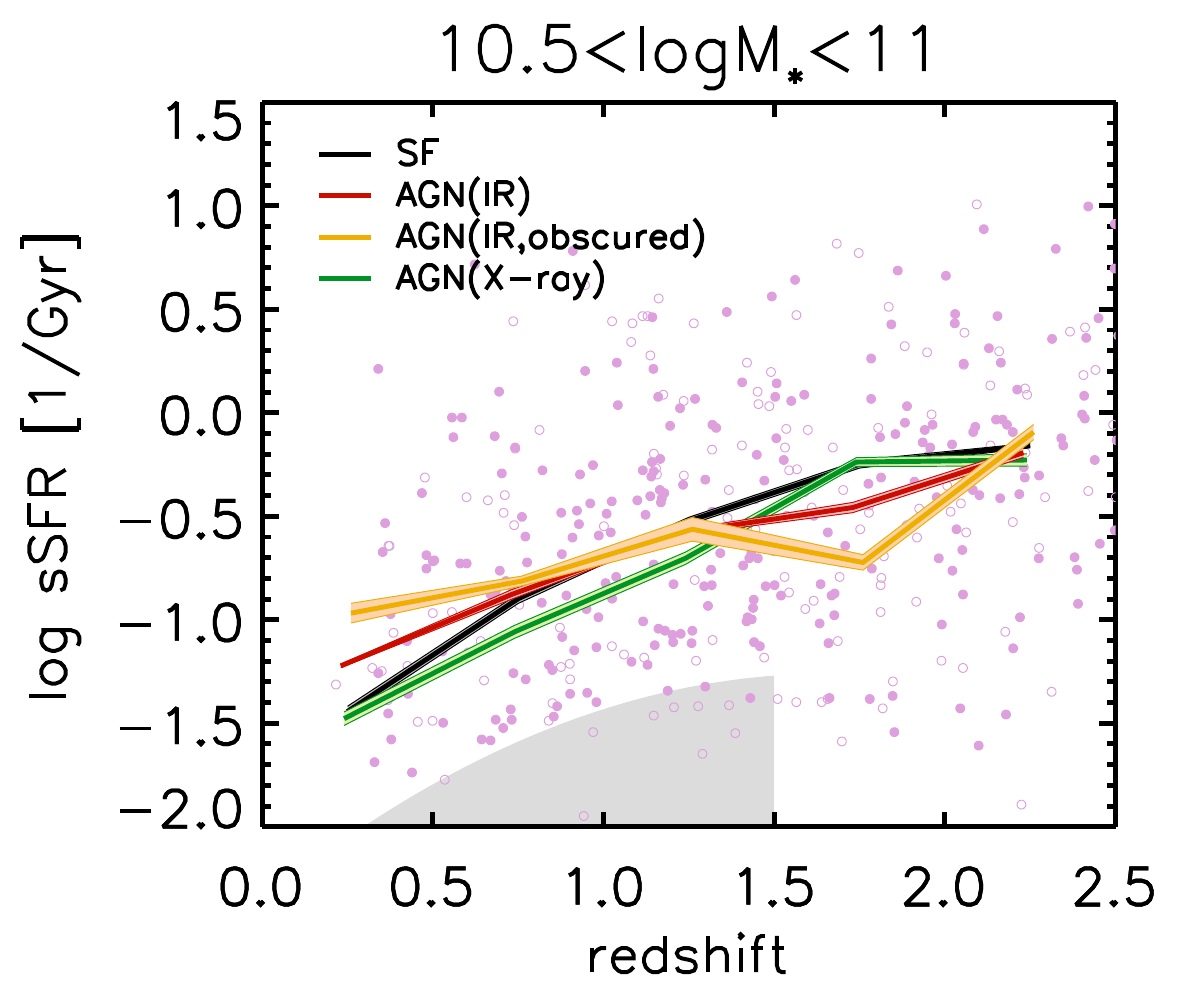} 
\includegraphics[width=1.00\columnwidth]{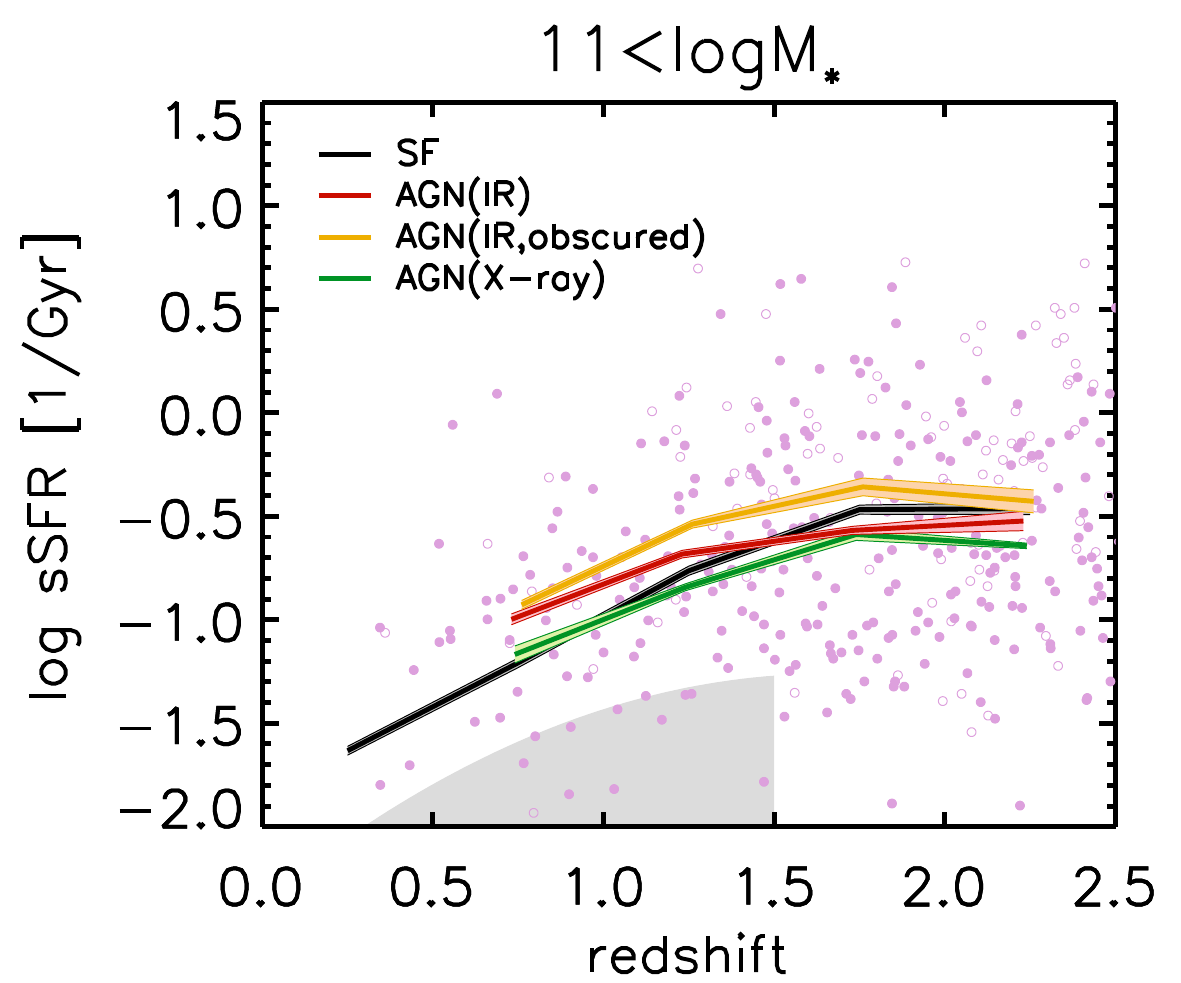} 
\caption[]{Specific SFR as a function of redshift in two different stellar mass bins. Here we show the median value of IR-AGN hosts (red), obscured IR-AGN hosts (orange), X-ray AGN hosts (green), and star-forming galaxies (black). The error bars are from bootstrapping. The plum dots show individual IR-AGN hosts. The open circles represent AGNs without X-ray detection. The shading region is the lower limit adopted from \citet{2015A&A...579A...2I}. The sSFR of obscured AGNs are similar or slightly larger than normal galaxies.}
\label{cosp_ms}
\end{figure*}

\begin{table}
\centering
\caption[]{Fitting parameters of the SFR to stellar mass plot (main sequence; see also Figure\,~\ref{cosp_ms_z}) in Equation \ref{eq6}. The uncertainties are estimated by bootstrapping (N=1000).}
\label{tab_3}
\begin{tabular}{ccc}
\hline
\hline
redshift & $a$ & $b$ \\ 
\hline 
 & \multicolumn{2}{c}{star-forming galaxies}\\
\cline{2-3}
$0.5<z<1.0$ & 0.34$\pm$0.02 & -2.86$\pm$0.24\\
$1.0<z<1.5$ & 0.42$\pm$0.02 & -3.30$\pm$0.24\\
$1.5<z<2.0$ & 0.34$\pm$0.02 & -2.16$\pm$0.26\\
$2.0<z<2.5$ & 0.22$\pm$0.04 & -0.80$\pm$0.45\\
\cline{2-3}
 & \multicolumn{2}{c}{IR-AGN hosts}\\
\cline{2-3}
$0.5<z<1.0$ & 0.53$\pm$0.18 & -4.76$\pm$1.93\\
$1.0<z<1.5$ & 0.50$\pm$0.18 & -4.08$\pm$1.93\\
$1.5<z<2.0$ & 0.47$\pm$0.13 & -3.66$\pm$1.36\\
$2.0<z<2.5$ & 0.25$\pm$0.19 & -0.96$\pm$2.01\\
\cline{2-3}
 & \multicolumn{2}{c}{IR-AGN hosts (obscured)}\\
\cline{2-3}
$0.5<z<1.0$ & 0.35$\pm$0.24 & -2.68$\pm$2.64\\
$1.0<z<1.5$ & 0.42$\pm$0.18 & -3.08$\pm$1.91\\
$1.5<z<2.0$ & 0.64$\pm$0.19 & -5.53$\pm$2.08\\
$2.0<z<2.5$ & 0.00$\pm$0.28 &  1.76$\pm$3.08\\
\cline{2-3}
 & \multicolumn{2}{c}{X-ray AGN hosts}\\
\cline{2-3}
$0.5<z<1.0$ & 0.59$\pm$0.10 & -5.57$\pm$1.08\\
$1.0<z<1.5$ & 0.27$\pm$0.08 & -1.79$\pm$0.91\\
$1.5<z<2.0$ & 0.40$\pm$0.09 & -2.82$\pm$0.95\\
$2.0<z<2.5$ & 0.24$\pm$0.14 & -0.99$\pm$1.48\\
\hline
\end{tabular} 
\end{table}

In order to understand the star formation in AGN hosts compared to the general galaxy population, we compare our sample with the star-forming sequence \citep[also call main sequence; e.g., ][]{2004MNRAS.351.1151B,2007A&A...468...33E,2007ApJ...660L..43N,2007ApJ...670..156D,2015A&A...579A...2I,2015A&A...575A..74S,2015ApJS..219....8C}. 
The lower star formation might imply AGN feedback, the higher star formation could imply that the stellar and black hole components are both growing at the same time, and similar star formation might indicate that the AGN does not affect significantly the star formation properties. 

Figure\,~\ref{cosp_ms_z} shows star formation sequence plots at different redshift bins, based on our SED fitting results. The black lines represent the star-forming galaxies which are not classified as IR-AGN, MAGPHYS AGN, or X-ray selected AGNs in our 24 $\mu$m sample. The dotted black lines define a rage of +/- 0.3 dex of our star-forming sample.

We defined our own star-forming sequence, rather than using any literature to avoid biases.
We fitted the median values of the star-forming sequence with a power law and find
\begin{equation}
\label{eq6}
\log \text{SFR}/(M_\odot/yr)=a \log M_* /(M_\odot) -b,
\end{equation}
where $a$ and $b$ are shown in Table\,~\ref{tab_3}.
The red lines represent IR-AGN sample, and  the orange lines show the obscured IR-AGN sample which are defined by our MIR color-color plot and obscuration as described in Section~\ref{sec2}.
The green lines show X-ray selected AGNs from Chandra Legacy Survey with $L_X$(2-10 keV)$>10^{42}$ ergs/s. 
Figure\,~\ref{cosp_ms} shows the specific star formation rate as a function of redshift for two stellar mass bins. We find that our IR-selected AGN hosts are not significantly or slightly above the star-forming sequence at $z<1.5$. 
In Figure\,~\ref{cosp_ms_z} and Figure\,~\ref{cosp_ms}, the most massive obscured IR-AGN hosts could be above the Main sequence at all redshifts, but most ($>$70\%) obscured  IR-AGN hosts at $10.5<\log M_*/M_\odot <11$ are below the star-forming sequence at $z>1.5$ .  
In general, X-ray selected AGN hosts are close to the star-forming sequence at all redshift and stellar mass bins, which implies that AGN host galaxies by both IR and X-ray selections are mostly on the main sequence.  

\subsection{Additional Parameter Constraints from SEDs}
\label{sec3_4}

\begin{figure*}[h]
\centering
\includegraphics[width=0.85\textwidth]{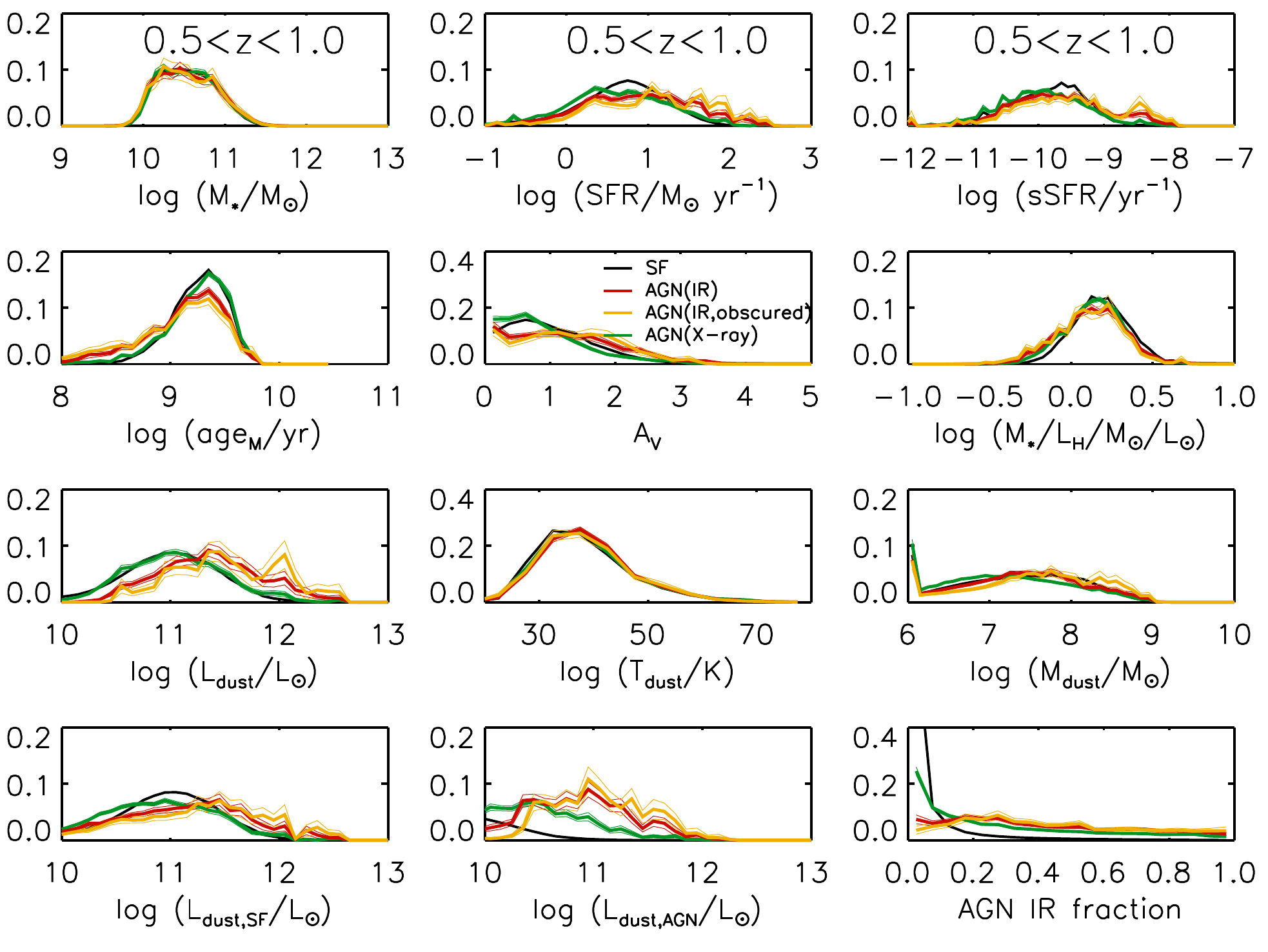}
\includegraphics[width=0.85\textwidth]{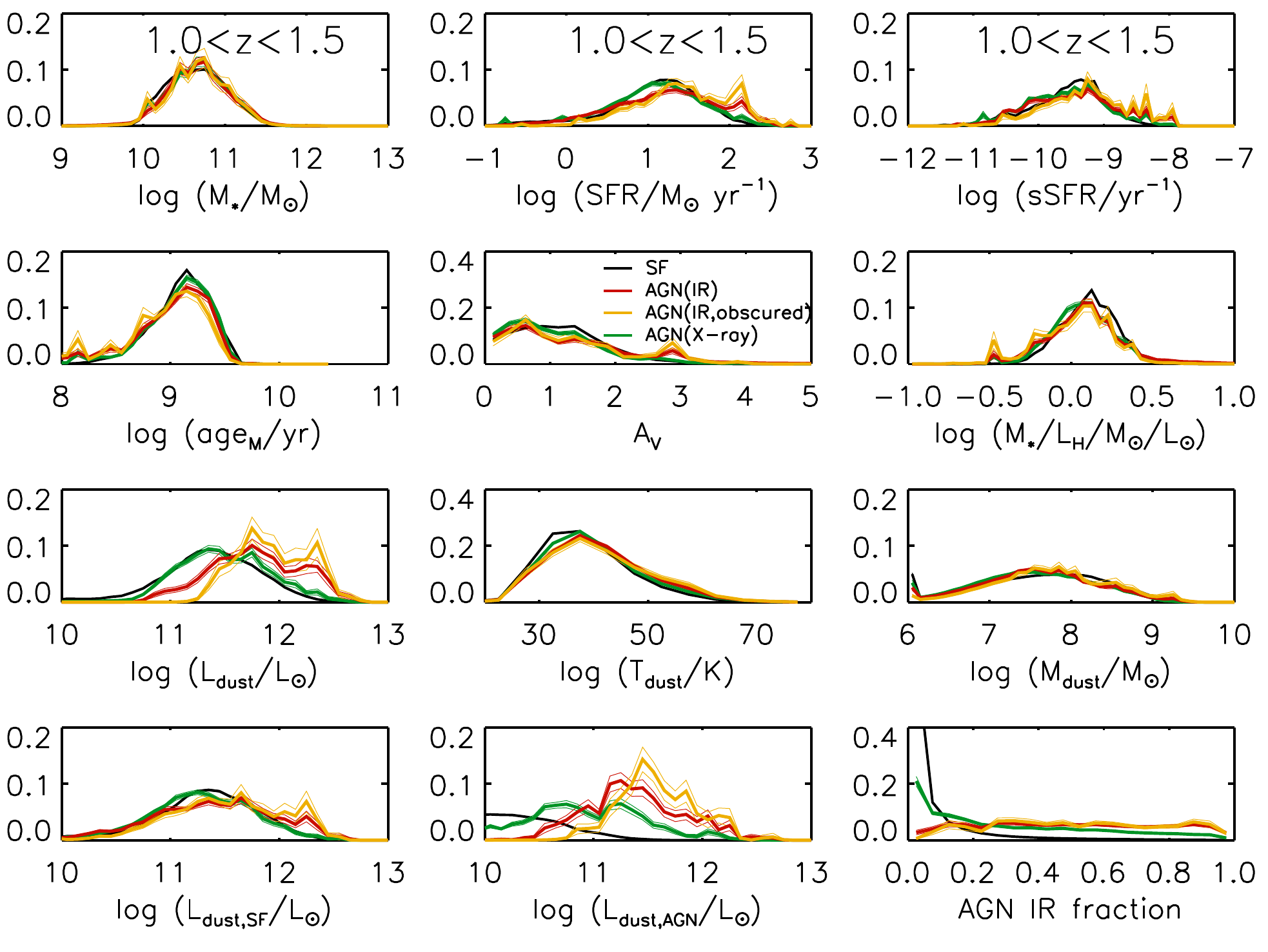}
\caption[]{Normalized stacked likelihood distributions of different physical parameters of mass matched star-forming galaxies (black line), IR-AGN hosts (red), obscured IR-AGN hosts (orange), and X-ray AGN hosts (green line) at $0.5<z<1.0$ and $1.0<z<1.5$. The thin lines are uncertainties estimated by bootstrapping. See \citet{2015ApJ...806..110D} for more details about these physical properties.}
\label{cosp_pdf1}
\end{figure*}
 \begin{figure*}
\centering
\includegraphics[width=0.85\textwidth]{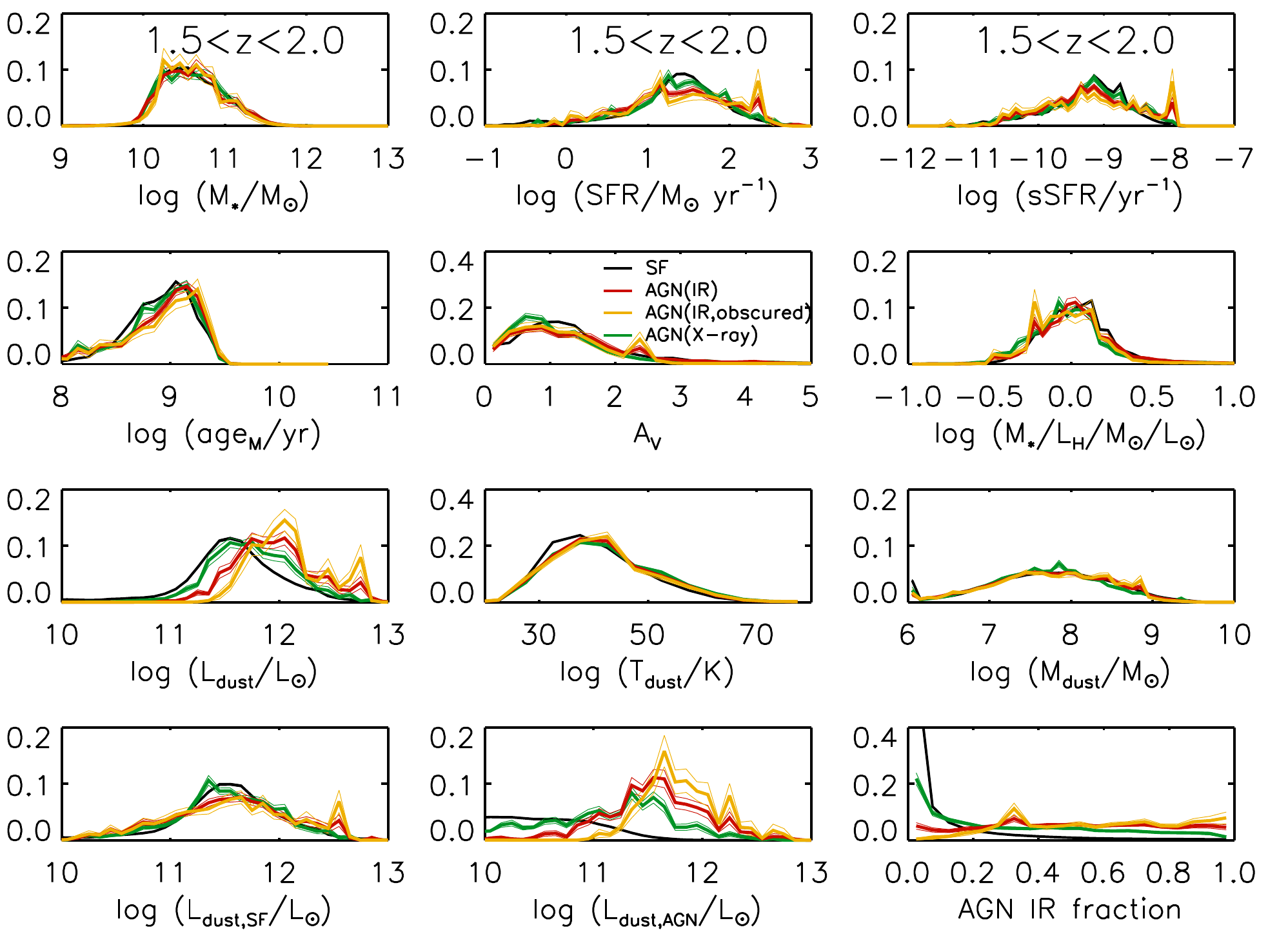}
\includegraphics[width=0.85\textwidth]{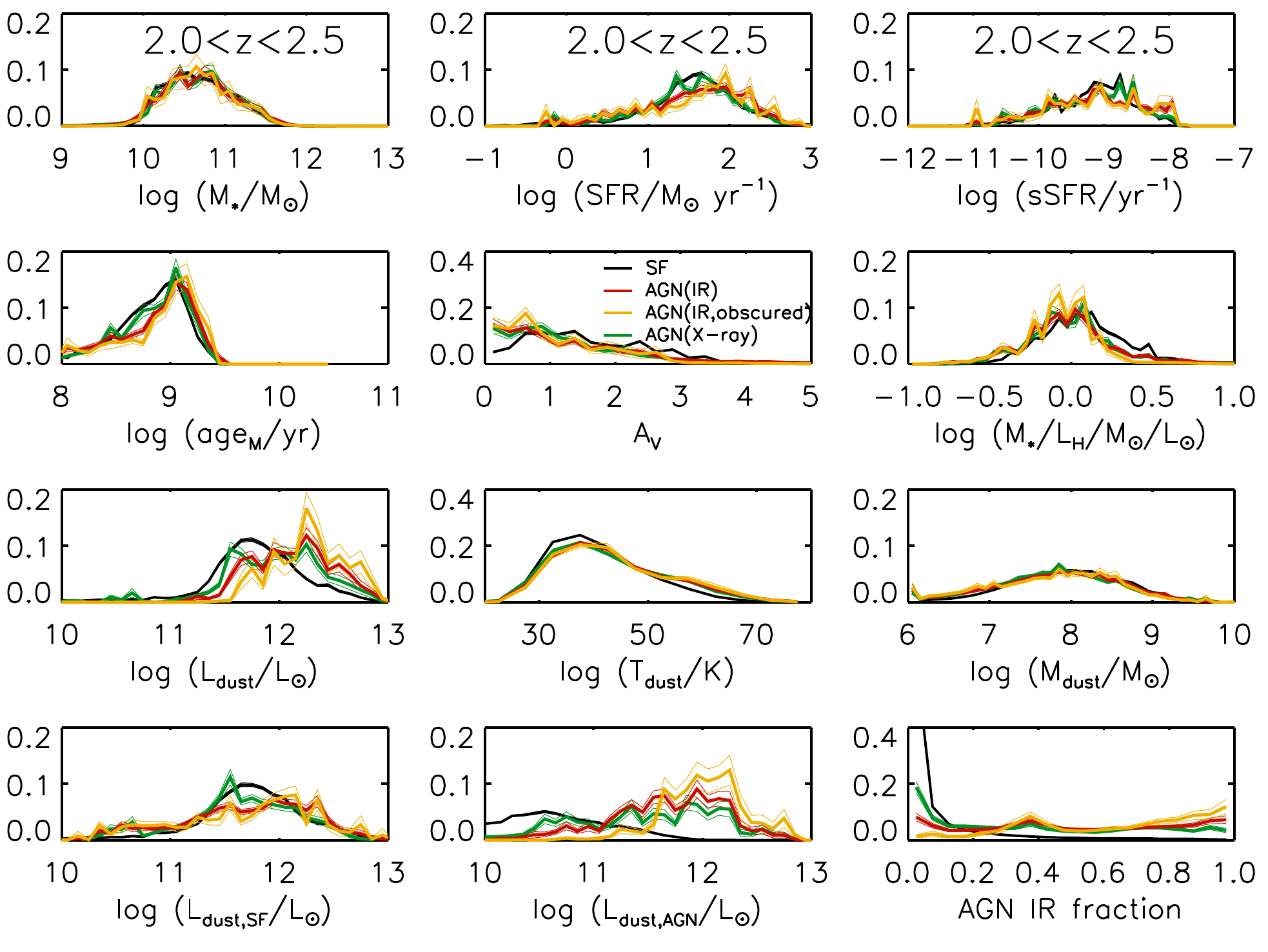} 
\caption[]{Normalized stacked likelihood distributions of different physical parameters of mass matched star-forming galaxies (black line), IR-AGN hosts (red), obscured IR-AGN hosts (orange), and X-ray AGN hosts (green line) at $1.5<z<2.0$ and $2.0<z<2.5$. The thin lines are uncertainties estimated by bootstrapping. See \citet{2015ApJ...806..110D} for more details about these physical properties.}
\label{cosp_pdf2}
\end{figure*}

In Figure\,~\ref{cosp_pdf1} and Figure\,~\ref{cosp_pdf2}, we show some key physical properties of their stacked likelihood distributions from SED fitting for mass-matched sample, according to the stellar mass distribution of star-forming galaxies (control sample). As discussed in Figure\,~\ref{cosp_ms_z} and \ref{cosp_ms}, we see slight differences of the SFRs and sSFRs. As a result, the mass-weighted ages of AGNs are younger than star-forming galaxies at $z\sim 1$ but older at $z\sim2$. 
The AGN fraction and luminosity of IR-AGNs are higher than X-ray selected AGNs and normal galaxies at all redshifts. In particular, the obscured IR-AGN sample shows the highest AGN luminousity. It is clear to see that the AGN IR fraction and AGN luminosity of the star-forming galaxies are significantly different from AGNs. 

\section{Morphological analysis}
\label{sec4}

In this section, we will show the influence of obscuration and AGN fraction on morphology. We use the HST/ACS $I$-band images to study a sample at $0.5<z<1.5$ and $\log M_*/M_\odot>10.5$. We focus on obscured IR-AGN hosts with $L_{IR,AGN}/L_{X,AGN} > 20$ as defined in Section \ref{sec2} and discussed in Section \ref{sec3} (orange lines in Figure\,~\ref{cosp_ms_z}, \ref{cosp_ms}, \ref{cosp_pdf1}, and \ref{cosp_pdf2}), which ensures that the optical imaging of the hosts is not affected by AGNs. 

\subsection{GALFIT}
\label{sec4_1}

\begin{figure}[h]
\centering
\includegraphics[width=1.0\columnwidth]{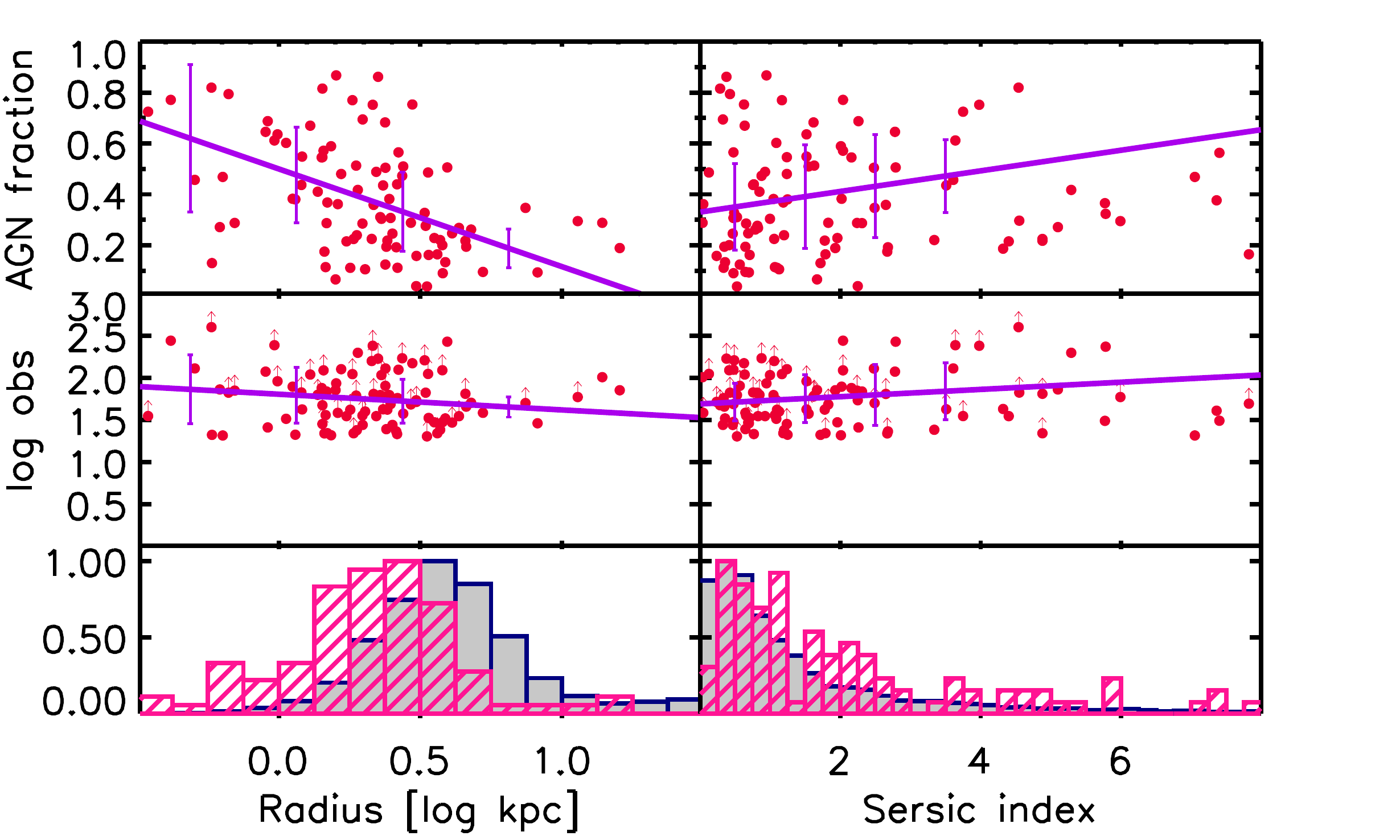}  
\caption[]{Radius and S\'ersic index of single S\'ersic fitting from GALFIT results at $0.5<z<1.5$ and $\log M_*/M_\odot>10.5$. Here we compare the GALFIT parameters with bolometric AGN fraction and obscuration for obscured IR-AGN hosts (red). We also show histograms of obscured IR-AGN hosts (pink) and normal star-forming galaxies (blue/gray). The purple lines show results of linear fitting and the error bars represent the standard deviation in each bin. We only consider detected sources in the linear fitting for obscuration.}
\label{cosp_galfit}
\end{figure}

\begin{figure}[h]
\centering
\includegraphics[width=1.0\columnwidth]{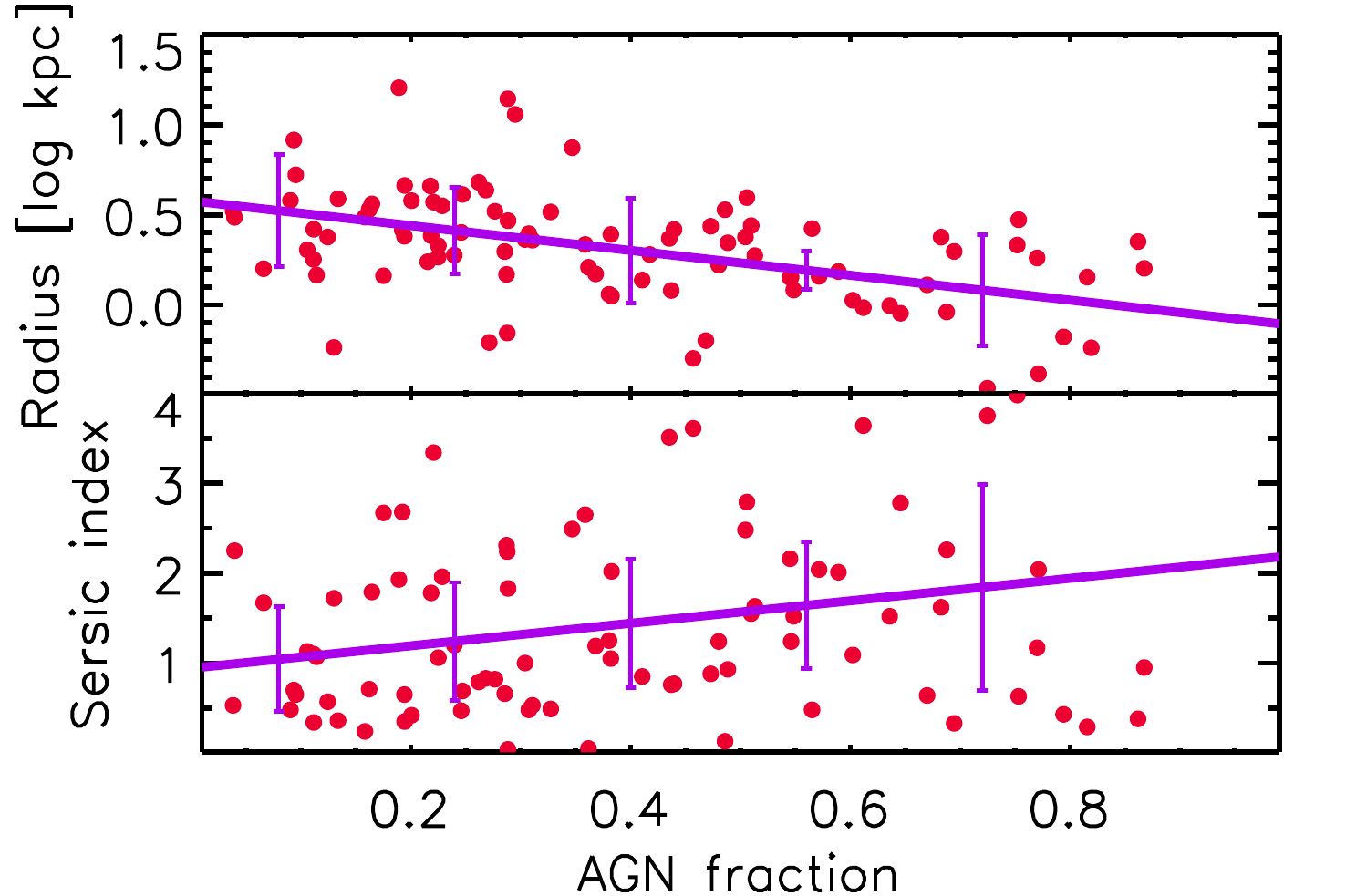} 
\caption[]{Here we show radius and S\'ersic index as a function of bolometric AGN fraction. The purple lines show results of linear fitting and the error bars represent the standard deviation in each bin. According to the linear correlation (purple line), a 20\% AGN contribution corresponds to a decreased radius by 24\% and an increased S\'ersic index by 18\%, and a 50\% AGN contribution corresponds to a decreased radius by 50\% and an increased S\'ersic index by 47\%.}
\label{cosp_galfity}
\end{figure}

We used both single S\'ersic profile and S\'ersic+PSF profile measured by GALFIT \citep{2002AJ....124..266P,2010AJ....139.2097P}. As described by \citet{2017MNRAS.466L.103C}, a PSF component is a negligible term, so we focus on single S\'ersic fitting results to compare with star-forming galaxies. In Figure\,~\ref{cosp_galfit}, we find that there is a slight trend between AGN fraction and radius/S\'ersic index. AGN hosts seem to be more compact (smaller in radius and larger in S\'ersic index) while the AGN fraction is higher. Moreover, AGN hosts are also more compact than normal star-forming galaxies. This is consistent with our recent finding about compact AGN hosts in \citet{2017MNRAS.466L.103C}, which suggested a possible indication of compaction of AGN hosts, and that a vast majority of obscured AGNs might reside in galaxies undergoing dynamical compaction.
We also check the relation between obscuration and GALFIT parameters in Figure\,~\ref{cosp_galfit} and find that there is almost no correlation (Pearson correlation coefficient $<$0.1). 

In Figure\,~\ref{cosp_galfity}, we found radius and S\'ersic index as functions of bolometric AGN fraction: $\log r=-(0.61\pm0.11) f + (0.53\pm0.05)$ and $n=(1.04\pm0.41) f + (1.10\pm0.16)$, where $r$ is the radius in kpc, $n$ is the S\'ersic index, and $f$ is the bolometric AGN fraction. According to these relations, a 20\% AGN contribution corresponds to a decreased radius by 24\% and an increased S\'ersic index by 18\%, and a 50\% AGN contribution corresponds to a decreased radius by 50\% and an increased S\'ersic index by 47\%.

\subsection{Nonparametric Morphology Method}
\label{sec4_2}

\begin{figure*}
\centering
\includegraphics[width=1.0\textwidth]{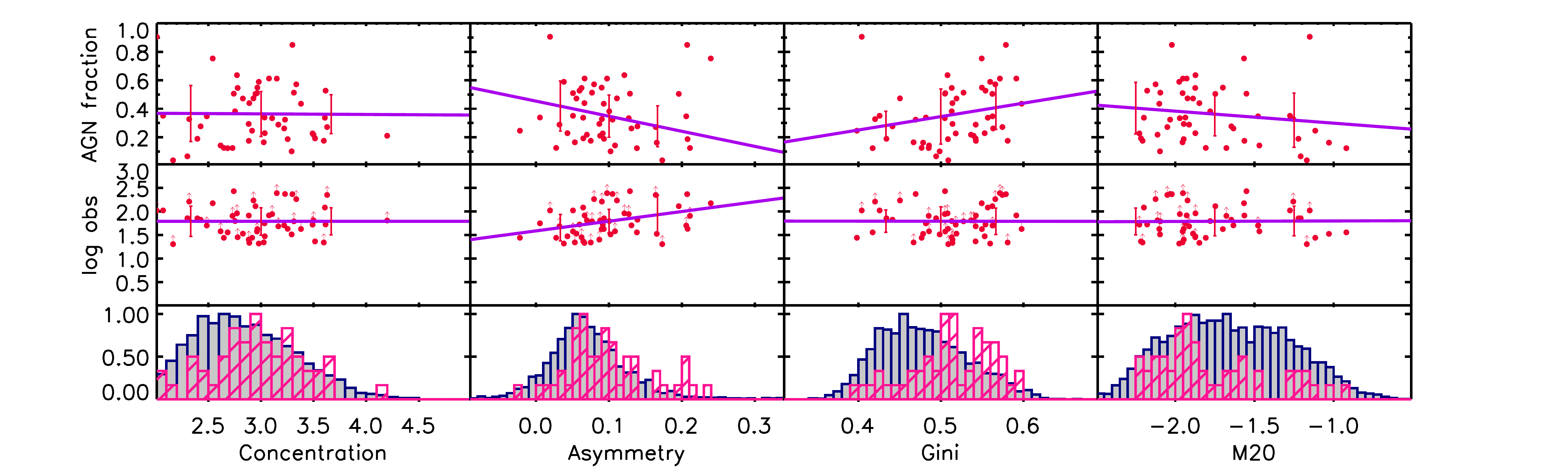} 
\includegraphics[width=1.0\textwidth]{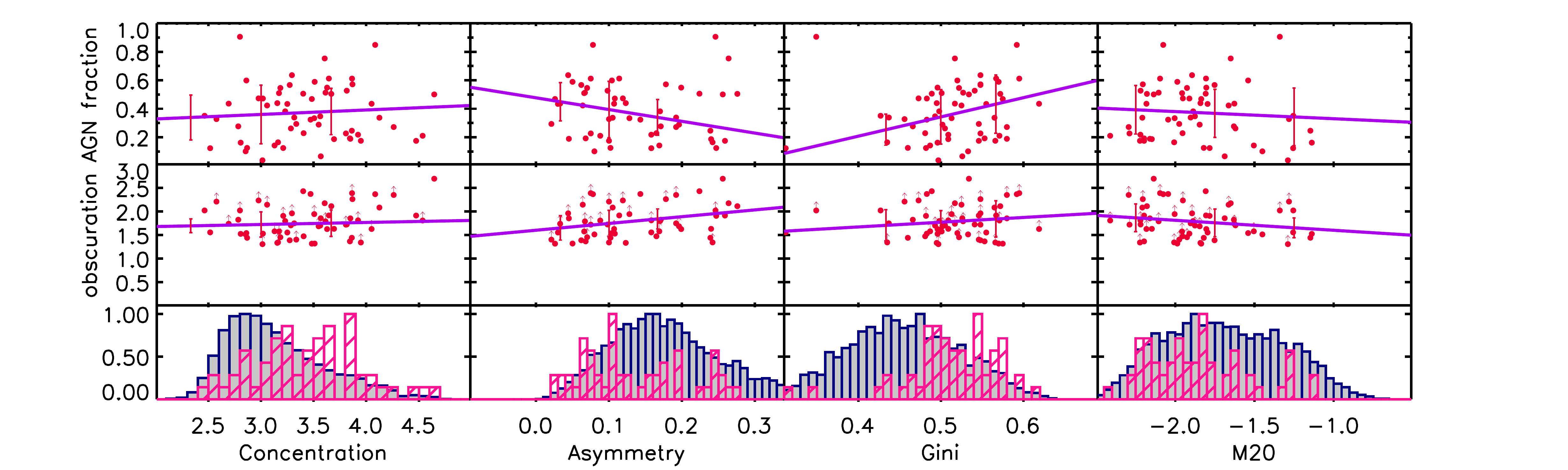} 
\caption[]{Morphology parameters: concentration, asymmetry, Gini and $M_{20}$  at $0.5<z<1.5$ and $\log M_*/M_\odot>10.5$. Here we compare the morphological parameters with bolometric AGN fraction and obscuration for obscured IR-AGN hosts (red). We also show histograms of obscured IR-AGN hosts (pink) and normal star-forming galaxies (blue/gray). The upper plot shows the value from Zurich Structure \& Morphology catalog \citep{2007ApJS..172..434S,2007ApJS..172..406S} and the lower plot shows the value from \citet{2007ApJS..172..270C} catalog. The purple lines show results of linear fitting and the error bars represent the standard deviation in each bin. We only consider detected sources in the linear fitting for obscuration. 
In general, AGN hosts are more compact and asymmetric compared with normal galaxies. AGN hosts also have slightly higher Gini coefficient and lower $M_{20}$ than star-forming galaxies.}
\label{cosp_nonpara}
\end{figure*}

\begin{figure*}
\centering
\includegraphics[width=1.0\columnwidth]{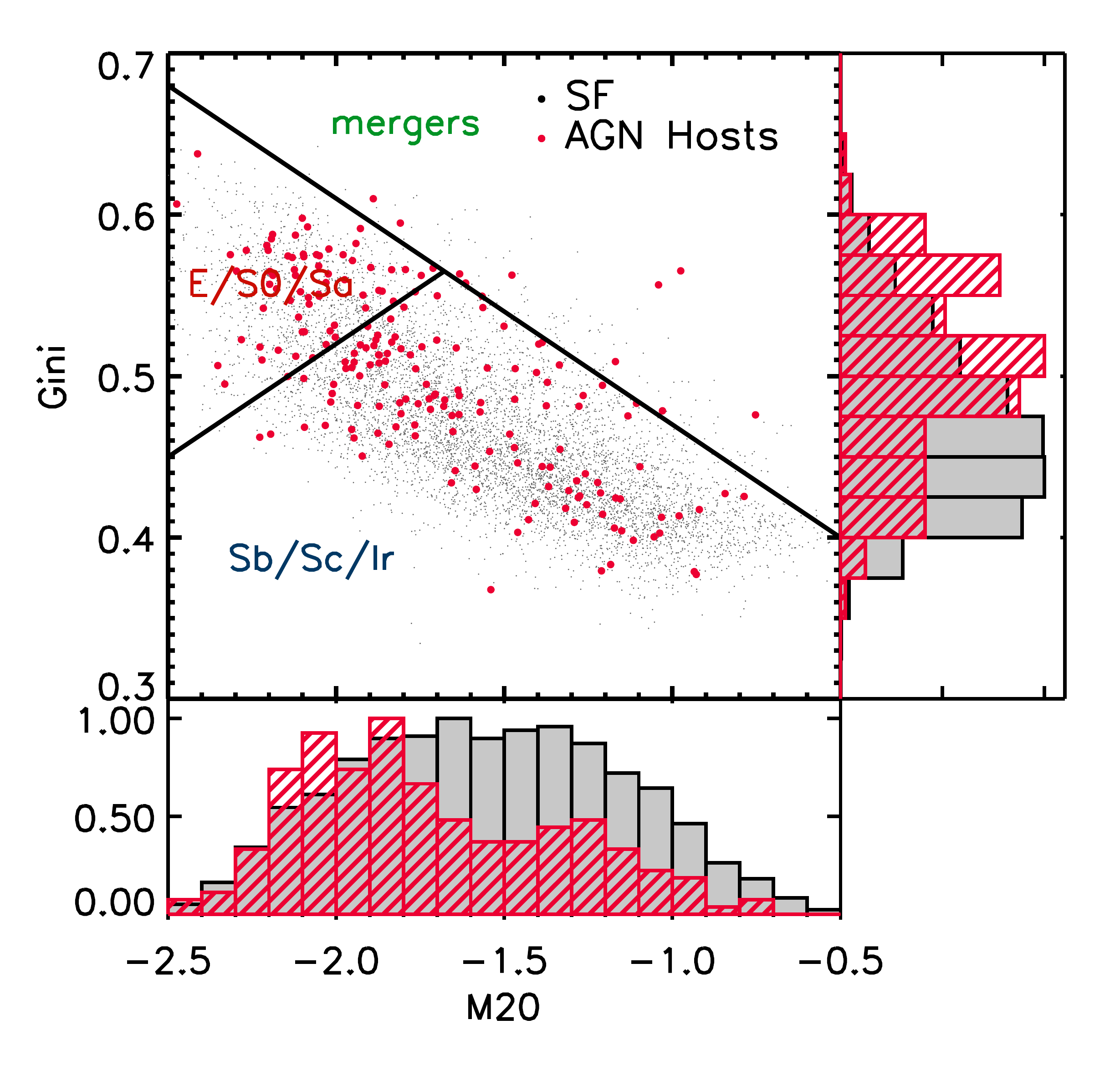} 
\includegraphics[width=1.0\columnwidth]{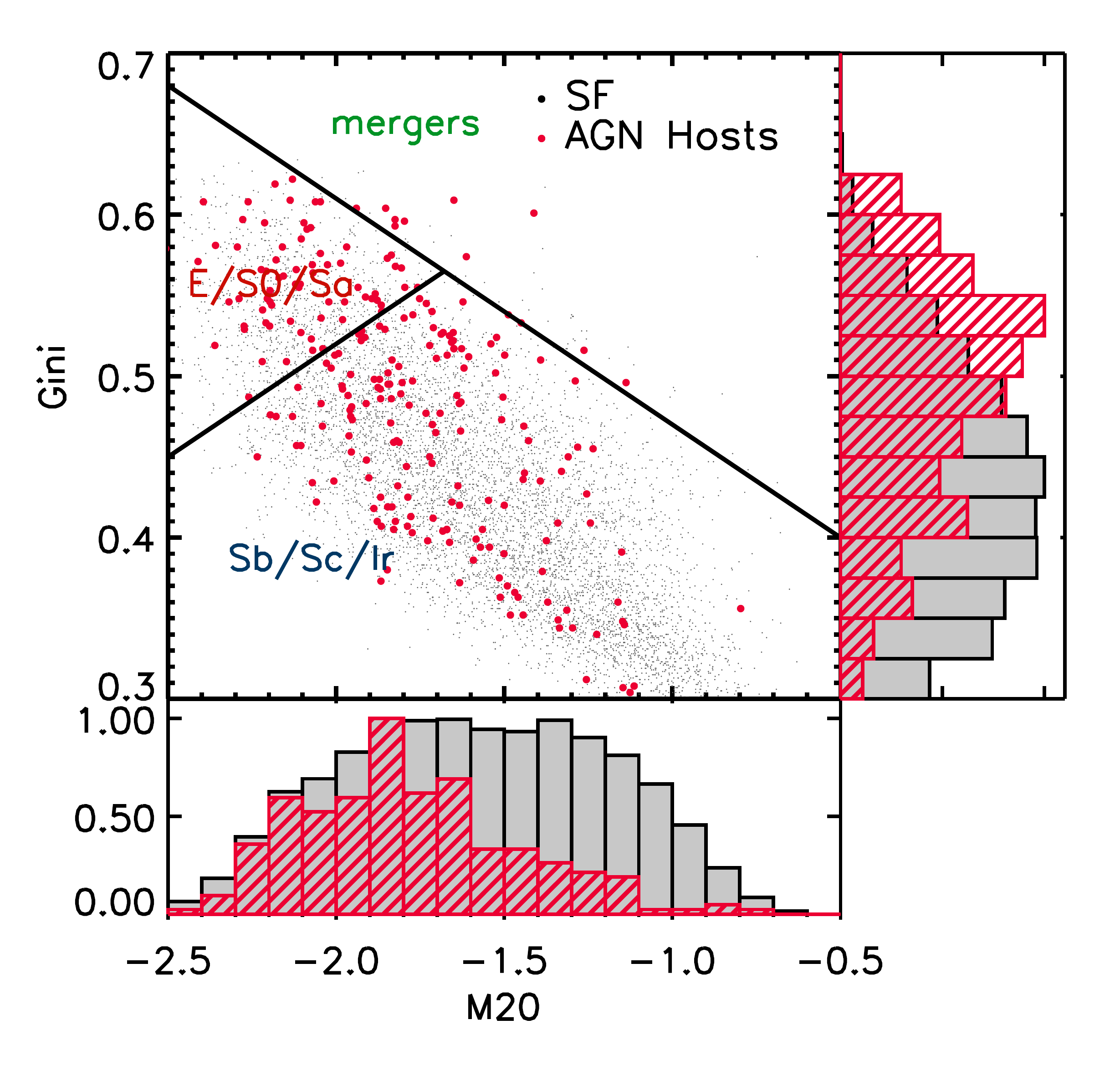} 
\caption[]{Gini to $M_{20}$ plot at $0.5<z<1.5$ and $\log M_*/M_\odot>10.5$. The left plot shows the value from Zurich Structure \& Morphology catalog \citep{2007ApJS..172..434S,2007ApJS..172..406S} and the right plot shows the value from \citet{2007ApJS..172..270C} catalog. The majority of AGN hosts are not merger-like according to the Gini-$M_{20}$ classification.}
\label{cosp_gm20}
\end{figure*}

\begin{figure*}[ht!]
\centering
\includegraphics[width=1.0\textwidth]{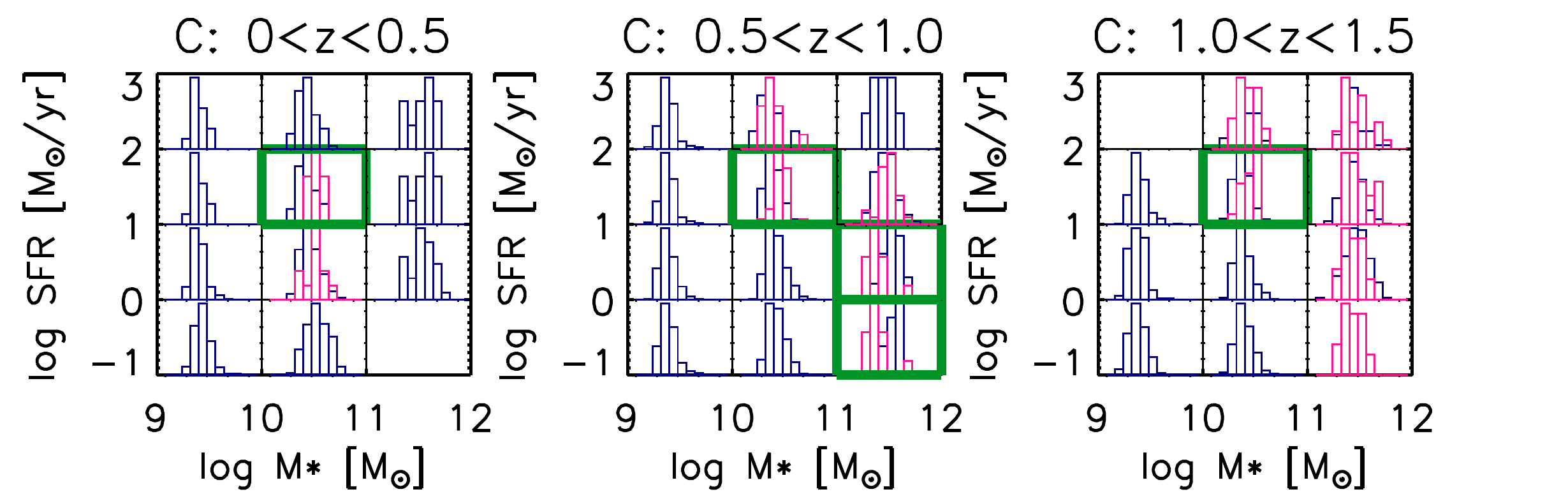} \\
\includegraphics[width=1.0\textwidth]{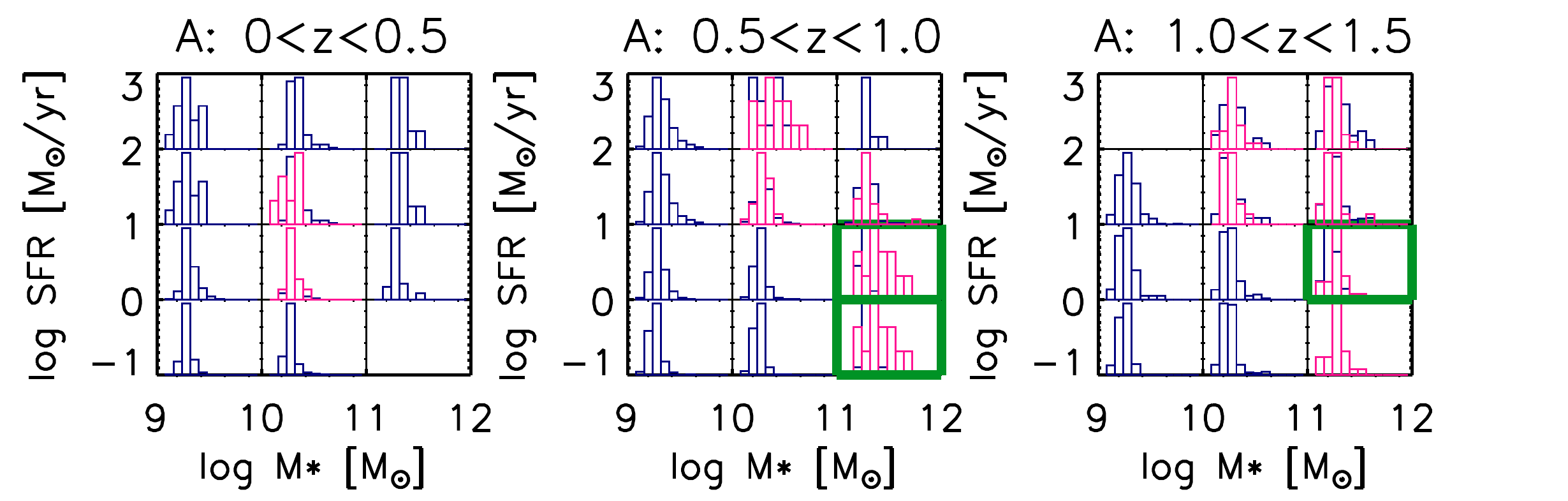} 
\caption[]{Morphology parameters (concentration and asymmetry) histogram and K-S tests for 12 bins of stellar mass and SFR, and 3 redshift ranges. The SFR and stellar mass plots are separated to 12 bins according to the boarders of subplot to show the histogram in each bins. The purple histogram represents AGN hosts and blue histogram represents normal star-forming galaxies. The thick (green) boxes highlight the distributions that are significantly different ($P_{K-S}<$0.05).}
\label{cosp_ks_morpho1}
\end{figure*}

\begin{figure*}[ht!]
\centering
\includegraphics[width=1.0\textwidth]{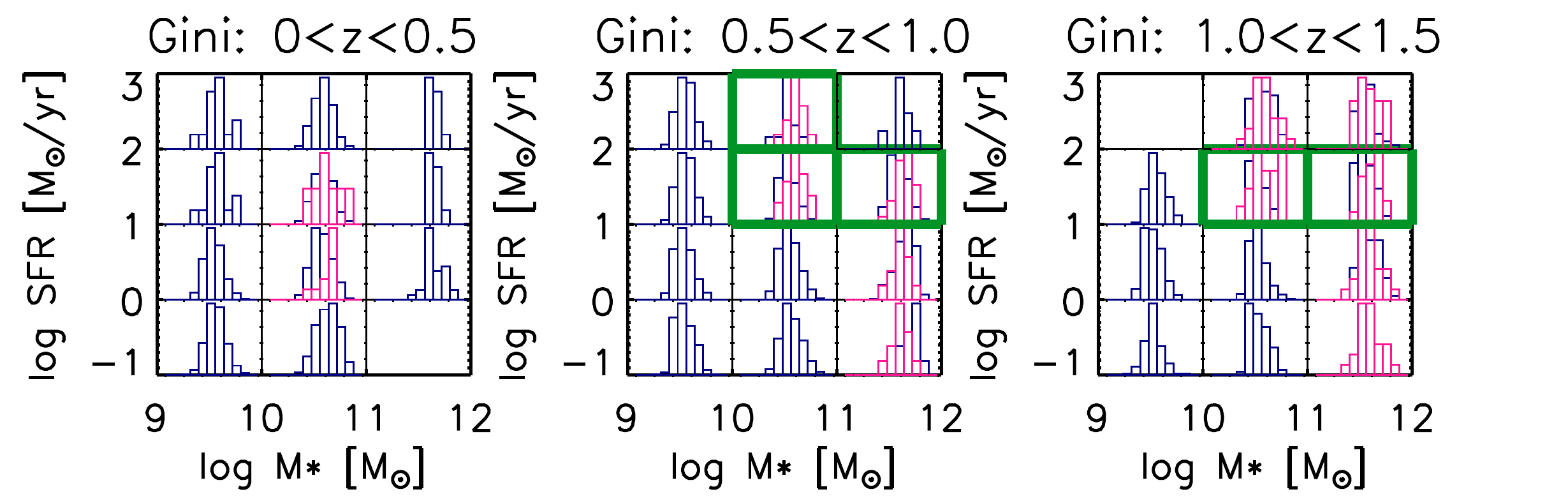} \\
\includegraphics[width=1.0\textwidth]{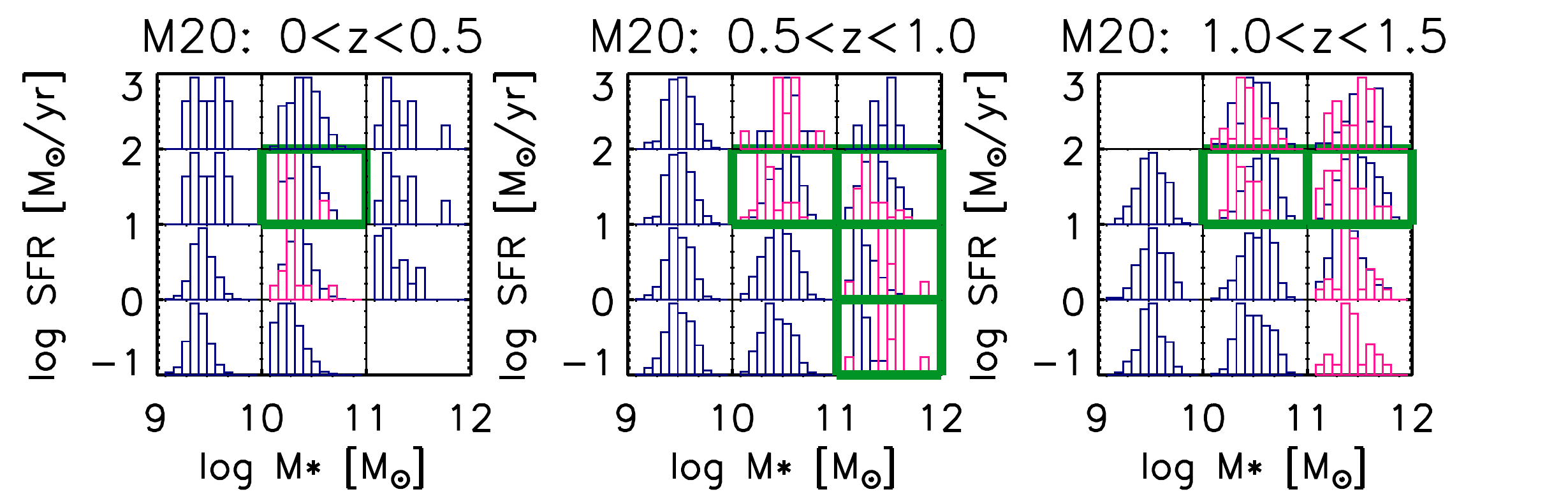} 
\caption[]{Morphology parameters (Gini and $M_{20}$) histogram and K-S tests for 12 bins of stellar mass and SFR, and 3 redshift ranges. The SFR and stellar mass plots are separated to 12 bins according to the boarders of subplot to show the histogram in each bins. The purple histogram represents AGN hosts and blue histogram represents normal star-forming galaxies. The thick (green) boxes highlight the distributions that are significantly different ($P_{K-S}<$0.05).}
\label{cosp_ks_morpho2}
\end{figure*}

We used the Zurich Structure \& Morphology catalog \citep{2007ApJS..172..434S,2007ApJS..172..406S} in the COSMOS field to investigate the relation between nonparametric methods and AGN fraction/obscuration in the upper plot of Figure\,~\ref{cosp_nonpara}. 
The four nonparametric measures are individual estimators of galaxy structures: concentration $C$, asymmetry $A$, Gini coefficient $G$, and second-order moment of the
brightest 20\% of galaxy pixels $M_{20}$ \citep[e.g.,][]{2003ApJS..147....1C,2003ApJ...588..218A,2004AJ....128..163L}. 
We also compared with the \citet{2007ApJS..172..270C} catalog in the lower plot. It is clear that AGN host galaxies (pink) are significantly different from star-forming galaxies (blue/gray). In general, AGN hosts are more compact and asymmetric compared with normal galaxies. AGN hosts also have slightly higher Gini coefficient and lower $M_{20}$ than star-forming galaxies but the majority of them are still not merger-like in the Gini-$M_{20}$ plot as shown in Figure\,~\ref{cosp_gm20}. The lines here are from \citet{2008ApJ...672..177L}, which was adopted for $0.2<z<1.2$ Extended Groth Strip galaxies. The merger fraction might evolve over redshifts \citep{2011ApJ...742..103L, 2016MNRAS.458..963P}, but our control sample also provides a fair comparison. Both catalogs show that AGN fraction follows the directions of the offset of structural parameters, that is, AGN fraction is higher while the hosts are more compact, asymmetric, and bulgier (higher Gini and lower $M_{20}$). Obscuration has no strong correlation except that obscured AGN hosts seem to be more symmetric. This suggests that the compactness is not sensitive to obscuration in our sample. 

Figure\,~\ref{cosp_ks_morpho1} and \ref{cosp_ks_morpho2}  show the comparison of nonparametric measures in the mass to SFR plot at different redshift bins. The Kolmologrov-Smirnov (K-S) tests suggest that morphology parameters of AGN hosts and normal star-forming galaxies are indistinguishable in many bins, but sill show significant differences ($P_{K-S}<$0.05) in several bins that we highlight with thick green boxes. 
Though the sample size is small in each bin, this implies that obscured AGN hosts do not necessarily have similar structures as normal galaxies. It might be also linked to our previous finding of compact AGN host galaxies, and provide possible constraints on future scenarios. 

\subsection{Visual Classification}
\label{sec4_3}

\begin{figure}
\centering
\includegraphics[width=1.00\columnwidth]{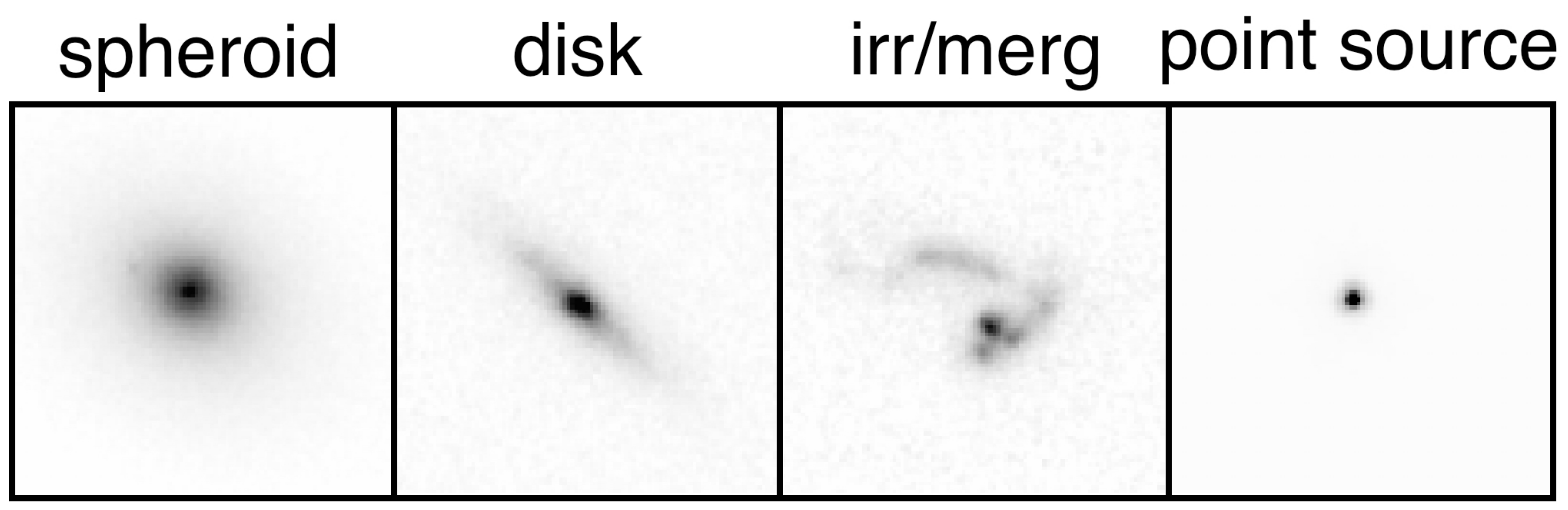} 
\caption[]{Example of morphology class of HST/ACS imaging in our visual classification: disk, spheroid, irregular/merger, and point source. The box size is 6''$\times$6''.}
\label{cosp_vis}
\end{figure}

\begin{figure}
\centering
\includegraphics[width=1.00\columnwidth]{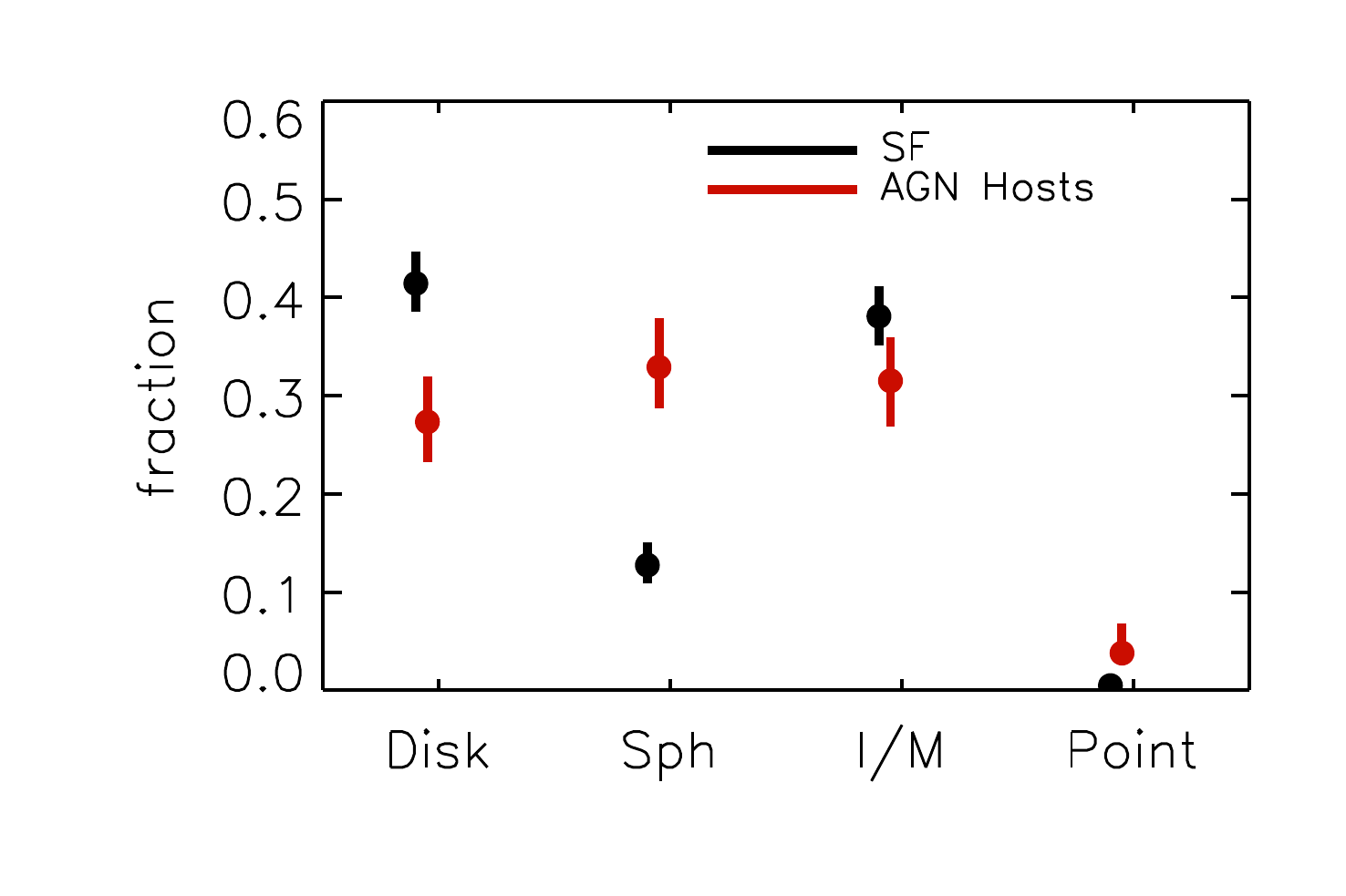} 
\caption[]{Visual classification for AGNs and star-forming galaxies at $0.5<z<1.5$ and $\log M_*/M_\odot>10.5$. The error bars represents the  68.3\% binomial confidence limits as described in \citet{2011PASA...28..128C}. The merger fraction of AGN host is not higher than normal star-forming galaxies.}
\label{cosp_visstat}
\end{figure}

\begin{figure}[h]
\centering
\includegraphics[width=1.00\columnwidth]{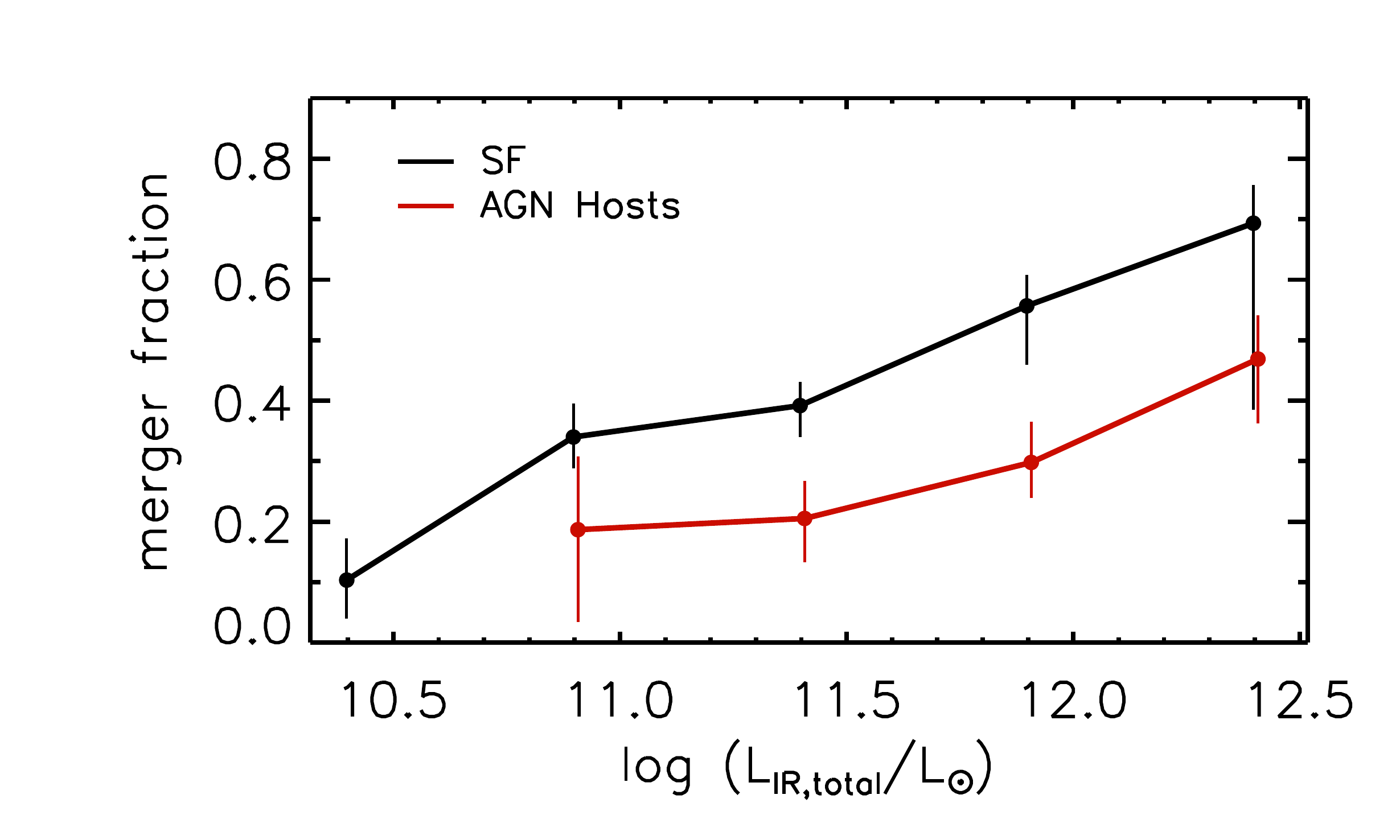}
\includegraphics[width=1.00\columnwidth]{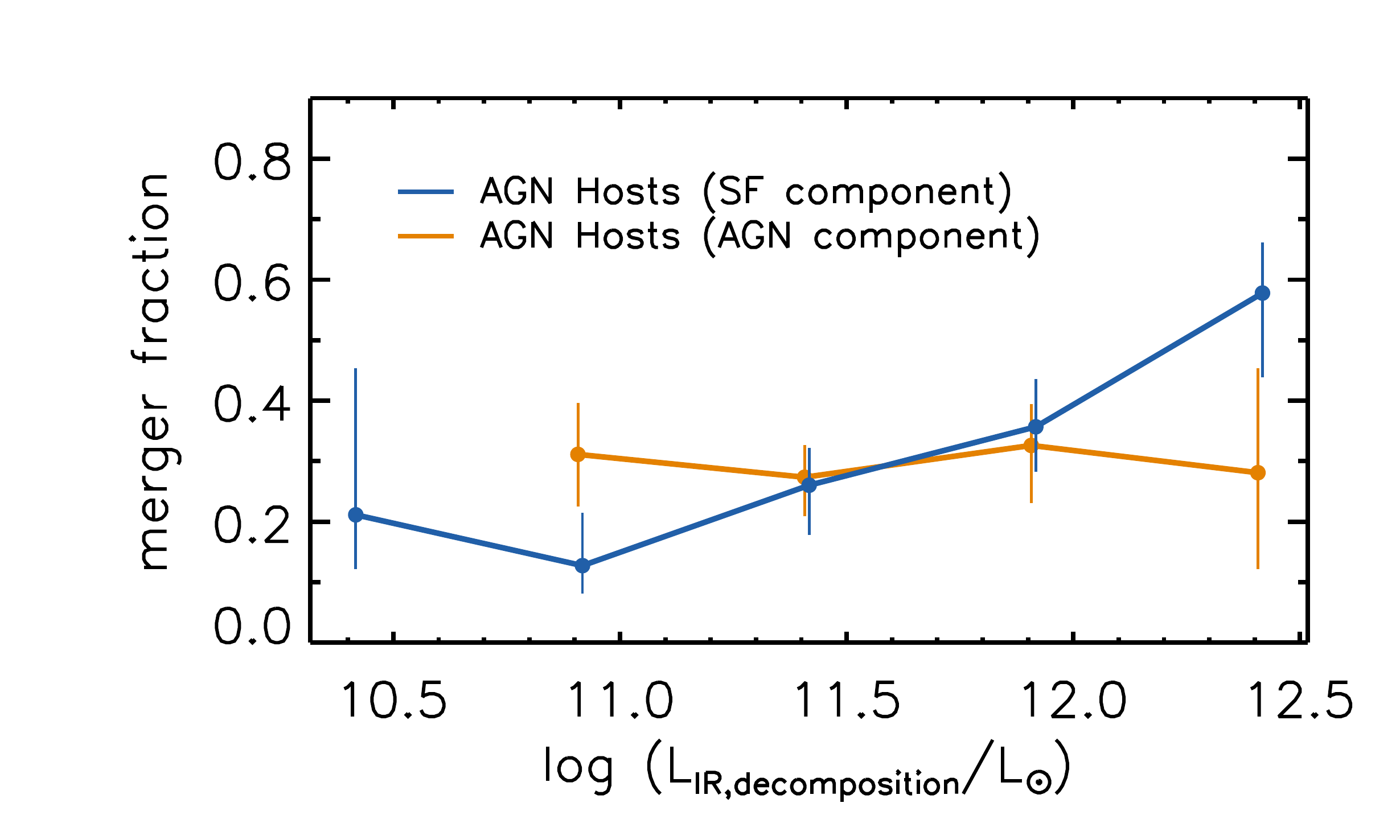}
\caption[]{Upper: Merger fraction from visual classification as a function of total infrared luminosity for IR-AGN hosts (red) and star-forming galaxies (black) at $0.5<z<1.5$ and $\log M_*/M_\odot>10.5$. Lower: Merger fraction in the IR-AGN population as a function of AGN (orange) and star-forming (blue) IR luminosity obtained from our SED decomposition. The error bars in each bin represents the  68.3\% binomial confidence limits as described in \citet{2011PASA...28..128C}. These two plots show that merger fraction depends on the total and star-forming infrared luminosity, rather than the decomposed AGN infrared luminosity.}
\label{cosp_mf}
\end{figure}

We separated all AGNs (infrared and X-ray selected) and randomly selected 1000 star-forming galaxies at $z<1.5$ and $\log M_*/M_\odot>10$ which we classify using four classes: disk, spheroid, irregular/merger, and point source (Figure\,~\ref{cosp_vis}). 
These classes are mutually exclusive, so the classification represents the dominant morphology.  
All the objects are examined by 6 classifiers (Y.-Y.C., Y.T., C.-F.L., J.-J.T., W.H.-W., N.F.). 
Following previous subsections, we only focus on the obscured IR-AGN and the control sample at $0.5<z<1.5$ and $\log M_*/M_\odot>10.5$ as our earlier work \citep{2017MNRAS.466L.103C}.
Figure\,~\ref{cosp_visstat} shows the classification results. The error bars in each class represents the  68.3\% (1-$\sigma$) confidence limits, derived by the method in \citet{2011PASA...28..128C}, which considers the estimation of confident intervals for a binomial population with a Bayesian approach.
This shows that  the spheroid fraction of AGNs ($\sim$ 30\%) is higher than star-forming galaxies ($\sim$ 10\%). The merger fraction of obscured AGNs ($\sim$ 30\%) is  lower than that of the control sample ($\sim$ 40\%). The sample size is limited to make small bins of redshift and stellar mass. Nevertheless, the results are consistent for objects at $0.5<z<1.5$ and $\log M_*/M_\odot>10.5$.

In the upper panel of Figure\,~\ref{cosp_mf}, we find a correlation between total infrared luminosity of obscured AGN and the merger fraction. For the most luminous AGNs ($\log (L_{IR}/L_\odot)\sim 12.5$), merger fraction can reach up to a fraction of 0.5. Moreover, the increase of the merger fraction occurs for the whole star-forming population, not only AGN hosts.
The lower panel of Figure\,~\ref{cosp_mf} shows that, from our SED decomposition, merger fraction has no dependence on the AGN infrared luminosity but does have dependence on the star-forming infrared luminosity. This implies that the infrared luminosity of star formation can be responsible for the disturbed features, rather than the infrared luminosity of AGN.
In general, our obscured AGN hosts have no strong disturbed features, which implies that the merger features is dominated by the total luminosity.
It might be more difficult to distinguish merger features of faint sources. Nevertheless, the increased merger rate only happens to the most luminous galaxies for both normal and AGN host galaxies.

\section{Discussion}
\label{sec5}

\subsection{How does star formation regulate AGN activity?}
\label{sec5_1}

In general, the offsets on the star-forming sequence between AGN hosts and star-forming galaxies are not significant. At $z<1.5$, our obscured AGN hosts are slightly above the main sequence, which seems to be consistent with previous findings of SFR enhancement in MIR selected galaxies \citep{2016MNRAS.458L..34E,2017ApJ...835...27A}. A possible explanation is that an abundant gas supply triggers enhanced star formation in an obscured AGN phase. Moreover, the sSFRs of massive obscured AGN hosts are also slightly higher than in the control sample up to $z\sim2.5$. 
However, lack of star-bursting host galaxies (i.e., galaxies above the star-forming sequence) of the obscured sample at $10.5<\log M_*/M_\odot<11$ leads a low sSFR at $z>1.5$. 
With the mass-matched sample in Section \ref{sec3_4}, we can see there is little difference on the star-formation-heated infrared luminosity among the AGNs and the control sample. Overall, obscured AGN host galaxies have slightly higher SFRs, sSFRs, and infrared luminosities of their star-forming component, which is consistent with the result in \citet{2013ApJ...764..176J}. 

Beside the (specific) star formation rate, other physical properties from SED fitting are slightly different (low significance level by K-S test) between AGN hosts and normal galaxies as shown in Section \ref{sec3_4}. In particular, the stellar mass weighted ages, attenuation, and infrared parameters of our obscured samples are significantly different ($P_{K-S}<$0.05) from normal star-forming galaxies. It is natural to see higher infrared luminosity and AGN fraction for our MIR-selected IR-AGNs.  At lower redshift, obscured AGN hosts seem slightly younger and more dust attenuated than star-forming galaxies. Owing to the small offsets, it is difficult to rule out or confirm a model in which mergers fuel a rapid starburst and a phase of obscured black hole growth, followed by an unobscured phase (e.g., \citet{1988ApJ...325...74S}, see \citet{2012NewAR..56...93A} for a review) due to the small offsets. However, at $z>1.5$, the stellar mass weighted ages of obscured AGN hosts are significantly different and larger than those of the control sample. 
It might be related to the previous results about the lifetime of strongly clustered obscured AGNs \citep{2011ApJ...731..117H,2014MNRAS.442.3443D,2017ApJ...835...36T}, but we still need further constraints on ages since ages derived from broad-band photometry are highly uncertain.
Moreover, our SFRs take the averaged star formation over the last $10^8$ years while AGNs can be luminous on scales as low as $10^5$ years. It might be a reason why it is difficult to see significant difference in location on the main sequence between AGN and normal galaxies.

\subsection{Are Obscured IR-AGNs different from X-ray selected AGNs?}
\label{sec5_2}

Compton-thick AGNs are hidden by extreme column densities ($N_H\gtrsim10^{24}$ cm$^{-2}$) because neutral gas can absorb X-ray photons. They are believed to provide important contribution to the overall cosmic energy budget as well as constraints on the co-evolution of AGN and galaxies. However, the identification of the heavily obscured AGNs is difficult, and a significant fraction of AGN are hidden by Compton-thick obscuration \citep[e.g., ][]{2003ApJ...598..886U,2009ApJ...696..110T}. Our AGN sample is based on an infrared color-color selection, rather than the widely used X-ray selection, and  we select obscured AGNs by the ratio between infrared and X-ray luminosity. It provides a sample of obscured AGNs which is complementary to a Compton-thick AGN sample selected purely by X-ray observations.

\citet{2017ApJ...841..102S} found that X-ray type-2 AGN hosts have similar SFRs compared to the normal star-forming galaxies. 
In this paper, we confirmed this result with our X-ray selected sample, and showed that IR-selected AGN hosts are not significant different from or slightly above the star-forming sequence. 
Besides, the distribution of physical parameters, such as stellar-mass weighted age, attenuation, and infrared properties, seems to be different for infrared selected AGNs, while the distribution of X-ray selected AGNs is close to that of normal galaxies. The differences between AGN types might be interpreted as different phases in the evolutionary sequence \citep{2008ApJS..175..356H}. \citet{2009ApJ...696..891H} showed that X-ray selected AGNs are preferentially found in the `green valley' and clustered similarly to normal galaxies, while infrared selected AGNs reside in slightly bluer, slightly less luminous galaxies than X-ray AGNs, and are weakly clustered. Moreover, \citet{2010ApJ...716..348B} discovered an obscured QSO at high-z caught in a transition stage from being starburst dominated to AGN dominated. Our results from a MIR selection sample imply that obscured AGNs can be in a different evolutionary stage than X-ray selected AGNs. 

According to the Gini-$M_{20}$ plot in Figure\,~\ref{cosp_gm20} and visual classification in Figure\,~\ref{cosp_mf}, we did not find a high merger rate in the whole obscured AGN sample. 
However, Figure\,~\ref{cosp_mf} showed an increasing merger fraction with infrared luminosity, which is consistent with the model prediction of \citet{2014ApJ...782....9H} and many previous observational results \citep[e.g.,][]{2011ApJ...730..125Z,2015ApJ...814..104K,2015A&A...578A.120L,2016ApJ...822L..32F}.
The high merger fraction of the most luminous obscured AGN hosts is consistent with recent results by HST/WFC3 imaging \citep[][Donley et al.]{2015ApJ...814..104K}, which suggested that Compton-thick AGNs are in a different phase of obscured supermassive black holes growth following a merger/interaction event. 
The little dependence of merger fraction on the infrared luminosity from the AGN component implies that the interaction event is more relevant to the total or star-forming infrared luminosity, rather than the AGN infrared luminosity. 
Moreover, we also find a correlation between merger fraction and total infrared luminosity for normal galaxies.
It is interesting that star-forming galaxies are more likely to be in a merger than AGN hosts at $\log (L_{IR}/L_\odot)>11$. A possible explanation is that our obscured AGNs are selected by infrared with intermediate luminosity. If we consider their compact features which will be discussed in the next subsection together, it may suggest that these obscured AGN hosts are in a special evolutionary stage. This finding can provide morphological constrains for future studies on different types of AGN host in galaxy evolution \citep[e.g.,][]{2012NewAR..56...93A,2014ApJ...783...40G}.
Our results also suggest that the whole obscured sample, including non X-ray detected sources, shows no stronger disturbed features compared to normal galaxies, and the increased merger fraction may only happen to the most luminous galaxies. 

\subsection{Are AGNs triggered by compaction?}
\label{sec5_3}

In terms of optical morphological interpretation, obscured AGN offer an advantage in avoidance of the PSF contribution to the optical image. 
This is consistent with earlier X-ray results without consideration of AGN contribution \citep{2005ApJ...627L..97G,2007ApJ...660L..19P}. More recent studies consider the central nucleus light and show similar \citep{2013A&A...549A..46B,2014MNRAS.439.3342V,2014ApJ...784L...9F} or flatter \citep{2011ApJ...743L..37S,2012ApJ...761...75S} radial light distribution of AGNs compare to inactive galaxies. However, recent results also show that the consideration of AGN contribution can be overestimated and it is difficult to separate the central bulge and the nucleus light \citep{2015A&A...573A..85R,2016MNRAS.458.2391B}. 
Therefore, obscured AGNs provide a robust sample to measure the optical light of host galaxies \citep{2017MNRAS.466L.103C}.
We noted that the majority of our AGN hosts are at $10.5<\log M_*/M_\odot<11$, and the most significant difference is also caused by these objects. 
The different sample selection and stellar mass range might explain the different results by \citet{2015A&A...573A..85R}, which showed the X-ray selected AGNs have slightly diskier light profiles than inactive galaxies at z$\sim$1 and a red central light enhancement at z$\sim$2. 
The matched stellar masses and redshifts ensure an unbiased selection in our sample. 
As a sanity check of the selection bias of 24 $\mu$m between galaxies with and without AGNs, we tested galaxies above a luminosity threshold at 24 $\mu$m by subtracting AGN contribution from the SEDs. This ensures that we remove a possible population of passive galaxies hosting AGNs, which are selected at 24 $\mu$m because of the accretion disk emission. We recalculated the size and S\'ersic index dependences with this sample, and still find very similar results as in Figure\,~\ref{cosp_galfit} and \ref{cosp_galfity}, that is, AGN host galaxies are more compact than normal galaxies.

From a morphological point of view, we investigated the two-dimensional surface brightness modeling and nonparametric methods. The obscured AGN host galaxies are smaller, more compact, more asymmetric, and more bulge-like than the control sample. At $z\sim$1, we found that a 20-50\% AGN contribution corresponds to a decreased size by 25-50\%. The correlation between AGN bolometric fraction and structural parameters implies that the AGN activity can be the cause of compactness. 
Besides, the lack of correlation between obscuration and structural parameters might be a hint that the morphological differences between AGN hosts and the control sample also happens to less obscured AGNs. However, we still have to be careful to consider the AGN contribution of unobscured sample in future works. Moreover, distribution of structural parameters show significant differences at several stellar mass, SFR, and redshift bins by K-S tests, and we also found higher fraction of spheroidal-like host galaxies compared to a control sample of star-forming galaxies by visual classification. This result confirmed our previous finding in \citet{2017MNRAS.466L.103C}, that is, obscured AGN hosts are more compact compared to the control sample.

In the compaction scenario \citep{2016MNRAS.457.2790T}, galaxies live through one or more blue nugget phases which a minimum in gas depletion time and a maximum in gas fraction are reached. As shown in Figure\,~\ref{cosp_ms_z} and \ref{cosp_ms}, the most massive and $z<1.5$ obscured AGNs are slightly above the main sequence, which is consistent with the scenario of the blue nugget phase. The lower sSFR at $10.5<\log M_*/M_\odot<11$ at higher redshifts can be explained by the hypothesis that the sSFR fluctuates down and up several times before it eventually quenches beyond the green valley \citep{2015MNRAS.450.2327Z}. Our results suggest that this scenario can also happen to z$\sim$1 obscured AGN hosts, especially those resulting from infrared selections. 
Since the fueling of central AGNs might be affected \citep{2015MNRAS.452.1502D}, our results of the physical properties, such as the stellar mass weighted age, extinction, and dust properties, could also provide further constraints on the compaction scenario. 

\section{Summary}
\label{sec6}

In this paper, we have provided a detailed study of infrared selected AGNs. We have found that the hosts of these AGNs are not exactly the same as normal star-forming galaxies, both from their physical properties and morphology. We confirm our previous structural results about obscured AGNs. Our main findings are as follows.

\begin{enumerate}

\item IR-AGNs and obscured IR-AGNs are located within or maybe slightly above the star-forming sequence at $z\sim 1$. 

\item Beside SFR, obscured AGNs show significantly different distributions for several physical parameters from SED fitting, such as stellar-mass weighted age, attenuation, and infrared properties. This suggests that obscured AGNs can be in a distinct evolutionary stage o X-ray selected AGNs. 

\item According to the correlation between bolometric AGN fraction and the structural parameters,  a 20\% AGN contribution corresponds to a decreased radius by 24\% and an increased S\'ersic index by 18\%, and a 50\% AGN contribution corresponds to a decreased radius by 50\% and an increased S\'ersic index by 47\%.

\item We do not find high merger rates in the whole obscured AGN samples, which includes non X-ray detected sources. However, the merger rate of the most luminous AGNs ($\log (L_{IR}/L_\odot)\sim 12.5$) can be up to $\sim 0.5$. The increasing disturbed features with infrared luminosity are consistent with previous finding on Compton-thick AGNs. We conclude that high merger fraction may only apply to the most luminous and heavily obscured AGN hosts, but not to the whole obscured AGN sample.

\item Merger fraction has no dependence on the AGN infrared luminosity derived from our SED decomposition. This implies that most obscured AGNs might be triggered by internal mechanisms, such as secular process, disk instabilities, and compaction.

\item We confirm our previous finding about compact obscured AGNs at z$\sim$1. Our results of sSFR show that massive obscured AGNs are slightly above the main sequence population at $z<2$. This implies that they might be in the blue nuggets phase, and affected by fluctuation in the compaction scenario. The differences of the physical properties between obscured AGNs and the control sample may provide further constraints on the compaction scenario.

\item We make publicly available the SED modeling results for all available objects in the COSMOS2015 catalog. 

\end{enumerate}

\acknowledgments
We thank the referee as well as D. Elbaz and A. Dekel for helpful comments and discussions.
We acknowledge financial support from Agence Nationale de la Recherche (contract \#ANR-12-JS05-0008-01), and the Ministry of Science and Technology of Taiwan grant (105-2112-M-001-029-MY3; Y.-Y.C., C.-F.L., W.-H.W).
We gratefully acknowledge the contributions of the entire COSMOS collaboration. The COSMOS team in France acknowledges support from the Centre National d'\'Etudes Spatiales.






\bibliographystyle{aasjournal}

\begin{thebibliography}{}
\expandafter\ifx\csname natexlab\endcsname\relax\def\natexlab#1{#1}\fi

\bibitem[{{Abraham} {et~al.}(2003){Abraham}, {van den Bergh}, \&
  {Nair}}]{2003ApJ...588..218A}
{Abraham}, R.~G., {van den Bergh}, S., \& {Nair}, P. 2003, \apj, 588, 218

\bibitem[{{Aird} {et~al.}(2012){Aird}, {Coil}, {Moustakas}, {Blanton},
  {Burles}, {Cool}, {Eisenstein}, {Smith}, {Wong}, \&
  {Zhu}}]{2012ApJ...746...90A}
{Aird}, J., {Coil}, A.~L., {Moustakas}, J., {et~al.} 2012, \apj, 746, 90

\bibitem[{{Alexander} \& {Hickox}(2012)}]{2012NewAR..56...93A}
{Alexander}, D.~M., \& {Hickox}, R.~C. 2012, New Astronomy Reviews, 56, 93

\bibitem[{{Azadi} {et~al.}(2017){Azadi}, {Coil}, {Aird}, {Reddy}, {Shapley},
  {Freeman}, {Kriek}, {Leung}, {Mobasher}, {Price}, {Sanders}, {Shivaei}, \&
  {Siana}}]{2017ApJ...835...27A}
{Azadi}, M., {Coil}, A.~L., {Aird}, J., {et~al.} 2017, \apj, 835, 27

\bibitem[{{B{\"o}hm} {et~al.}(2013){B{\"o}hm}, {Wisotzki}, {Bell}, {Jahnke},
  {Wolf}, {Bacon}, {Barden}, {Gray}, {Hoeppe}, {Jogee}, {McIntosh}, {Peng},
  {Robaina}, {Balogh}, {Barazza}, {Caldwell}, {Heymans}, {H{\"a}u{\ss}ler},
  {van Kampen}, {Lane}, {Meisenheimer}, {S{\'a}nchez}, {Taylor}, \&
  {Zheng}}]{2013A&A...549A..46B}
{B{\"o}hm}, A., {Wisotzki}, L., {Bell}, E.~F., {et~al.} 2013, \aap, 549, A46

\bibitem[{{Bongiorno} {et~al.}(2014){Bongiorno}, {Maiolino}, {Brusa},
  {Marconi}, {Piconcelli}, {Lamastra}, {Cano-D{\'{\i}}az}, {Schulze},
  {Magnelli}, {Vignali}, {Fiore}, {Menci}, {Cresci}, {La Franca}, \&
  {Merloni}}]{2014MNRAS.443.2077B}
{Bongiorno}, A., {Maiolino}, R., {Brusa}, M., {et~al.} 2014, \mnras, 443, 2077

\bibitem[{{Bongiorno} {et~al.}(2016){Bongiorno}, {Schulze}, {Merloni},
  {Zamorani}, {Ilbert}, {La Franca}, {Peng}, {Piconcelli}, {Mainieri},
  {Silverman}, {Brusa}, {Fiore}, {Salvato}, \&
  {Scoville}}]{2016A&A...588A..78B}
{Bongiorno}, A., {Schulze}, A., {Merloni}, A., {et~al.} 2016, \aap, 588, A78

\bibitem[{Breiman {et~al.}(1984)Breiman, Friedman, Olshen, \& Stone}]{cart84}
Breiman, L., Friedman, J., Olshen, R., \& Stone, C. 1984, {Classification and
  Regression Trees} (Monterey, CA: Wadsworth and Brooks)

\bibitem[{{Brightman} {et~al.}(2014){Brightman}, {Nandra}, {Salvato}, {Hsu},
  {Aird}, \& {Rangel}}]{2014MNRAS.443.1999B}
{Brightman}, M., {Nandra}, K., {Salvato}, M., {et~al.} 2014, \mnras, 443, 1999

\bibitem[{{Brinchmann} {et~al.}(2004){Brinchmann}, {Charlot}, {White},
  {Tremonti}, {Kauffmann}, {Heckman}, \& {Brinkmann}}]{2004MNRAS.351.1151B}
{Brinchmann}, J., {Charlot}, S., {White}, S.~D.~M., {et~al.} 2004, \mnras, 351,
  1151

\bibitem[{{Bruce} {et~al.}(2016){Bruce}, {Dunlop}, {Mortlock}, {Kocevski},
  {McGrath}, \& {Rosario}}]{2016MNRAS.458.2391B}
{Bruce}, V.~A., {Dunlop}, J.~S., {Mortlock}, A., {et~al.} 2016, \mnras, 458,
  2391

\bibitem[{{Brusa} {et~al.}(2010){Brusa}, {Civano}, {Comastri}, {Miyaji},
  {Salvato}, {Zamorani}, {Cappelluti}, {Fiore}, {Hasinger}, {Mainieri},
  {Merloni}, {Bongiorno}, {Capak}, {Elvis}, {Gilli}, {Hao}, {Jahnke},
  {Koekemoer}, {Ilbert}, {Le Floc'h}, {Lusso}, {Mignoli}, {Schinnerer},
  {Silverman}, {Treister}, {Trump}, {Vignali}, {Zamojski}, {Aldcroft},
  {Aussel}, {Bardelli}, {Bolzonella}, {Cappi}, {Caputi}, {Contini},
  {Finoguenov}, {Fruscione}, {Garilli}, {Impey}, {Iovino}, {Iwasawa},
  {Kampczyk}, {Kartaltepe}, {Kneib}, {Knobel}, {Kovac}, {Lamareille},
  {Leborgne}, {Le Brun}, {Le Fevre}, {Lilly}, {Maier}, {McCracken}, {Pello},
  {Peng}, {Perez-Montero}, {de Ravel}, {Sanders}, {Scodeggio}, {Scoville},
  {Tanaka}, {Taniguchi}, {Tasca}, {de la Torre}, {Tresse}, {Vergani}, \&
  {Zucca}}]{2010ApJ...716..348B}
{Brusa}, M., {Civano}, F., {Comastri}, A., {et~al.} 2010, \apj, 716, 348

\bibitem[{{Brusa} {et~al.}(2015){Brusa}, {Bongiorno}, {Cresci}, {Perna},
  {Marconi}, {Mainieri}, {Maiolino}, {Salvato}, {Lusso}, {Santini}, {Comastri},
  {Fiore}, {Gilli}, {La Franca}, {Lanzuisi}, {Lutz}, {Merloni}, {Mignoli},
  {Onori}, {Piconcelli}, {Rosario}, {Vignali}, \&
  {Zamorani}}]{2015MNRAS.446.2394B}
{Brusa}, M., {Bongiorno}, A., {Cresci}, G., {et~al.} 2015, \mnras, 446, 2394

\bibitem[{{Cameron}(2011)}]{2011PASA...28..128C}
{Cameron}, E. 2011, \pasa, 28, 128

\bibitem[{{Cassata} {et~al.}(2007){Cassata}, {Guzzo}, {Franceschini},
  {Scoville}, {Capak}, {Ellis}, {Koekemoer}, {McCracken}, {Mobasher},
  {Renzini}, {Ricciardelli}, {Scodeggio}, {Taniguchi}, \&
  {Thompson}}]{2007ApJS..172..270C}
{Cassata}, P., {Guzzo}, L., {Franceschini}, A., {et~al.} 2007, \apjs, 172, 270

\bibitem[{{Ceverino} {et~al.}(2010){Ceverino}, {Dekel}, \&
  {Bournaud}}]{2010MNRAS.404.2151C}
{Ceverino}, D., {Dekel}, A., \& {Bournaud}, F. 2010, \mnras, 404, 2151

\bibitem[{{Chabrier}(2003)}]{2003PASP..115..763C}
{Chabrier}, G. 2003, \pasp, 115, 763

\bibitem[{{Chang} {et~al.}(2015){Chang}, {van der Wel}, {da Cunha}, \&
  {Rix}}]{2015ApJS..219....8C}
{Chang}, Y.-Y., {van der Wel}, A., {da Cunha}, E., \& {Rix}, H.-W. 2015, \apjs,
  219, 8

\bibitem[{{Chang} {et~al.}(2017){Chang}, {Le Floc'h}, {Juneau}, {da Cunha},
  {Salvato}, {Civano}, {Marchesi}, {Gabor}, {Ilbert}, {Laigle}, {McCracken},
  {Hsieh}, \& {Capak}}]{2017MNRAS.466L.103C}
{Chang}, Y.-Y., {Le Floc'h}, E., {Juneau}, S., {et~al.} 2017, \mnras, 466, L103

\bibitem[{{Cisternas} {et~al.}(2011){Cisternas}, {Jahnke}, {Inskip},
  {Kartaltepe}, {Koekemoer}, {Lisker}, {Robaina}, {Scodeggio}, {Sheth},
  {Trump}, {Andrae}, {Miyaji}, {Lusso}, {Brusa}, {Capak}, {Cappelluti},
  {Civano}, {Ilbert}, {Impey}, {Leauthaud}, {Lilly}, {Salvato}, {Scoville}, \&
  {Taniguchi}}]{2011ApJ...726...57C}
{Cisternas}, M., {Jahnke}, K., {Inskip}, K.~J., {et~al.} 2011, \apj, 726, 57

\bibitem[{{Civano} {et~al.}(2016){Civano}, {Marchesi}, {Comastri}, {Urry},
  {Elvis}, {Cappelluti}, {Puccetti}, {Brusa}, {Zamorani}, {Hasinger},
  {Aldcroft}, {Alexander}, {Allevato}, {Brunner}, {Capak}, {Finoguenov},
  {Fiore}, {Fruscione}, {Gilli}, {Glotfelty}, {Griffiths}, {Hao}, {Harrison},
  {Jahnke}, {Kartaltepe}, {Karim}, {LaMassa}, {Lanzuisi}, {Miyaji}, {Ranalli},
  {Salvato}, {Sargent}, {Scoville}, {Schawinski}, {Schinnerer}, {Silverman},
  {Smolcic}, {Stern}, {Toft}, {Trakhenbrot}, {Treister}, \&
  {Vignali}}]{2016ApJ...819...62C}
{Civano}, F., {Marchesi}, S., {Comastri}, A., {et~al.} 2016, \apj, 819, 62

\bibitem[{{Conselice}(2003)}]{2003ApJS..147....1C}
{Conselice}, C.~J. 2003, \apjs, 147, 1

\bibitem[{{Coppin} {et~al.}(2010){Coppin}, {Pope}, {Men{\'e}ndez-Delmestre},
  {Alexander}, {Dunlop}, {Egami}, {Gabor}, {Ibar}, {Ivison}, {Austermann},
  {Blain}, {Chapman}, {Clements}, {Dunne}, {Dye}, {Farrah}, {Hughes},
  {Mortier}, {Page}, {Rowan-Robinson}, {Scott}, {Simpson}, {Smail}, {Swinbank},
  {Vaccari}, \& {Yun}}]{2010ApJ...713..503C}
{Coppin}, K., {Pope}, A., {Men{\'e}ndez-Delmestre}, K., {et~al.} 2010, \apj,
  713, 503

\bibitem[{{da Cunha} {et~al.}(2008){da Cunha}, {Charlot}, \&
  {Elbaz}}]{2008MNRAS.388.1595D}
{da Cunha}, E., {Charlot}, S., \& {Elbaz}, D. 2008, \mnras, 388, 1595

\bibitem[{{da Cunha} {et~al.}(2015){da Cunha}, {Walter}, {Smail}, {Swinbank},
  {Simpson}, {Decarli}, {Hodge}, {Weiss}, {van der Werf}, {Bertoldi},
  {Chapman}, {Cox}, {Danielson}, {Dannerbauer}, {Greve}, {Ivison}, {Karim}, \&
  {Thomson}}]{2015ApJ...806..110D}
{da Cunha}, E., {Walter}, F., {Smail}, I.~R., {et~al.} 2015, \apj, 806, 110

\bibitem[{{Daddi} {et~al.}(2007){Daddi}, {Dickinson}, {Morrison}, {Chary},
  {Cimatti}, {Elbaz}, {Frayer}, {Renzini}, {Pope}, {Alexander}, {Bauer},
  {Giavalisco}, {Huynh}, {Kurk}, \& {Mignoli}}]{2007ApJ...670..156D}
{Daddi}, E., {Dickinson}, M., {Morrison}, G., {et~al.} 2007, \apj, 670, 156

\bibitem[{{Dale} {et~al.}(2014){Dale}, {Helou}, {Magdis}, {Armus},
  {D{\'{\i}}az-Santos}, \& {Shi}}]{2014ApJ...784...83D}
{Dale}, D.~A., {Helou}, G., {Magdis}, G.~E., {et~al.} 2014, \apj, 784, 83

\bibitem[{{Dekel} {et~al.}(2009){Dekel}, {Birnboim}, {Engel}, {Freundlich},
  {Goerdt}, {Mumcuoglu}, {Neistein}, {Pichon}, {Teyssier}, \&
  {Zinger}}]{2009Natur.457..451D}
{Dekel}, A., {Birnboim}, Y., {Engel}, G., {et~al.} 2009, \nat, 457, 451

\bibitem[{{Del Moro} {et~al.}(2016){Del Moro}, {Alexander}, {Bauer}, {Daddi},
  {Kocevski}, {McIntosh}, {Stanley}, {Brandt}, {Elbaz}, {Harrison}, {Luo},
  {Mullaney}, \& {Xue}}]{2016MNRAS.456.2105D}
{Del Moro}, A., {Alexander}, D.~M., {Bauer}, F.~E., {et~al.} 2016, \mnras, 456,
  2105

\bibitem[{{DiPompeo} {et~al.}(2014){DiPompeo}, {Myers}, {Hickox}, {Geach}, \&
  {Hainline}}]{2014MNRAS.442.3443D}
{DiPompeo}, M.~A., {Myers}, A.~D., {Hickox}, R.~C., {Geach}, J.~E., \&
  {Hainline}, K.~N. 2014, \mnras, 442, 3443

\bibitem[{{Donley} {et~al.}(2012){Donley}, {Koekemoer}, {Brusa}, {Capak},
  {Cardamone}, {Civano}, {Ilbert}, {Impey}, {Kartaltepe}, {Miyaji}, {Salvato},
  {Sanders}, {Trump}, \& {Zamorani}}]{2012ApJ...748..142D}
{Donley}, J.~L., {Koekemoer}, A.~M., {Brusa}, M., {et~al.} 2012, \apj, 748, 142

\bibitem[{{Dubois} {et~al.}(2015){Dubois}, {Volonteri}, {Silk}, {Devriendt},
  {Slyz}, \& {Teyssier}}]{2015MNRAS.452.1502D}
{Dubois}, Y., {Volonteri}, M., {Silk}, J., {et~al.} 2015, \mnras, 452, 1502

\bibitem[{{Elbaz} {et~al.}(2007){Elbaz}, {Daddi}, {Le Borgne}, {Dickinson},
  {Alexander}, {Chary}, {Starck}, {Brandt}, {Kitzbichler}, {MacDonald},
  {Nonino}, {Popesso}, {Stern}, \& {Vanzella}}]{2007A&A...468...33E}
{Elbaz}, D., {Daddi}, E., {Le Borgne}, D., {et~al.} 2007, \aap, 468, 33

\bibitem[{{Ellison} {et~al.}(2016){Ellison}, {Teimoorinia}, {Rosario}, \&
  {Mendel}}]{2016MNRAS.458L..34E}
{Ellison}, S.~L., {Teimoorinia}, H., {Rosario}, D.~J., \& {Mendel}, J.~T. 2016,
  \mnras, 458, L34

\bibitem[{{Fan} {et~al.}(2014){Fan}, {Fang}, {Chen}, {Li}, {Lv}, {Knudsen}, \&
  {Kong}}]{2014ApJ...784L...9F}
{Fan}, L., {Fang}, G., {Chen}, Y., {et~al.} 2014, \apjl, 784, L9

\bibitem[{{Fan} {et~al.}(2016){Fan}, {Han}, {Fang}, {Gao}, {Zhang}, {Jiang},
  {Wu}, {Yang}, \& {Li}}]{2016ApJ...822L..32F}
{Fan}, L., {Han}, Y., {Fang}, G., {et~al.} 2016, \apjl, 822, L32

\bibitem[{{Fiore} {et~al.}(2009){Fiore}, {Puccetti}, {Brusa}, {Salvato},
  {Zamorani}, {Aldcroft}, {Aussel}, {Brunner}, {Capak}, {Cappelluti}, {Civano},
  {Comastri}, {Elvis}, {Feruglio}, {Finoguenov}, {Fruscione}, {Gilli},
  {Hasinger}, {Koekemoer}, {Kartaltepe}, {Ilbert}, {Impey}, {Le Floc'h},
  {Lilly}, {Mainieri}, {Martinez-Sansigre}, {McCracken}, {Menci}, {Merloni},
  {Miyaji}, {Sanders}, {Sargent}, {Schinnerer}, {Scoville}, {Silverman},
  {Smolcic}, {Steffen}, {Santini}, {Taniguchi}, {Thompson}, {Trump}, {Vignali},
  {Urry}, \& {Yan}}]{2009ApJ...693..447F}
{Fiore}, F., {Puccetti}, S., {Brusa}, M., {et~al.} 2009, \apj, 693, 447

\bibitem[{{Gabor} {et~al.}(2009){Gabor}, {Impey}, {Jahnke}, {Simmons}, {Trump},
  {Koekemoer}, {Brusa}, {Cappelluti}, {Schinnerer}, {Smol{\v c}i{\'c}},
  {Salvato}, {Rhodes}, {Mobasher}, {Capak}, {Massey}, {Leauthaud}, \&
  {Scoville}}]{2009ApJ...691..705G}
{Gabor}, J.~M., {Impey}, C.~D., {Jahnke}, K., {et~al.} 2009, \apj, 691, 705

\bibitem[{{Georgakakis} {et~al.}(2008){Georgakakis}, {Nandra}, {Yan},
  {Willner}, {Lotz}, {Pierce}, {Cooper}, {Laird}, {Koo}, {Barmby}, {Newman},
  {Primack}, \& {Coil}}]{2008MNRAS.385.2049G}
{Georgakakis}, A., {Nandra}, K., {Yan}, R., {et~al.} 2008, \mnras, 385, 2049

\bibitem[{{Goulding} {et~al.}(2014){Goulding}, {Forman}, {Hickox}, {Jones},
  {Murray}, {Paggi}, {Ashby}, {Coil}, {Cooper}, {Huang}, {Kraft}, {Newmerger
  fraction depends on the total and star-forming infrared luminosity, rather
  than the decomposed AGN infrared luminosity.man}, {Weiner}, \&
  {Willner}}]{2014ApJ...783...40G}
{Goulding}, A.~D., {Forman}, W.~R., {Hickox}, R.~C., {et~al.} 2014, \apj, 783,
  40

\bibitem[{{Grogin} {et~al.}(2005){Grogin}, {Conselice}, {Chatzichristou},
  {Alexander}, {Bauer}, {Hornschemeier}, {Jogee}, {Koekemoer}, {Laidler},
  {Livio}, {Lucas}, {Paolillo}, {Ravindranath}, {Schreier}, {Simmons}, \&
  {Urry}}]{2005ApJ...627L..97G}
{Grogin}, N.~A., {Conselice}, C.~J., {Chatzichristou}, E., {et~al.} 2005,
  \apjl, 627, L97

\bibitem[{{Harrison} {et~al.}(2012){Harrison}, {Alexander}, {Mullaney},
  {Altieri}, {Coia}, {Charmandaris}, {Daddi}, {Dannerbauer}, {Dasyra}, {Del
  Moro}, {Dickinson}, {Hickox}, {Ivison}, {Kartaltepe}, {Le Floc'h}, {Leiton},
  {Magnelli}, {Popesso}, {Rovilos}, {Rosario}, \&
  {Swinbank}}]{2012ApJ...760L..15H}
{Harrison}, C.~M., {Alexander}, D.~M., {Mullaney}, J.~R., {et~al.} 2012, \apjl,
  760, L15

\bibitem[{{Hern{\'a}n-Caballero} {et~al.}(2015){Hern{\'a}n-Caballero},
  {Alonso-Herrero}, {Hatziminaoglou}, {Spoon}, {Ramos Almeida}, {D{\'{\i}}az
  Santos}, {H{\"o}nig}, {Gonz{\'a}lez-Mart{\'{\i}}n}, \&
  {Esquej}}]{2015ApJ...803..109H}
{Hern{\'a}n-Caballero}, A., {Alonso-Herrero}, A., {Hatziminaoglou}, E.,
  {et~al.} 2015, \apj, 803, 109

\bibitem[{{Hickox} {et~al.}(2014){Hickox}, {Mullaney}, {Alexander}, {Chen},
  {Civano}, {Goulding}, \& {Hainline}}]{2014ApJ...782....9H}
{Hickox}, R.~C., {Mullaney}, J.~R., {Alexander}, D.~M., {et~al.} 2014, \apj,
  782, 9

\bibitem[{{Hickox} {et~al.}(2009){Hickox}, {Jones}, {Forman}, {Murray},
  {Kochanek}, {Eisenstein}, {Jannuzi}, {Dey}, {Brown}, {Stern}, {Eisenhardt},
  {Gorjian}, {Brodwin}, {Narayan}, {Cool}, {Kenter}, {Caldwell}, \&
  {Anderson}}]{2009ApJ...696..891H}
{Hickox}, R.~C., {Jones}, C., {Forman}, W.~R., {et~al.} 2009, \apj, 696, 891

\bibitem[{{Hickox} {et~al.}(2011){Hickox}, {Myers}, {Brodwin}, {Alexander},
  {Forman}, {Jones}, {Murray}, {Brown}, {Cool}, {Kochanek}, {Dey}, {Jannuzi},
  {Eisenstein}, {Assef}, {Eisenhardt}, {Gorjian}, {Stern}, {Le Floc'h},
  {Caldwell}, {Goulding}, \& {Mullaney}}]{2011ApJ...731..117H}
{Hickox}, R.~C., {Myers}, A.~D., {Brodwin}, M., {et~al.} 2011, \apj, 731, 117

\bibitem[{{Hopkins} {et~al.}(2008){Hopkins}, {Hernquist}, {Cox}, \& {Kere{\v
  s}}}]{2008ApJS..175..356H}
{Hopkins}, P.~F., {Hernquist}, L., {Cox}, T.~J., \& {Kere{\v s}}, D. 2008,
  \apjs, 175, 356

\bibitem[{{Ilbert} {et~al.}(2009){Ilbert}, {Capak}, {Salvato}, {Aussel},
  {McCracken}, {Sanders}, {Scoville}, {Kartaltepe}, {Arnouts}, {Le Floc'h},
  {Mobasher}, {Taniguchi}, {Lamareille}, {Leauthaud}, {Sasaki}, {Thompson},
  {Zamojski}, {Zamorani}, {Bardelli}, {Bolzonella}, {Bongiorno}, {Brusa},
  {Caputi}, {Carollo}, {Contini}, {Cook}, {Coppa}, {Cucciati}, {de la Torre},
  {de Ravel}, {Franzetti}, {Garilli}, {Hasinger}, {Iovino}, {Kampczyk},
  {Kneib}, {Knobel}, {Kovac}, {Le Borgne}, {Le Brun}, {F{\`e}vre}, {Lilly},
  {Looper}, {Maier}, {Mainieri}, {Mellier}, {Mignoli}, {Murayama}, {Pell{\`o}},
  {Peng}, {P{\'e}rez-Montero}, {Renzini}, {Ricciardelli}, {Schiminovich},
  {Scodeggio}, {Shioya}, {Silverman}, {Surace}, {Tanaka}, {Tasca}, {Tresse},
  {Vergani}, \& {Zucca}}]{2009ApJ...690.1236I}
{Ilbert}, O., {Capak}, P., {Salvato}, M., {et~al.} 2009, \apj, 690, 1236

\bibitem[{{Ilbert} {et~al.}(2013){Ilbert}, {McCracken}, {Le F{\`e}vre},
  {Capak}, {Dunlop}, {Karim}, {Renzini}, {Caputi}, {Boissier}, {Arnouts},
  {Aussel}, {Comparat}, {Guo}, {Hudelot}, {Kartaltepe}, {Kneib}, {Krogager},
  {Le Floc'h}, {Lilly}, {Mellier}, {Milvang-Jensen}, {Moutard}, {Onodera},
  {Richard}, {Salvato}, {Sanders}, {Scoville}, {Silverman}, {Taniguchi},
  {Tasca}, {Thomas}, {Toft}, {Tresse}, {Vergani}, {Wolk}, \&
  {Zirm}}]{2013A&A...556A..55I}
{Ilbert}, O., {McCracken}, H.~J., {Le F{\`e}vre}, O., {et~al.} 2013, \aap, 556,
  A55

\bibitem[{{Ilbert} {et~al.}(2015){Ilbert}, {Arnouts}, {Le Floc'h}, {Aussel},
  {Bethermin}, {Capak}, {Hsieh}, {Kajisawa}, {Karim}, {Le F{\`e}vre}, {Lee},
  {Lilly}, {McCracken}, {Michel-Dansac}, {Moutard}, {Renzini}, {Salvato},
  {Sanders}, {Scoville}, {Sheth}, {Silverman}, {Smol{\v c}i{\'c}}, {Taniguchi},
  \& {Tresse}}]{2015A&A...579A...2I}
{Ilbert}, O., {Arnouts}, S., {Le Floc'h}, E., {et~al.} 2015, \aap, 579, A2

\bibitem[{{Juneau} {et~al.}(2013){Juneau}, {Dickinson}, {Bournaud},
  {Alexander}, {Daddi}, {Mullaney}, {Magnelli}, {Kartaltepe}, {Hwang},
  {Willner}, {Coil}, {Rosario}, {Trump}, {Weiner}, {Willmer}, {Cooper},
  {Elbaz}, {Faber}, {Frayer}, {Kocevski}, {Laird}, {Monkiewicz}, {Nandra},
  {Newman}, {Salim}, \& {Symeonidis}}]{2013ApJ...764..176J}
{Juneau}, S., {Dickinson}, M., {Bournaud}, F., {et~al.} 2013, \apj, 764, 176

\bibitem[{{Kartaltepe} {et~al.}(2012){Kartaltepe}, {Dickinson}, {Alexander},
  {Bell}, {Dahlen}, {Elbaz}, {Faber}, {Lotz}, {McIntosh}, {Wiklind}, {Altieri},
  {Aussel}, {Bethermin}, {Bournaud}, {Charmandaris}, {Conselice}, {Cooray},
  {Dannerbauer}, {Dav{\'e}}, {Dunlop}, {Dekel}, {Ferguson}, {Grogin}, {Hwang},
  {Ivison}, {Kocevski}, {Koekemoer}, {Koo}, {Lai}, {Leiton}, {Lucas}, {Lutz},
  {Magdis}, {Magnelli}, {Morrison}, {Mozena}, {Mullaney}, {Newman}, {Pope},
  {Popesso}, {van der Wel}, {Weiner}, \& {Wuyts}}]{2012ApJ...757...23K}
{Kartaltepe}, J.~S., {Dickinson}, M., {Alexander}, D.~M., {et~al.} 2012, \apj,
  757, 23

\bibitem[{{Kirkpatrick} {et~al.}(2015){Kirkpatrick}, {Pope}, {Sajina},
  {Roebuck}, {Yan}, {Armus}, {D{\'{\i}}az-Santos}, \&
  {Stierwalt}}]{2015ApJ...814....9K}
{Kirkpatrick}, A., {Pope}, A., {Sajina}, A., {et~al.} 2015, \apj, 814, 9

\bibitem[{{Kirkpatrick} {et~al.}(2012){Kirkpatrick}, {Pope}, {Alexander},
  {Charmandaris}, {Daddi}, {Dickinson}, {Elbaz}, {Gabor}, {Hwang}, {Ivison},
  {Mullaney}, {Pannella}, {Scott}, {Altieri}, {Aussel}, {Bournaud}, {Buat},
  {Coia}, {Dannerbauer}, {Dasyra}, {Kartaltepe}, {Leiton}, {Lin}, {Magdis},
  {Magnelli}, {Morrison}, {Popesso}, \& {Valtchanov}}]{2012ApJ...759..139K}
{Kirkpatrick}, A., {Pope}, A., {Alexander}, D.~M., {et~al.} 2012, \apj, 759,
  139

\bibitem[{{Kirkpatrick} {et~al.}(2013){Kirkpatrick}, {Pope}, {Charmandaris},
  {Daddi}, {Elbaz}, {Hwang}, {Pannella}, {Scott}, {Altieri}, {Aussel}, {Coia},
  {Dannerbauer}, {Dasyra}, {Dickinson}, {Kartaltepe}, {Leiton}, {Magdis},
  {Magnelli}, {Popesso}, \& {Valtchanov}}]{2013ApJ...763..123K}
{Kirkpatrick}, A., {Pope}, A., {Charmandaris}, V., {et~al.} 2013, \apj, 763,
  123

\bibitem[{{Kocevski} {et~al.}(2012){Kocevski}, {Faber}, {Mozena}, {Koekemoer},
  {Nandra}, {Rangel}, {Laird}, {Brusa}, {Wuyts}, {Trump}, {Koo}, {Somerville},
  {Bell}, {Lotz}, {Alexander}, {Bournaud}, {Conselice}, {Dahlen}, {Dekel},
  {Donley}, {Dunlop}, {Finoguenov}, {Georgakakis}, {Giavalisco}, {Guo},
  {Grogin}, {Hathi}, {Juneau}, {Kartaltepe}, {Lucas}, {McGrath}, {McIntosh},
  {Mobasher}, {Robaina}, {Rosario}, {Straughn}, {van der Wel}, \&
  {Villforth}}]{2012ApJ...744..148K}
{Kocevski}, D.~D., {Faber}, S.~M., {Mozena}, M., {et~al.} 2012, \apj, 744, 148

\bibitem[{{Kocevski} {et~al.}(2015){Kocevski}, {Brightman}, {Nandra},
  {Koekemoer}, {Salvato}, {Aird}, {Bell}, {Hsu}, {Kartaltepe}, {Koo}, {Lotz},
  {McIntosh}, {Mozena}, {Rosario}, \& {Trump}}]{2015ApJ...814..104K}
{Kocevski}, D.~D., {Brightman}, M., {Nandra}, K., {et~al.} 2015, \apj, 814, 104

\bibitem[{{Lacy} {et~al.}(2004){Lacy}, {Storrie-Lombardi}, {Sajina},
  {Appleton}, {Armus}, {Chapman}, {Choi}, {Fadda}, {Fang}, {Frayer},
  {Heinrichsen}, {Helou}, {Im}, {Marleau}, {Masci}, {Shupe}, {Soifer},
  {Surace}, {Teplitz}, {Wilson}, \& {Yan}}]{2004ApJS..154..166L}
{Lacy}, M., {Storrie-Lombardi}, L.~J., {Sajina}, A., {et~al.} 2004, \apjs, 154,
  166

\bibitem[{{Laigle} {et~al.}(2016){Laigle}, {McCracken}, {Ilbert}, {Hsieh},
  {Davidzon}, {Capak}, {Hasinger}, {Silverman}, {Pichon}, {Coupon}, {Aussel},
  {Le Borgne}, {Caputi}, {Cassata}, {Chang}, {Civano}, {Dunlop}, {Fynbo},
  {Kartaltepe}, {Koekemoer}, {Le F{\'e}vre}, {Le Floc'h}, {Leauthaud}, {Lilly},
  {Lin}, {Marchesi}, {Milvang-Jensen}, {Salvato}, {Sanders}, {Scoville},
  {Smolcic}, {Stockmann}, {Taniguchi}, {Tasca}, {Toft}, {Vaccari}, \&
  {Zabl}}]{2016ApJS..224...24L}
{Laigle}, C., {McCracken}, H.~J., {Ilbert}, O., {et~al.} 2016, \apjs, 224, 24

\bibitem[{{Lanzuisi} {et~al.}(2015{\natexlab{a}}){Lanzuisi}, {Ranalli},
  {Georgantopoulos}, {Georgakakis}, {Delvecchio}, {Akylas}, {Berta},
  {Bongiorno}, {Brusa}, {Cappelluti}, {Civano}, {Comastri}, {Gilli},
  {Gruppioni}, {Hasinger}, {Iwasawa}, {Koekemoer}, {Lusso}, {Marchesi},
  {Mainieri}, {Merloni}, {Mignoli}, {Piconcelli}, {Pozzi}, {Rosario},
  {Salvato}, {Silverman}, {Trakhtenbrot}, {Vignali}, \&
  {Zamorani}}]{2015A&A...573A.137L}
{Lanzuisi}, G., {Ranalli}, P., {Georgantopoulos}, I., {et~al.}
  2015{\natexlab{a}}, \aap, 573, A137

\bibitem[{{Lanzuisi} {et~al.}(2015{\natexlab{b}}){Lanzuisi}, {Perna},
  {Delvecchio}, {Berta}, {Brusa}, {Cappelluti}, {Comastri}, {Gilli},
  {Gruppioni}, {Mignoli}, {Pozzi}, {Vietri}, {Vignali}, \&
  {Zamorani}}]{2015A&A...578A.120L}
{Lanzuisi}, G., {Perna}, M., {Delvecchio}, I., {et~al.} 2015{\natexlab{b}},
  \aap, 578, A120

\bibitem[{{Le Floc'h} {et~al.}(2009){Le Floc'h}, {Aussel}, {Ilbert},
  {Riguccini}, {Frayer}, {Salvato}, {Arnouts}, {Surace}, {Feruglio},
  {Rodighiero}, {Capak}, {Kartaltepe}, {Heinis}, {Sheth}, {Yan}, {McCracken},
  {Thompson}, {Sanders}, {Scoville}, \& {Koekemoer}}]{2009ApJ...703..222L}
{Le Floc'h}, E., {Aussel}, H., {Ilbert}, O., {et~al.} 2009, \apj, 703, 222

\bibitem[{{Lotz} {et~al.}(2011){Lotz}, {Jonsson}, {Cox}, {Croton}, {Primack},
  {Somerville}, \& {Stewart}}]{2011ApJ...742..103L}
{Lotz}, J.~M., {Jonsson}, P., {Cox}, T.~J., {et~al.} 2011, \apj, 742, 103

\bibitem[{{Lotz} {et~al.}(2004){Lotz}, {Primack}, \&
  {Madau}}]{2004AJ....128..163L}
{Lotz}, J.~M., {Primack}, J., \& {Madau}, P. 2004, \aj, 128, 163

\bibitem[{{Lotz} {et~al.}(2008){Lotz}, {Davis}, {Faber}, {Guhathakurta},
  {Gwyn}, {Huang}, {Koo}, {Le Floc'h}, {Lin}, {Newman}, {Noeske}, {Papovich},
  {Willmer}, {Coil}, {Conselice}, {Cooper}, {Hopkins}, {Metevier}, {Primack},
  {Rieke}, \& {Weiner}}]{2008ApJ...672..177L}
{Lotz}, J.~M., {Davis}, M., {Faber}, S.~M., {et~al.} 2008, \apj, 672, 177

\bibitem[{{Lusso} {et~al.}(2011){Lusso}, {Comastri}, {Vignali}, {Zamorani},
  {Treister}, {Sanders}, {Bolzonella}, {Bongiorno}, {Brusa}, {Civano}, {Gilli},
  {Mainieri}, {Nair}, {Aller}, {Carollo}, {Koekemoer}, {Merloni}, \&
  {Trump}}]{2011A&A...534A.110L}
{Lusso}, E., {Comastri}, A., {Vignali}, C., {et~al.} 2011, \aap, 534, A110

\bibitem[{{Lusso} {et~al.}(2013){Lusso}, {Hennawi}, {Comastri}, {Zamorani},
  {Richards}, {Vignali}, {Treister}, {Schawinski}, {Salvato}, \&
  {Gilli}}]{2013ApJ...777...86L}
{Lusso}, E., {Hennawi}, J.~F., {Comastri}, A., {et~al.} 2013, \apj, 777, 86

\bibitem[{{Mahoro} {et~al.}(2017){Mahoro}, {Povi{\'c}}, \&
  {Nkundabakura}}]{2017arXiv170700254M}
{Mahoro}, A., {Povi{\'c}}, M., \& {Nkundabakura}, P. 2017, ArXiv e-prints,
  arXiv:1707.00254

\bibitem[{{Marchesi} {et~al.}(2016){Marchesi}, {Civano}, {Elvis}, {Salvato},
  {Brusa}, {Comastri}, {Gilli}, {Hasinger}, {Lanzuisi}, {Miyaji}, {Treister},
  {Urry}, {Vignali}, {Zamorani}, {Allevato}, {Cappelluti}, {Cardamone},
  {Finoguenov}, {Griffiths}, {Karim}, {Laigle}, {LaMassa}, {Jahnke}, {Ranalli},
  {Schawinski}, {Schinnerer}, {Silverman}, {Smolcic}, {Suh}, \&
  {Trakhtenbrot}}]{2016ApJ...817...34M}
{Marchesi}, S., {Civano}, F., {Elvis}, M., {et~al.} 2016, \apj, 817, 34

\bibitem[{{Mendez} {et~al.}(2013){Mendez}, {Coil}, {Aird}, {Diamond-Stanic},
  {Moustakas}, {Blanton}, {Cool}, {Eisenstein}, {Wong}, \&
  {Zhu}}]{2013ApJ...770...40M}
{Mendez}, A.~J., {Coil}, A.~L., {Aird}, J., {et~al.} 2013, \apj, 770, 40

\bibitem[{{Merloni} {et~al.}(2014){Merloni}, {Bongiorno}, {Brusa}, {Iwasawa},
  {Mainieri}, {Magnelli}, {Salvato}, {Berta}, {Cappelluti}, {Comastri},
  {Fiore}, {Gilli}, {Koekemoer}, {Le Floc'h}, {Lusso}, {Lutz}, {Miyaji},
  {Pozzi}, {Riguccini}, {Rosario}, {Silverman}, {Symeonidis}, {Treister},
  {Vignali}, \& {Zamorani}}]{2014MNRAS.437.3550M}
{Merloni}, A., {Bongiorno}, A., {Brusa}, M., {et~al.} 2014, \mnras, 437, 3550

\bibitem[{{Messias} {et~al.}(2012){Messias}, {Afonso}, {Salvato}, {Mobasher},
  \& {Hopkins}}]{2012ApJ...754..120M}
{Messias}, H., {Afonso}, J., {Salvato}, M., {Mobasher}, B., \& {Hopkins}, A.~M.
  2012, \apj, 754, 120

\bibitem[{{Messias} {et~al.}(2014){Messias}, {Afonso}, {Salvato}, {Mobasher},
  \& {Hopkins}}]{2014A&A...562A.144M}
{Messias}, H., {Afonso}, J.~M., {Salvato}, M., {Mobasher}, B., \& {Hopkins},
  A.~M. 2014, \aap, 562, A144

\bibitem[{{Mullaney} {et~al.}(2011){Mullaney}, {Alexander}, {Goulding}, \&
  {Hickox}}]{2011MNRAS.414.1082M}
{Mullaney}, J.~R., {Alexander}, D.~M., {Goulding}, A.~D., \& {Hickox}, R.~C.
  2011, \mnras, 414, 1082

\bibitem[{{Mullaney} {et~al.}(2012){Mullaney}, {Pannella}, {Daddi},
  {Alexander}, {Elbaz}, {Hickox}, {Bournaud}, {Altieri}, {Aussel}, {Coia},
  {Dannerbauer}, {Dasyra}, {Dickinson}, {Hwang}, {Kartaltepe}, {Leiton},
  {Magdis}, {Magnelli}, {Popesso}, {Valtchanov}, {Bauer}, {Brandt}, {Del Moro},
  {Hanish}, {Ivison}, {Juneau}, {Luo}, {Lutz}, {Sargent}, {Scott}, \&
  {Xue}}]{2012MNRAS.419...95M}
{Mullaney}, J.~R., {Pannella}, M., {Daddi}, E., {et~al.} 2012, \mnras, 419, 95

\bibitem[{{Noeske} {et~al.}(2007){Noeske}, {Weiner}, {Faber}, {Papovich},
  {Koo}, {Somerville}, {Bundy}, {Conselice}, {Newman}, {Schiminovich}, {Le
  Floc'h}, {Coil}, {Rieke}, {Lotz}, {Primack}, {Barmby}, {Cooper}, {Davis},
  {Ellis}, {Fazio}, {Guhathakurta}, {Huang}, {Kassin}, {Martin}, {Phillips},
  {Rich}, {Small}, {Willmer}, \& {Wilson}}]{2007ApJ...660L..43N}
{Noeske}, K.~G., {Weiner}, B.~J., {Faber}, S.~M., {et~al.} 2007, \apjl, 660,
  L43

\bibitem[{Pedregosa {et~al.}(2011)Pedregosa, Varoquaux, Gramfort, Michel,
  Thirion, Grisel, Blondel, Prettenhofer, Weiss, Dubourg, Vanderplas, Passos,
  Cournapeau, Brucher, Perrot, \& Duchesnay}]{scikit-learn}
Pedregosa, F., Varoquaux, G., Gramfort, A., {et~al.} 2011, Journal of Machine
  Learning Research, 12, 2825

\bibitem[{{Peng} {et~al.}(2002){Peng}, {Ho}, {Impey}, \&
  {Rix}}]{2002AJ....124..266P}
{Peng}, C.~Y., {Ho}, L.~C., {Impey}, C.~D., \& {Rix}, H.-W. 2002, \aj, 124, 266

\bibitem[{{Peng} {et~al.}(2010){Peng}, {Ho}, {Impey}, \&
  {Rix}}]{2010AJ....139.2097P}
---. 2010, \aj, 139, 2097

\bibitem[{{Perna} {et~al.}(2015){Perna}, {Brusa}, {Salvato}, {Cresci},
  {Lanzuisi}, {Berta}, {Delvecchio}, {Fiore}, {Lutz}, {Le Floc'h}, {Mainieri},
  \& {Riguccini}}]{2015A&A...583A..72P}
{Perna}, M., {Brusa}, M., {Salvato}, M., {et~al.} 2015, \aap, 583, A72

\bibitem[{{Peth} {et~al.}(2016){Peth}, {Lotz}, {Freeman}, {McPartland},
  {Mortazavi}, {Snyder}, {Barro}, {Grogin}, {Guo}, {Hemmati}, {Kartaltepe},
  {Kocevski}, {Koekemoer}, {McIntosh}, {Nayyeri}, {Papovich}, {Primack}, \&
  {Simons}}]{2016MNRAS.458..963P}
{Peth}, M.~A., {Lotz}, J.~M., {Freeman}, P.~E., {et~al.} 2016, \mnras, 458, 963

\bibitem[{{Pierce} {et~al.}(2007){Pierce}, {Lotz}, {Laird}, {Lin}, {Nandra},
  {Primack}, {Faber}, {Barmby}, {Park}, {Willner}, {Gwyn}, {Koo}, {Coil},
  {Cooper}, {Georgakakis}, {Koekemoer}, {Noeske}, {Weiner}, \&
  {Willmer}}]{2007ApJ...660L..19P}
{Pierce}, C.~M., {Lotz}, J.~M., {Laird}, E.~S., {et~al.} 2007, \apjl, 660, L19

\bibitem[{{Polletta} {et~al.}(2007){Polletta}, {Tajer}, {Maraschi},
  {Trinchieri}, {Lonsdale}, {Chiappetti}, {Andreon}, {Pierre}, {Le F{\`e}vre},
  {Zamorani}, {Maccagni}, {Garcet}, {Surdej}, {Franceschini}, {Alloin},
  {Shupe}, {Surace}, {Fang}, {Rowan-Robinson}, {Smith}, \&
  {Tresse}}]{2007ApJ...663...81P}
{Polletta}, M., {Tajer}, M., {Maraschi}, L., {et~al.} 2007, \apj, 663, 81

\bibitem[{{Prieto} {et~al.}(2010){Prieto}, {Reunanen}, {Tristram}, {Neumayer},
  {Fernandez-Ontiveros}, {Orienti}, \& {Meisenheimer}}]{2010MNRAS.402..724P}
{Prieto}, M.~A., {Reunanen}, J., {Tristram}, K.~R.~W., {et~al.} 2010, \mnras,
  402, 724

\bibitem[{{Richards} {et~al.}(2006){Richards}, {Lacy}, {Storrie-Lombardi},
  {Hall}, {Gallagher}, {Hines}, {Fan}, {Papovich}, {Vanden Berk}, {Trammell},
  {Schneider}, {Vestergaard}, {York}, {Jester}, {Anderson}, {Budav{\'a}ri}, \&
  {Szalay}}]{2006ApJS..166..470R}
{Richards}, G.~T., {Lacy}, M., {Storrie-Lombardi}, L.~J., {et~al.} 2006, \apjs,
  166, 470

\bibitem[{{Rosario} {et~al.}(2013){Rosario}, {Santini}, {Lutz}, {Netzer},
  {Bauer}, {Berta}, {Magnelli}, {Popesso}, {Alexander}, {Brandt}, {Genzel},
  {Maiolino}, {Mullaney}, {Nordon}, {Saintonge}, {Tacconi}, \&
  {Wuyts}}]{2013ApJ...771...63R}
{Rosario}, D.~J., {Santini}, P., {Lutz}, D., {et~al.} 2013, \apj, 771, 63

\bibitem[{{Rosario} {et~al.}(2015){Rosario}, {McIntosh}, {van der Wel},
  {Kartaltepe}, {Lang}, {Santini}, {Wuyts}, {Lutz}, {Rafelski}, {Villforth},
  {Alexander}, {Bauer}, {Bell}, {Berta}, {Brandt}, {Conselice}, {Dekel},
  {Faber}, {Ferguson}, {Genzel}, {Grogin}, {Kocevski}, {Koekemoer}, {Koo},
  {Lotz}, {Magnelli}, {Maiolino}, {Mozena}, {Mullaney}, {Papovich}, {Popesso},
  {Tacconi}, {Trump}, {Avadhuta}, {Bassett}, {Bell}, {Bernyk}, {Bournaud},
  {Cassata}, {Cheung}, {Croton}, {Donley}, {DeGroot}, {Guedes}, {Hathi},
  {Herrington}, {Hilton}, {Lai}, {Lani}, {Martig}, {McGrath}, {Mutch},
  {Mortlock}, {McPartland}, {O'Leary}, {Peth}, {Pillepich}, {Poole}, {Snyder},
  {Straughn}, {Telford}, {Tonini}, \& {Wandro}}]{2015A&A...573A..85R}
{Rosario}, D.~J., {McIntosh}, D.~H., {van der Wel}, A., {et~al.} 2015, \aap,
  573, A85

\bibitem[{{Salvato} {et~al.}(2011){Salvato}, {Ilbert}, {Hasinger}, {Rau},
  {Civano}, {Zamorani}, {Brusa}, {Elvis}, {Vignali}, {Aussel}, {Comastri},
  {Fiore}, {Le Floc'h}, {Mainieri}, {Bardelli}, {Bolzonella}, {Bongiorno},
  {Capak}, {Caputi}, {Cappelluti}, {Carollo}, {Contini}, {Garilli}, {Iovino},
  {Fotopoulou}, {Fruscione}, {Gilli}, {Halliday}, {Kneib}, {Kakazu},
  {Kartaltepe}, {Koekemoer}, {Kovac}, {Ideue}, {Ikeda}, {Impey}, {Le Fevre},
  {Lamareille}, {Lanzuisi}, {Le Borgne}, {Le Brun}, {Lilly}, {Maier},
  {Manohar}, {Masters}, {McCracken}, {Messias}, {Mignoli}, {Mobasher}, {Nagao},
  {Pello}, {Puccetti}, {Perez-Montero}, {Renzini}, {Sargent}, {Sanders},
  {Scodeggio}, {Scoville}, {Shopbell}, {Silvermann}, {Taniguchi}, {Tasca},
  {Tresse}, {Trump}, \& {Zucca}}]{2011ApJ...742...61S}
{Salvato}, M., {Ilbert}, O., {Hasinger}, G., {et~al.} 2011, \apj, 742, 61

\bibitem[{{Sanders} {et~al.}(1988){Sanders}, {Soifer}, {Elias}, {Madore},
  {Matthews}, {Neugebauer}, \& {Scoville}}]{1988ApJ...325...74S}
{Sanders}, D.~B., {Soifer}, B.~T., {Elias}, J.~H., {et~al.} 1988, \apj, 325, 74

\bibitem[{{Santini} {et~al.}(2012){Santini}, {Rosario}, {Shao}, {Lutz},
  {Maiolino}, {Alexander}, {Altieri}, {Andreani}, {Aussel}, {Bauer}, {Berta},
  {Bongiovanni}, {Brandt}, {Brusa}, {Cepa}, {Cimatti}, {Daddi}, {Elbaz},
  {Fontana}, {F{\"o}rster Schreiber}, {Genzel}, {Grazian}, {Le Floc'h},
  {Magnelli}, {Mainieri}, {Nordon}, {P{\'e}rez Garcia}, {Poglitsch}, {Popesso},
  {Pozzi}, {Riguccini}, {Rodighiero}, {Salvato}, {Sanchez-Portal}, {Sturm},
  {Tacconi}, {Valtchanov}, \& {Wuyts}}]{2012A&A...540A.109S}
{Santini}, P., {Rosario}, D.~J., {Shao}, L., {et~al.} 2012, \aap, 540, A109

\bibitem[{{Sargent} {et~al.}(2007){Sargent}, {Carollo}, {Lilly}, {Scarlata},
  {Feldmann}, {Kampczyk}, {Koekemoer}, {Scoville}, {Kneib}, {Leauthaud},
  {Massey}, {Rhodes}, {Tasca}, {Capak}, {McCracken}, {Porciani}, {Renzini},
  {Taniguchi}, {Thompson}, \& {Sheth}}]{2007ApJS..172..434S}
{Sargent}, M.~T., {Carollo}, C.~M., {Lilly}, S.~J., {et~al.} 2007, \apjs, 172,
  434

\bibitem[{{Scarlata} {et~al.}(2007){Scarlata}, {Carollo}, {Lilly}, {Sargent},
  {Feldmann}, {Kampczyk}, {Porciani}, {Koekemoer}, {Scoville}, {Kneib},
  {Leauthaud}, {Massey}, {Rhodes}, {Tasca}, {Capak}, {Maier}, {McCracken},
  {Mobasher}, {Renzini}, {Taniguchi}, {Thompson}, {Sheth}, {Ajiki}, {Aussel},
  {Murayama}, {Sanders}, {Sasaki}, {Shioya}, \&
  {Takahashi}}]{2007ApJS..172..406S}
{Scarlata}, C., {Carollo}, C.~M., {Lilly}, S., {et~al.} 2007, \apjs, 172, 406

\bibitem[{{Schawinski} {et~al.}(2011){Schawinski}, {Urry}, {Treister},
  {Simmons}, {Natarajan}, \& {Glikman}}]{2011ApJ...743L..37S}
{Schawinski}, K., {Urry}, M., {Treister}, E., {et~al.} 2011, \apjl, 743, L37

\bibitem[{{Schreiber} {et~al.}(2015){Schreiber}, {Pannella}, {Elbaz},
  {B{\'e}thermin}, {Inami}, {Dickinson}, {Magnelli}, {Wang}, {Aussel}, {Daddi},
  {Juneau}, {Shu}, {Sargent}, {Buat}, {Faber}, {Ferguson}, {Giavalisco},
  {Koekemoer}, {Magdis}, {Morrison}, {Papovich}, {Santini}, \&
  {Scott}}]{2015A&A...575A..74S}
{Schreiber}, C., {Pannella}, M., {Elbaz}, D., {et~al.} 2015, \aap, 575, A74

\bibitem[{{Scoville} {et~al.}(2007){Scoville}, {Aussel}, {Brusa}, {Capak},
  {Carollo}, {Elvis}, {Giavalisco}, {Guzzo}, {Hasinger}, {Impey}, {Kneib},
  {LeFevre}, {Lilly}, {Mobasher}, {Renzini}, {Rich}, {Sanders}, {Schinnerer},
  {Schminovich}, {Shopbell}, {Taniguchi}, \& {Tyson}}]{2007ApJS..172....1S}
{Scoville}, N., {Aussel}, H., {Brusa}, M., {et~al.} 2007, \apjs, 172, 1

\bibitem[{{Simmons} {et~al.}(2012){Simmons}, {Urry}, {Schawinski}, {Cardamone},
  \& {Glikman}}]{2012ApJ...761...75S}
{Simmons}, B.~D., {Urry}, C.~M., {Schawinski}, K., {Cardamone}, C., \&
  {Glikman}, E. 2012, \apj, 761, 75

\bibitem[{{Stanley} {et~al.}(2015){Stanley}, {Harrison}, {Alexander},
  {Swinbank}, {Aird}, {Del Moro}, {Hickox}, \&
  {Mullaney}}]{2015MNRAS.453..591S}
{Stanley}, F., {Harrison}, C.~M., {Alexander}, D.~M., {et~al.} 2015, \mnras,
  453, 591

\bibitem[{{Stern} {et~al.}(2005){Stern}, {Eisenhardt}, {Gorjian}, {Kochanek},
  {Caldwell}, {Eisenstein}, {Brodwin}, {Brown}, {Cool}, {Dey}, {Green},
  {Jannuzi}, {Murray}, {Pahre}, \& {Willner}}]{2005ApJ...631..163S}
{Stern}, D., {Eisenhardt}, P., {Gorjian}, V., {et~al.} 2005, \apj, 631, 163

\bibitem[{{Suh} {et~al.}(2017){Suh}, {Civano}, {Hasinger}, {Lusso}, {Lanzuisi},
  {Marchesi}, {Trakhtenbrot}, {Allevato}, {Cappelluti}, {Capak}, {Elvis},
  {Griffiths}, {Laigle}, {Lira}, {Riguccini}, {Rosario}, {Salvato},
  {Schawinski}, \& {Vignali}}]{2017ApJ...841..102S}
{Suh}, H., {Civano}, F., {Hasinger}, G., {et~al.} 2017, \apj, 841, 102

\bibitem[{{Tacchella} {et~al.}(2016){Tacchella}, {Dekel}, {Carollo},
  {Ceverino}, {DeGraf}, {Lapiner}, {Mandelker}, \& {Primack
  Joel}}]{2016MNRAS.457.2790T}
{Tacchella}, S., {Dekel}, A., {Carollo}, C.~M., {et~al.} 2016, \mnras, 457,
  2790

\bibitem[{{Toba} {et~al.}(2017){Toba}, {Nagao}, {Kajisawa}, {Oogi}, {Akiyama},
  {Ikeda}, {Coupon}, {Strauss}, {Wang}, {Tanaka}, {Niida}, {Imanishi}, {Lee},
  {Matsuhara}, {Matsuoka}, {Onoue}, {Terashima}, {Ueda}, {Harikane},
  {Komiyama}, {Miyazaki}, {Noboriguchi}, \& {Usuda}}]{2017ApJ...835...36T}
{Toba}, Y., {Nagao}, T., {Kajisawa}, M., {et~al.} 2017, \apj, 835, 36

\bibitem[{{Treister} {et~al.}(2012){Treister}, {Schawinski}, {Urry}, \&
  {Simmons}}]{2012ApJ...758L..39T}
{Treister}, E., {Schawinski}, K., {Urry}, C.~M., \& {Simmons}, B.~D. 2012,
  \apjl, 758, L39

\bibitem[{{Treister} {et~al.}(2009){Treister}, {Urry}, \&
  {Virani}}]{2009ApJ...696..110T}
{Treister}, E., {Urry}, C.~M., \& {Virani}, S. 2009, \apj, 696, 110

\bibitem[{{Ueda} {et~al.}(2003){Ueda}, {Akiyama}, {Ohta}, \&
  {Miyaji}}]{2003ApJ...598..886U}
{Ueda}, Y., {Akiyama}, M., {Ohta}, K., \& {Miyaji}, T. 2003, \apj, 598, 886

\bibitem[{{Villforth} {et~al.}(2014){Villforth}, {Hamann}, {Rosario},
  {Santini}, {McGrath}, {van der Wel}, {Chang}, {Guo}, {Dahlen}, {Bell},
  {Conselice}, {Croton}, {Dekel}, {Faber}, {Grogin}, {Hamilton}, {Hopkins},
  {Juneau}, {Kartaltepe}, {Kocevski}, {Koekemoer}, {Koo}, {Lotz}, {McIntosh},
  {Mozena}, {Somerville}, \& {Wild}}]{2014MNRAS.439.3342V}
{Villforth}, C., {Hamann}, F., {Rosario}, D.~J., {et~al.} 2014, \mnras, 439,
  3342

\bibitem[{{Villforth} {et~al.}(2017){Villforth}, {Hamilton}, {Pawlik},
  {Hewlett}, {Rowlands}, {Herbst}, {Shankar}, {Fontana}, {Hamann}, {Koekemoer},
  {Pforr}, {Trump}, \& {Wuyts}}]{2017MNRAS.466..812V}
{Villforth}, C., {Hamilton}, T., {Pawlik}, M.~M., {et~al.} 2017, \mnras, 466,
  812

\bibitem[{{Virani} {et~al.}(2000){Virani}, {De Robertis}, \&
  {VanDalfsen}}]{2000AJ....120.1739V}
{Virani}, S.~N., {De Robertis}, M.~M., \& {VanDalfsen}, M.~L. 2000, \aj, 120,
  1739

\bibitem[{{Zamojski} {et~al.}(2011){Zamojski}, {Yan}, {Dasyra}, {Sajina},
  {Surace}, {Heckman}, \& {Helou}}]{2011ApJ...730..125Z}
{Zamojski}, M., {Yan}, L., {Dasyra}, K., {et~al.} 2011, \apj, 730, 125

\bibitem[{{Zolotov} {et~al.}(2015){Zolotov}, {Dekel}, {Mandelker}, {Tweed},
  {Inoue}, {DeGraf}, {Ceverino}, {Primack}, {Barro}, \&
  {Faber}}]{2015MNRAS.450.2327Z}
{Zolotov}, A., {Dekel}, A., {Mandelker}, N., {et~al.} 2015, \mnras, 450, 2327

\end{thebibliography}


\listofchanges

\end{document}